\documentclass[12pt,a4paper]{article}
\pdfoutput=1
\usepackage{amsmath}
\usepackage{amssymb}
\usepackage{mathrsfs}  
\usepackage{braket}

\usepackage[affil-it]{authblk}

\usepackage[utf8]{inputenc}
\usepackage{url}
\usepackage{hyperref}
\usepackage{graphicx}
\usepackage[dvipsnames]{xcolor}

\renewcommand{\topfraction}{1.0}
\renewcommand{\bottomfraction}{1.0}
\renewcommand{\textfraction}{0.0}
\usepackage[T1,T2A]{fontenc}
\newcommand {\blue} {\color{blue}}

\newcommand {\red} {\color{red}}
\newcommand {\magenta} {\color{Plum}}
\definecolor{NewYellow}{cmyk}{0.2,0.2,1,0.2}
\newcommand {\yellow} {\color{NewYellow}}
\definecolor{NewGreen}{rgb}{0,0.4,0}
\newcommand {\green} {\color{NewGreen}}

\newcommand{\MET}{\textbf{\textit{E}$_{\rm T}^{\rm miss}$}}
\usepackage{subfigure}

\begin{document}

\renewcommand{\topfraction}{1.0}
\renewcommand{\bottomfraction}{1.0}
\renewcommand{\textfraction}{0.0}

\title{Anatomy of the Inert Two Higgs Doublet Model in the light of the LHC and non-LHC Dark Matter Searches}

\author[1,2]{Alexander Belyaev}
\author[3]{Giacomo Cacciapaglia}
\author[4]{Igor P. Ivanov}
\author[2,5]{Felipe Rojas-Abatte}
\author[1,2]{Marc Thomas}

\affil[1]{Rutherford Appleton Laboratory, Didcot, United Kingdom}
\affil[2]{University of Southampton, Southampton, United Kingdom}
\affil[3]{Univ Lyon, Universit\'e Lyon 1, CNRS/IN2P3, IPNL, F-69622, Villeurbanne, France }
\affil[4]{CFTP, Instituto Superior T\'ecnico, Universidade de Lisboa, Lisbon, Portugal}
\affil[5]{Universidad T\'ecnica Federico Santa Mar\'ia, Valpara\'iso, Chile}

\maketitle
\vspace{0.8cm}
\begin{abstract}
The inert Two Higgs Doublet Model (i2HDM) is a theoretically well-motivated  example  of a
minimal consistent Dark Matter (DM)  model  which provides mono-jet, mono-$Z$, mono-Higgs and
Vector-Boson-Fusion+$\MET{}$ signatures at the LHC, complemented by signals in direct and
indirect DM search experiments. In this paper we have performed a detailed analysis of the  
constraints in the full 5D  parameter space of the i2HDM, coming from perturbativity, unitarity, electroweak
precision data, Higgs data from the LHC, DM relic density, direct/indirect DM detection and LHC
mono-jet analysis, as well as  implications of experimental LHC studies on disappearing charged
tracks relevant to high DM mass region. We demonstrate the complementarity of the above
constraints and present projections for future LHC data and direct DM detection
experiments to probe further i2HDM parameter space. The model is implemented into the CalcHEP and
micrOMEGAs packages, which are publicly available at the HEPMDB database, and is ready for a further
exploration in the context of the LHC, relic density and DM direct detection.

\end{abstract}

\newpage
\tableofcontents
\newpage

\section{Introduction}

The evidence for dark matter (DM) is well-established from several independent cosmological observations, 
including galactic rotation curves, cosmic microwave background fits of the WMAP and PLANCK data, 
gravitational lensing, large scale structure of the Universe, as well as interacting galaxy clusters such as the Bullet Cluster.
Despite these large-scale evidences, the microscopic nature of the DM particles remains unknown,
since no experiment so far has been able to claim their detection in the laboratory and probe their properties.
Potentially, DM can be produced at the LHC and probed in the DM direct detection (DD) underground experiments.
The fundamental importance and vast experimental opportunities make the search for and investigation of DM 
one of the key goals in astroparticle physics and high energy physics (HEP), worthy of the intense efforts undertaken by the physics community.

At the other end of the length scale, the Standard Model (SM) of particle physics recently demonstrated its vitality once again.
The scalar boson with mass $m_H \approx 125$ GeV found at the LHC \cite{Chatrchyan:2012xdj,Aad:2012tfa} 
closely resembles, in all its manifestations,
the SM Higgs boson. Since the SM cannot be the ultimate theory, many constructions beyond the SM (BSM) 
have been put forth, at different levels of sophistication. Yet, without direct experimental confirmation,
none of them can be named the true theory beyond the SM.

One way the particle theory community can respond to this situation is to propose simple, fully calculable, renormalizable BSM models 
with viable DM candidates. We do not know yet which  of these models (if any) corresponds to reality,
but all models of this kind offer an excellent opportunity to gain insight into the intricate interplay among various
astrophysical and collider constraints. We call here these models Minimal Consistent Dark Matter (MCDM) models.
MCDM models which can be  viewed as  toy models, are self-consistent and can be easily be incorporated into a bigger BSM model.
Because of these attractive features, MCDM models can be considered as the next step beyond DM Effective Field Theory (EFT) (see e.g. \cite{Fox:2011pm,Rajaraman:2011wf,Goodman:2010ku,Bai:2010hh,Beltran:2010ww,Goodman:2010yf,Fox:2011fx,Shoemaker:2011vi,Fox:2012ru,Haisch:2012kf,Busoni:2013lha,Busoni2014a,Belyaev:2016pxe}) and simplified DM models (see e.g. \cite{Buchmueller:2013dya,Busoni:2014sya,Busoni:2014haa,Buchmueller:2014yoa,Buckley:2014fba,Abdallah:2015ter,Abdallah:2014hon,Abercrombie:2015wmb}).

In this paper, we explore, in the light of the recent collider, astroparticle and DD DM experimental  data, 
the inert Two-Higgs Doublet Model (i2HDM), also known as the Inert doublet model.
This model is easily doable with analytic calculations, its parameter space is relatively small and can be strongly 
constrained by the present and future data.
The model leads to a variety of collider signatures, and, in spite of many years of investigation, not all of them have yet been fully and properly explored.
It is the goal of the present paper to investigate in fine detail the present constraints and the impact of the future LHC 
and DD DM data on the parameter space of this model.


The i2HDM \cite{Deshpande:1977rw,Ma:2006km,Barbieri:2006dq,LopezHonorez:2006gr} is a minimalistic extension of the SM 
with a second scalar doublet $\phi_2$ possessing the same quantum numbers as the SM Higgs doublet $\phi_1$
but with no direct coupling to fermions (the inert doublet). 
This construction is protected by the discrete $Z_2$ symmetry under which $\phi_2$ is odd and all the other fields are even. 
The scalar Lagrangian is
  \begin{equation}
  \mathcal{L} = |D_{\mu}\phi_1|^2 + |D_{\mu}\phi_2|^2 -V(\phi_1,\phi_2)
  \textrm{.}
  \end{equation}
with the potential $V$ containing all scalar interactions compatible with the $Z_2$ symmetry:
\begin{eqnarray}
  V &=& -m_1^2 (\phi_1^{\dagger}\phi_1) - m_2^2 (\phi_2^{\dagger}\phi_2) + \lambda_1 (\phi_1^{\dagger}\phi_1)^2 + \lambda_2 (\phi_2^{\dagger}\phi_2)^2    \nonumber  \\
  &+&  \lambda_3(\phi_1^{\dagger}\phi_1)(\phi_2^{\dagger}\phi_2) 
  + \lambda_4(\phi_2^{\dagger}\phi_1)(\phi_1^{\dagger}\phi_2) 
  + \frac{\lambda_5}{2}\left[(\phi_1^{\dagger}\phi_2)^2 + (\phi_2^{\dagger}\phi_1)^2 \right]\,.\label{potential}
\end{eqnarray}
All free parameters here are real,\footnote{Even if we started with a complex $\lambda_5$, 
we could redefine the second doublet via a global phase rotation, which would render $\lambda_5$ real
without affecting any other part of the Lagrangian.} which precludes the $CP$-violation in the scalar sector.
There is a large part of the parameter space, in which only the first, SM-like doublet, 
acquires the vacuum expectation value (vev).
In the notation $\langle\phi_i^0\rangle = v_i/\sqrt{2}$, this inert minimum corresponds to
$v_1 = v$, $v_2 = 0$.
In the unitary gauge, the doublets are expanded near the minimum as
\begin{equation}
\phi_1=\frac{1}{\sqrt{2}}
\begin{pmatrix}
0\\
v+H 
\end{pmatrix}
  \qquad
  \phi_2= \frac{1}{\sqrt{2}}
\begin{pmatrix}
 \sqrt{2}{h^+} \\
 h_1 + ih_2
\end{pmatrix}
\end{equation}
The $Z_2$ symmetry is still conserved by the vacuum state, which forbids direct coupling of any single inert field to the SM fields
and it stabilizes the lightest inert boson against decay. 
Pairwise interactions of the inert scalars with the gauge-bosons and with the SM-like Higgs $H$ are still possible,
which gives rise to various i2HDM signatures at colliders and in the DM detection experiments.


The idea that the symmetry-protected second Higgs doublet naturally produces a scalar dark matter candidate
was first mentioned more that 30 years ago \cite{Deshpande:1977rw}.
However, the real interest in phenomenological consequences of the i2HDM woke up in mid-2000 and intensified
in the last few years.
Its simplicity, predictive power, rich yet manageable parameter space, makes it an ideal playground for checking 
its compatibility with the DM relic density, with the results of the direct and indirect DM searches,
and with collider searches of various BSM signals.

Assuming that the lightest inert scalar is the only DM candidate,
one typically finds that the low-mass region, below about 50 GeV, is excluded 
by the relic density constraints coupled with the LHC constraints on the invisible Higgs decay 
\cite{Krawczyk:2013jta,Ilnicka:2015jba,Diaz:2015pyv}.
The funnel region, with the DM mass close to $M_H/2$, the intermediate, 100--500 GeV, 
and the high mass regions are still compatible with data and lead to interesting predictions at colliders.
Additional theoretical constraints on the parameter space and DM candidate properties
can be deduced from assumptions of full stability of the i2HDM up to the PLANCK scale
 \cite{Chakrabarty:2015yia,Khan:2015ipa} or of multi-doublet Higgs inflation \cite{Gong:2012ri}.
The i2HDM can also produce signals for direct \cite{Arina:2009um} and indirect DM search experiments via heavy inert scalar annihilation,
which can be detectable via $\gamma$-rays \cite{Modak:2015uda,Queiroz:2015utg,Garcia-Cely:2015khw}
or via its neutrino \cite{Agrawal:2008xz,Andreas:2009hj} and cosmic-ray signals \cite{Nezri:2009jd}.

The i2HDM can also have interesting cosmological consequences.
Being an example of 2HDM, it possesses a rich vacuum structure, which evolves at high temperatures 
\cite{Turok:1991uc,Cottingham:1995cj,Ginzburg:2009dp}.
This opens up the possibility within i2HDM that the early Universe, while cooling down, 
went through a sequence of phase transitions including strong first-order phase transitions 
\cite{Ginzburg:2010wa,Chowdhury:2011ga,Borah:2012pu,Gil:2012ya,Dorsch:2013wja,Cline:2013bln,Blinov:2015vma}.
Such analyses are capable of restricting the parameter space;
for example, the recent study \cite{Blinov:2015vma} showed that combining the strong 
first-order phase transition with other astroparticle and collider constraints gives preference to the funnel region.

There has also been a number of studies on collider signatures of the i2HDM. 
They focus either on specific processes such as SM-like Higgs decays to $\gamma\gamma$ and $\gamma Z$ 
\cite{Arhrib:2012ia,Swiezewska:2012eh,Krawczyk:2013jta,Krawczyk:2013pea},
multilepton production plus missing transverse momentum ($\MET{}$) \cite{Miao:2010rg,Gustafsson:2012aj,Hashemi:2016wup}
with as many as five leptons \cite{Datta:2016nfz},
dijet$+\MET{}$ \cite{Poulose:2016lvz} and dileptons accompanied with dijets \cite{Hashemi:2016wup}.
Other works present combined analyses of astroparticle and collider constraints 
\cite{Goudelis:2013uca,Arhrib:2013ela,Ilnicka:2015jba,Blinov:2015qva,Alves:2016bib,Datta:2016nfz}.
Comparing the i2HDM predictions with the electroweak precision data, 
the measured SM-like Higgs properties, the non-observation
of long-lived charged particles and other exotic signals, and finally the astroparticle
observations, allows one to significantly restrict the i2HDM parameter space.
The recent analysis \cite{Ilnicka:2015jba} gave a detailed account of these constraints.
For specific benchmark points or benchmark planes in the surviving parameter space, 
it predicted the cross section of pair production of inert scalars followed by various modes of their decay.
As for the specific signatures of the i2HDM at the LHC, dileptons and mono-$Z$ signals were mentioned.
An earlier analysis \cite{Arhrib:2013ela} investigated multilepton, multijet, mono-$Z$,
and several channels for the mono-jet with large $\MET{}$.
Ref.~\cite{Goudelis:2013uca} took into account one-loop corrections to the masses
and, for a part of the numerical scans, included the additional theoretical constraint that 
the perturbativity and stability be satisfied up to a large scale $\Lambda$.
The version of i2HDM equipped with Peccei-Quinn $U(1)$ symmetry spontaneously broken to $Z_2$ 
was investigated in \cite{Alves:2016bib}. Here, dark matter acquires a second component, the axion, 
which changes the DM phenomenology.
It is also possible to hunt for i2HDM at the future colliders, via searching for new scalars
and reconstructing the potential \cite{Aoki:2013lhm} or by accurately measuring the SM-like Higgs couplings
and deducing patterns of the deviations from the SM \cite{Kanemura:2016sos}.

In the present work, to these many 
studies on the i2HDM, we add:
\begin{itemize}
\item detailed combined analysis of the i2HDM model in its full five-dimensional (5D)  parameter space, taking
into account perturbativity and unitarity,
LEP and electroweak precision data, Higgs data from the LHC, DM relic density, direct/indirect DM detection 
complemented by realistic (beyond-the-parton-level) LHC mono-jet analysis at the LHC;
\item
quantitative exploration of the surviving regions of parameters, including very fine details  and 
qualitatively new region not seen in previous studies,
which is enabled by our extensive numerical scans;
\item
a combination of different processes giving the LHC mono-jet signatures: 
those with direct DM pair production and those with associate production of DM with another scalar with a close mass
from the inert multiplet;
\item 
implication of experimental LHC studies on disappearing
charged tracks relevant to high ($\simeq 500$ GeV) DM mass region;
\item 
separate, equally detailed analyses for the assumptions of the DM relic density being fitted to the PLANCK results
or under-abundant, allowing thus for additional allowed regions of the parameter space.
\end{itemize}
All these points above are in close focus of the present paper where we have performed a comprehensive  
scan and study of the  full parameter space of the i2HDM model. 
In addition we have performed an independent implementation and validation of the model in two gauges 
including Higgs-gluon-gluon and Higgs-photon-photon effective couplings, and we made it public together with the LanHEP model source.

The paper is organised as follows. 
In Sect.~2 we discuss the i2HDM model parameter space, implementation, 
theoretical constraints as well as constraints from LEP and electroweak precision data.
In Sect.~3 we discuss results of a comprehensive scan of the i2HDM parameter space
and combined constraints considering both the cases when the relic density is ``just right"
and agrees with the PLANCK results and when it is under-abundant.
In this section we also present the reach of  LHC studies in the high DM mass region using
results on disappearing charged tracks.
In Sect.~4 we present results on future projections of the LHC and DM DD experiments
in combination with all previous constraints. Finally, in Sect.~5 we draw our conclusions.
\section{i2HDM: parameter space, model implementation,  theoretical and experimental  constraints}

\subsection{Constraints from the Higgs potential}\label{sec:2a}

In order to represent a viable model, the potential (\ref{potential}) must be bounded from below
and must have a neutral, not charge-breaking vacuum.
The former requirement leads to the well-known restrictions on the free parameters of the model:
\begin{equation}
\lambda_1>0,\ \  \lambda_2>0,\ \ 
2\sqrt{\lambda_1\lambda_2}+\lambda_3>0,\ \ 
2\sqrt{\lambda_1\lambda_2}+\lambda_3+\lambda_4 - |\lambda_5|>0\,.
\end{equation}
The absence of the charge-breaking vacuum is guaranteed if one assumes
\begin{equation}
\lambda_4 - |\lambda_5| < 0\,. \label{lam45-condition}
\end{equation}
This is a sufficient but not necessary condition for the vacuum to be neutral.
A neutral vacuum can also be achieved for positive $\lambda_4 - |\lambda_5|$ with appropriate
$m_1^2$ and $m_2^2$. However in this case the lightest DM candidate will be the charged scalar.
Condition (\ref{lam45-condition}) avoids this situation as well.

Once these restrictions are applied, the vacuum is neutral, and one can calculate the masses of the physical Higgs bosons.
In addition to the SM-like scalar $H$, one gets charged $h^\pm$ and neutral $h_1, h_2$ scalars. 
It is well known that the two neutral scalars of the i2HDM have opposite $CP$-parities, but it is impossible 
to unambigously assign which of them is $CP$-even and which is $CP$-odd.
In the absence of any suitable vertex, the model has two $CP$-symmetries, $h_1 \to h_1, h_2 \to -h_2$ and
$h_1 \to -h_1, h_2 \to h_2$, which get interchanged upon basis change $\phi_2 \to i \phi_2$. 
Either can be used as ``\textit{the $CP$-symmetry}'' of the model, 
making the specification of the $CP$ properties of $h_1$ and $h_2$ a basis dependent statement. 
Therefore, we denote the two neutral inert scalar masses as $M_{h_1} < M_{h_2}$, without specifying which is scalar and pseudoscalar.
The masses of the physical scalars are 
\begin{eqnarray}
&&M_H^2 = 2 \lambda_1 v^2 = 2m_1^2\,,\qquad
M_{h^+}^2 = {1\over 2} \lambda_3 v^2 - m_2^2\,,\nonumber\\
&&M_{h_1}^2 = {1\over 2}(\lambda_3 +\lambda_4 - |\lambda_5|) v^2 - m_2^2\,,\qquad
M_{h_2}^2 = {1\over 2}(\lambda_3 +\lambda_4 + |\lambda_5|) v^2 - m_2^2 \ >M_{h_1}^2\,.\label{masses}
\end{eqnarray}
The mass differences, written as
\begin{equation}
M_{h_2}^2 - M_{h_1}^2 = |\lambda_5| v^2\,,\quad 
M_{h^+}^2 - M_{h_1}^2 = -(\lambda_4 - |\lambda_5|) v^2/2\,,
\end{equation}
highlight the role of the parameters $\lambda_4$ and $\lambda_5$ and are consistent with (\ref{lam45-condition}).
It should also be stressed that the parameters $\lambda_1$ and $m_1^2$ correspond to the Higgs potential in the SM, and can thus be fixed by the values of the VEV and Higgs mass.

One also notices that the sign of $\lambda_5$ is phenomenologically irrelevant: 
flipping the sign of $\lambda_5$ would only lead to swapping the $CP$-parities
of the inert neutral scalars, which are unobservable anyway.
In order to eliminate double-counting, we make the standard choice of $\lambda_5 < 0$,
and introduce the shorthand notation $\lambda_{345}=\lambda_3+\lambda_4+\lambda_5$.
The latter parameter plays an important phenomenological role, as it governs the Higgs-DM interaction vertex $H h_1 h_1$.
For future convenience, we also introduce the shorthand notation
\begin{equation}
\tilde\lambda_{345} \equiv \lambda_3+\lambda_4-\lambda_5 = \lambda_{345} + 2|\lambda_5| = 
\lambda_{345} + \frac{2(M_{h_2}^2-M_{h_1}^2)}{v^2}\,,
\label{tildelam345}
\end{equation}
which is not a new free parameter and is the combination which governs, in particular, 
the $Hh_2h_2$ coupling as well as the quartic coupling of $h_1$ to the longitudinal $Z$-bosons
$h_1 h_1 Z_L Z_L$.

With all these conventions, we describe the five dimensional parameter space of i2HDM
with the following phenomenologically relevant variables: 
\begin{equation}
\label{eq:model-parameters}
M_{h_1}\,,\quad M_{h_2} > M_{h_1}\,,\quad M_{h^+} > M_{h_1}\,, \quad \lambda_2 > 0\,,\quad \lambda_{345} > -2\sqrt{\lambda_1\lambda_2}\,.
\end{equation}

\begin{figure}[htb]
\subfigure[$R > 1$]{\includegraphics[width=0.3\textwidth]{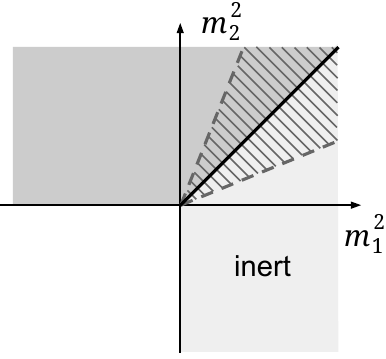}\hspace{5mm}}
\subfigure[$0 < R < 1$]{\includegraphics[width=0.3\textwidth]{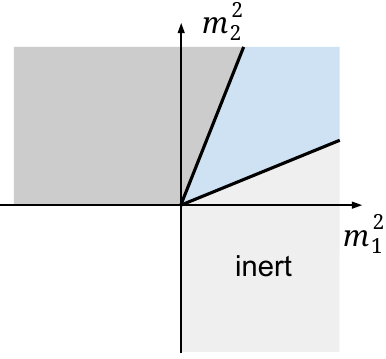}\hspace{5mm}}
\subfigure[$-1 < R < 0$]{\includegraphics[width=0.3\textwidth]{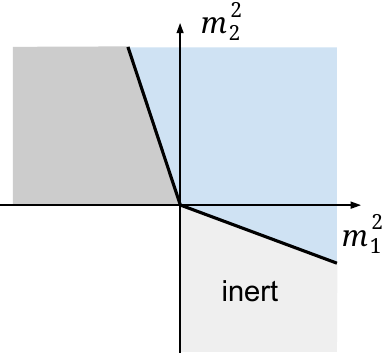}}
\caption{Restrictions on the $(m_1^2,\, m_2^2)$ plane coming from the requirement that
the inert vacuum is the deepest minimum of the potential. The three cases correspond to (a) $R > 1$,
(b) $0 < R < 1$, (c) $-1 < R < 0$. Light and dark grey correspond to models 
with an inert $v_1 = v,\, v_2 = 0$ and a pseudoinert $v_1 = 0,\, v_2 = v$ vacuum,
respectively, while the blue region in between corresponds to the mixed vacuum,
when both $v_1$ and $v_2$ are non-zero. The dashed region in the left plot indicates coexistence
of the inert and pseudoinert minima at different depths.}
\label{fig:parameter-space}
\end{figure}

Another set of theoretical constraints comes from the symmetry breaking patterns in i2HDM \cite{Deshpande:1977rw}
and from the fact that the potential can have two minima at different depths.
Following \cite{Ginzburg:2010wa},
we introduce $R=\lambda_{345}/2\sqrt{\lambda_1\lambda_2}$,
which satisfies $R> -1$.
Requiring that the inert vacuum corresponds to the global minimum
leads to the following conditions on the parameters of the potential, apart from $m_1^2 > 0$:
\begin{eqnarray}
\label{eq:condition1}
m_2^2 < \frac{\lambda_{345}}{2\lambda_1} m_1^2 = R\, \sqrt{\frac{\lambda_2}{\lambda_1}} m_1^2 \,,&\mbox{if}& |R| < 1\,,\nonumber\\ 
m_2^2 < \sqrt{\frac{\lambda_2}{\lambda_1}} m_1^2 \,, &\mbox{if}& R > 1\,.
\end{eqnarray}
In Fig.~\ref{fig:parameter-space} we visualise these restrictions on the $(m_1^2,\, m_2^2)$ plane
for the three choices of $R$.
The inert, $v_1 = v$, $v_2=0$, and pseudoinert, $v_1 = 0$, $v_2 = v$, vacua can coexist only when $R>1$,
which is shown by the dashed region in Fig.~\ref{fig:parameter-space} (a).
For $R > 1$, the second line in Eq.~(\ref{eq:condition1}) is a stronger condition
than the first line and it guarantees that the inert minimum is the deepest one.
This condition is shown in Fig.~\ref{fig:parameter-space} (a) by the solid black line.

Rewriting  conditions~(\ref{eq:condition1}) for the physical parameters 
we get  the constraint on the Higgs potential in the following compact final form:
\begin{eqnarray}
\label{eq:scalar-pot1}
&&\mbox{the trivial one, } M_{h_1}^2 > 0 \mbox{ for } |R|<1, \\
&&\mbox{and}\nonumber\\
\label{eq:scalar-pot2}
&&M_{h_1}^2 > 
  (\lambda_{345}/2\sqrt{\lambda_1\lambda_2}-1) \sqrt{\lambda_1\lambda_2} v^2
=  (R-1) \sqrt{\lambda_1\lambda_2} v^2 \mbox{ for } R>1,
\end{eqnarray}
where $\lambda_1 \approx 0.129$ is fixed as in the Standard Model by the Higgs mass (\ref{masses}). 
The latter condition places an upper bound on $\lambda_{345}$ for a given DM mass $M_{h_1}$.

\subsection{Model implementation}
We have implemented the i2HDM into the CalcHEP package~\cite{CALCHEP}
with the help of the LanHEP program~\cite{Semenov:1998eb,Semenov:2008jy}
for automatic Feynman rules derivation. The effective $Hgg$ and $H\gamma\gamma$ vertices were included and the model was cross-checked in two different gauges
to ensure a correct, gauge invariant implementation.
It is publicly available at the
High Energy Physics Model Data-Base (HEPMDB) \cite{Brooijmans:2012yi} at
\url{http://hepmdb.soton.ac.uk/hepmdb:0715.0187}
together with the LanHEP source of the model.
The model is implemented in terms of the five independent parameters
defined in Eq.~(\ref{eq:model-parameters}), consisting of three physical masses
and two couplings.
We found this choice the most convenient for exploration of i2HDM phenomenology
and constraints of its parameter space.
We should stress that the $M_{h_1}$ and $M_{h_2}$ parameters 
conveniently  define the mass order of the two neutral inert states 
$independently$ of their $CP$ properties. 
This choice is especially convenient and relevant for collider
phenomenology since, as we discussed above, one can not assign (or determine) the CP parity of each neutral inert scalar.

To explore the phenomenology of the i2HDM we need to consider other constraints on its parameter space
in addition to those coming from vacuum stability which we discussed above.

\subsection{Constraints from perturbativity and unitarity}

%
%

The first requirement we impose on the quartic couplings in (\ref{potential}) is that their values are such that perturbative calculations can be trusted in the model.  The most effective way is to impose perturbative unitarity on all the scattering processes involving the scalars.
Following~\cite{Arhrib:2012ia}, we impose this condition on the full scattering matrix, which leads to the following bounds on combinations of couplings $e_i$:
\begin{equation}
\label{eq:unit}
|e_i|\leq 8\pi\,,
\end{equation}
where
$e_{1,2}= \lambda_3 \pm \lambda_4, \ \ e_{3,4}=\lambda_3 \pm \lambda_5, \ \ \
e_{5,6}= \lambda_3 + 2\lambda_4 \pm 3\lambda_5, \ \ \  e_{7,8} = -\lambda_1-\lambda_2 \pm \sqrt{(\lambda_1-\lambda_2)^2+\lambda_4^2}$, $e_{9,10} = -3\lambda_1 - 3\lambda_2 \pm \sqrt{9(\lambda_1 -\lambda_2)^2 + (2\lambda_3+\lambda_4)^2}, \ \ \ 
 e_{11,12} = -\lambda_1 - \lambda_2 \pm \sqrt{(\lambda_1 - \lambda_2)^2 + \lambda_5^2}
$.
The parameter $\lambda_1$ is fixed by SM-Higgs mass and the vacuum expectation value.
One can verify that the constraints given by Eq.~(\ref{eq:unit}) imply that all quartic couplings in (\ref{potential}) are bound to be smaller than $8\pi$, thus within the perturbative regime.
The perturbativity constraints can also be used to find upper bounds on the two input couplings we defined in the previous section, i.e. $\lambda_2$ and $\lambda_{345}$. From $e_{10}$ one finds:
\begin{equation}
\lambda_2<\lambda_2^{max} < 4\pi/3,
\end{equation}
where $\lambda_2^{max}$ is a function of  model parameters,
while from $e_5 = 3 \lambda_{345} - (2 \lambda_3 + \lambda_4)$, combined with $e_{10}$ in the limit $\lambda_2 = 0$, we obtain an upper bound for $\lambda_{345}$:
\begin{equation}
- 1.47 \simeq -2\sqrt{\lambda_1 (4\pi/3)} < -2\sqrt{\lambda_1\lambda_2^{max}}<\lambda_{345}\lessapprox {2\over 3} \times 8\pi - \lambda_1,
\label{eq:l345min}
\end{equation}
where we expanded at leading order in the small coupling $\lambda_1$, and the lower bound comes from the stability of the potential.
This limit, derived from the constraints on $e_5$ and $e_{10}$ is not actually the most stringent one:
in the limit of $\lambda_2 \to 0$ we have found  that the biggest value for $\lambda_{345}$
is realised in the  $|\lambda_{4,5}| \to 0$ limit when $\lambda_3 \simeq  4\pi$ 
and respectively  $\lambda_{345} \simeq 4\pi$.
After expansion in  the small coupling $\lambda_1$, the upper limit on  $\lambda_{345}$ in the small $\lambda_2 $ limit reads as 
\begin{equation}
\lambda_{345}\lessapprox 4\pi-{3\over 2}\lambda_1
\label{eq:l345max}
\end{equation}
while for finite  $\lambda_2 $ the limit can be found numerically.
{We would like to  note that the limit from perturbative unitarity and perturbativity
given by Eqs.(\ref{eq:unit})-(\ref{eq:l345max}) we are using in our study is consistent with  that implemented in the 2HDMC code~\cite{Eriksson:2009ws}.}

One should also stress  that the vacuum stability condition given by  Eq.(\ref{eq:scalar-pot2})
sets an important constraint on the  maximum value of $\lambda_{345}$
in the small  $M_{h_1}$ region (which is the region of our special interest
because of the collider phenomenology constraints as we discuss below).
This can be seen from Eq.(\ref{eq:scalar-pot2}) which can be written 
as:
\begin{equation}
\lambda_{345}< 
2\left(\frac{M_{h_1}^2}{v^2}+
\sqrt{\lambda_1\lambda_2^{max}}\right)
\label{eq:l345-vacuum-stab}
\end{equation}

\begin{figure}[htb]
\centering
\includegraphics[width=0.7\textwidth]{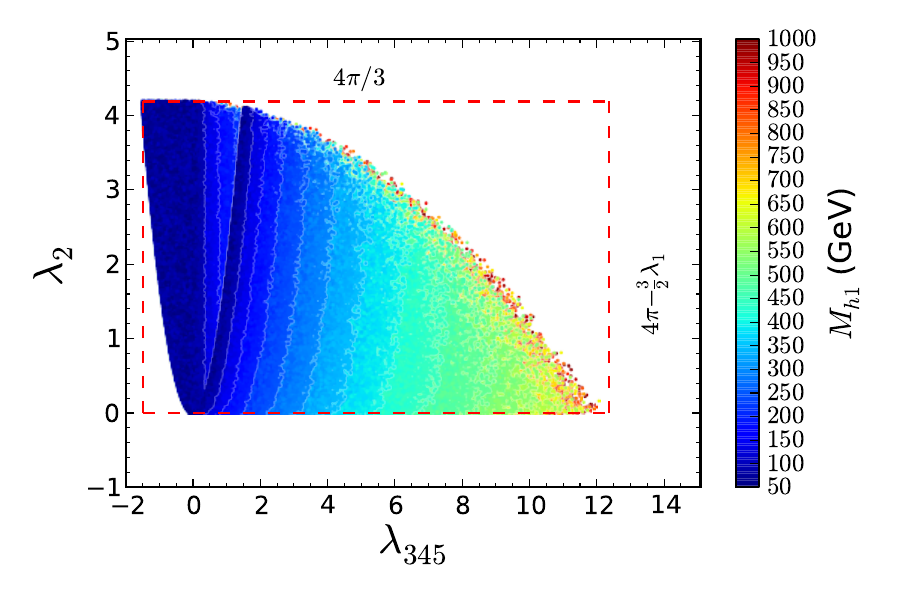}
\caption{The part of the ($\lambda_{345},\lambda_2$) parameter space 
allowed by the unitarity, perturbativity and scalar potential constraints.\label{fig:par-space1}}
\end{figure}

In Fig.\ref{fig:par-space1},
we present viable parameter space in the ($\lambda_{345},\lambda_2$) plane
after 
constraints from Eq.~(\ref{eq:unit}) as well constraints from 
scalar potential given by Eqs.~(\ref{eq:scalar-pot1}),~(\ref{eq:scalar-pot2}),~(\ref{eq:l345-vacuum-stab}).
To produce this plot we have performed the wide random scan to cover  the full five-dimensional  parameter space of the model,  with the following chosen 
range for the  model parameters:
\begin{eqnarray}
10\mbox{ GeV} < &M_{h1,h2,h^+}& < 1000 \mbox{ GeV} \nonumber\\
0< &\lambda_2& < {4\pi\over 3} \nonumber\\
-1.47 < &\lambda_{345}& < 4\pi 
\label{eq:scan-limits}
\end{eqnarray}
The colour map in Fig.~\ref{fig:par-space1} presents the values for the third essential parameter,
the DM candidate mass $M_{h_1}$,
with points  of smaller values of $M_{h_1}$ on the top of points with larger   $M_{h_1}$ values.
From this figure one can observe a non-trivial shape of the allowed parameter space in the 
($\lambda_{345},\lambda_2$) plane defined by the constraints mentioned above. In particular,
for small  $M_{h_1}$ values, the upper limit on $\lambda_{345}$
comes from Eq.(\ref{eq:l345-vacuum-stab}) which restricts $H h_1 h_1$ coupling 
$\lambda_{345}$ to be not very large.
The value of $\lambda_2^{max}$ entering there can be found in general only numerically.

\subsection{Constraints from LEP and electroweak precision data}

Very strong constraints on the i2HDM arise from precision data and searches from LEP experiments. First of all,
the model should respect the  precise measurements of the W and Z widths
which lead to the following lower limit on the odd scalar masses:
\begin{eqnarray}
\label{eq:constr-widths}
&& M_{h_1} + M_{h^{+}} > M_{W^{+}} \quad , \quad M_{h_2} + M_{h^{+}} > M_{W^{+}} \nonumber\\
&& M_{h_1} + M_{h_2} > M_{Z} \quad , \quad 2M_{h^{+}} > M_{Z}
\end{eqnarray}
to make sure that $\Gamma({W^{+} \rightarrow h_1 h^{+},h_2h^{+}})$ and $\Gamma({Z \rightarrow h_1h_2,h^+h^-})$ decay channels are kinematically 
forbidden.

\begin{figure}[htb]
\centering
\hspace*{-0.3cm}\subfigure[]{\includegraphics[width=0.55\textwidth]{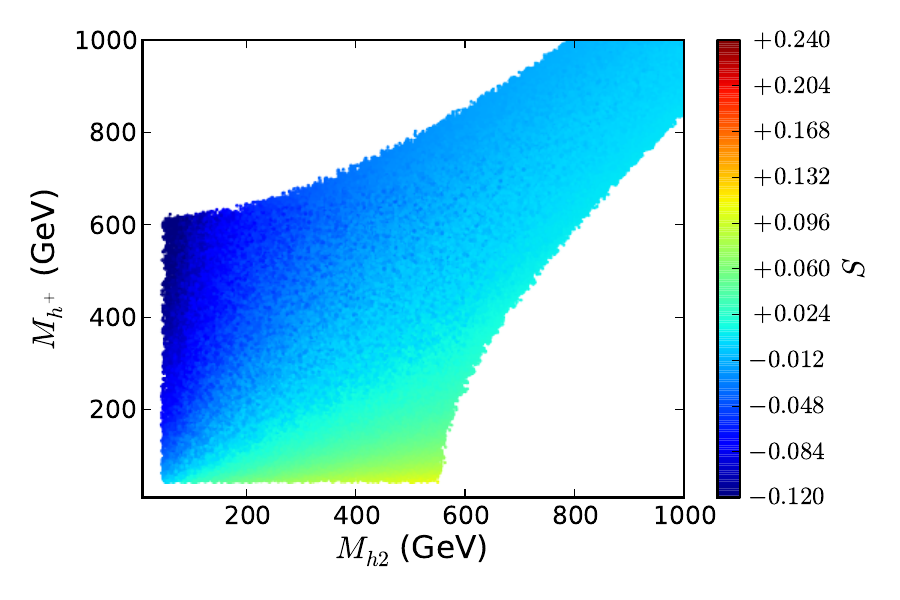}}%
\hspace*{-0.3cm}\subfigure[]{\includegraphics[width=0.55\textwidth]{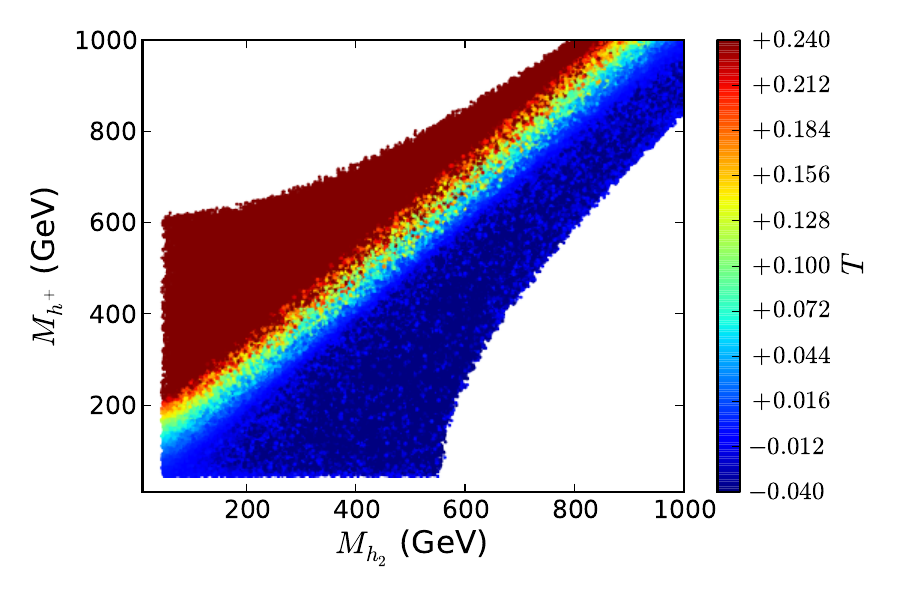}}\\
\subfigure[]{\includegraphics[width=0.55\textwidth]{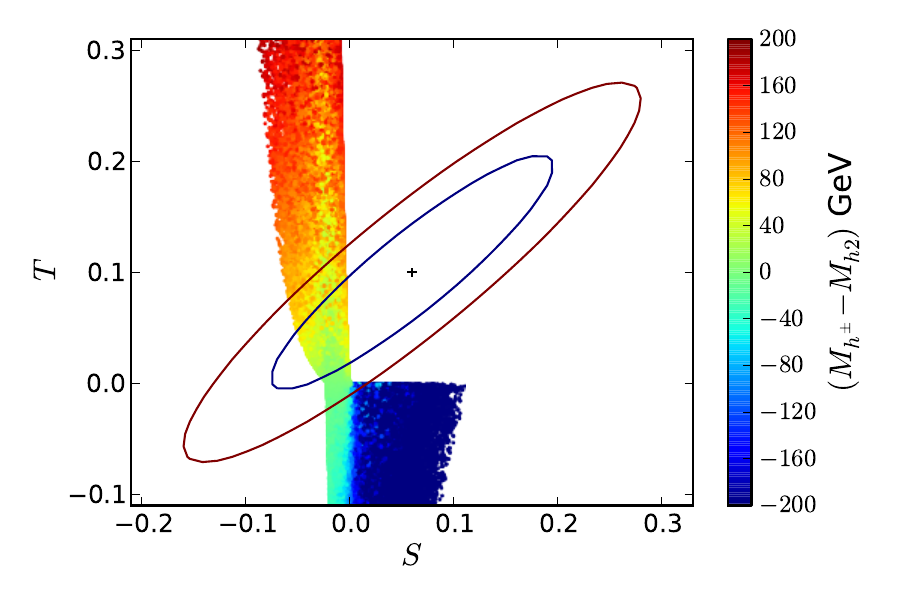}}
\caption{
Effect of the  $S$ and $T$ constraints on the $M_{h^+}-M_{h_2}$ mass difference:
(a) and (b) show the 
colour map of the $S$ and  $T$ parameters respectively 
in the $(M_{h^+},M_{h_2})$ plane; (c) shows  the colour map
of the $M_{h^+}-M_{h_2}$ split in the ($S,T$) plane together with the
65\% and 95\% exclusion contours
based on the $\chi^2$ ($S,T$) characterisation for two degrees of freedom.} \label{fig:s-t-u}
\end{figure}

While studying the phenomenology of the i2HDM, we should also make sure that 
Electroweak Precision Test (EWPT) data is respected.
As we know, EWPT can be  expressed in terms of three measurable quantities, called S, T, and U, that parameterise contributions
from beyond standard model physics  to electroweak radiative corrections~\cite{PhysRevD.46.381}.
The contribution to the S and T parameters~\cite{Barbieri:2006dq} can be written as
\begin{equation}
S = 
\frac{1}{72\pi}\frac{1}{(x_2^2-x_1^2)^3}
\left[ 
x_2^6 f_a(x_2) -x_1^6 f_a(x_1)
+ 9 x_2^2 x_1^2( x_2^2 f_b(x_2) - x_1^2 f_b(x_1)
\right]
\end{equation}
where $x_1=M_{h_1}/M_{h^+}, x_2=M_{h_2}/M_{h^+}$, $f_a(x) = -5+12\log(x), f_b(x)=3-4\log(x)$
and
\begin{equation}
T = \frac{1}{32\pi^2\alpha v^2}\left[f_c(M_{h^{+}}^2,M_{h_2}^2) + f_c(M_{h^{+}}^2,M_{h_1}^2) - f_c(M_{h_2}^2,M_{h_1}^2)\right]
\end{equation}
where the function $f_c(x,y)$ is defined by
\begin{equation*}
f_c(x,y) = 
\begin{cases}
\frac{x+y}{2}-\frac{xy}{x-y}\log{\left(\frac{x}{y}\right)}, & x\neq y\\
0, & x = y
\end{cases}
\end{equation*}
We have written the contributions to $S$ and $T$ in a form which demonstrates explicitly their symmetry
with respect to swapping $h_1 \leftrightarrow h_2$, pointing again to the fact that one can not distinguish their CP properties.
With $U$ fixed to be zero, the central values of $S$ and $T$, assuming a SM Higgs boson mass of $m_h$ = 125 GeV, are given by~\cite{Baak:2014ora}:
\begin{equation}
S = 0.06 \pm 0.09 ,\qquad T = 0.1 \pm 0.07
\label{eq:ewpt}
\end{equation}
with correlation coefficient +0.91.
The effect of the constraints on $S$ and $T$ is presented in  Fig.\ref{fig:s-t-u}, where panels a) and b)  present the 
colour map of the $S$ and  $T$ parameters respectively 
in the $(M_{h^+},M_{h_2})$ plane.
One can see that the $T$ variable is more  sensitive than $S$
to this mass split, thus only modest splits
are allowed by EWPT data.
Finally, Fig.\ref{fig:s-t-u} c) presents  the colour map
of the $M_{h^+}-M_{h_2}$  split in the ($S,T$) plane together with the
65\% and 95\% exclusion contours, based on a $\chi^2$ with  two degrees of freedom.
  One can see that
EWPT data prefer a modest positive $M_{h^+}-M_{h_2}$ mass split 
 below  about $100$ GeV, which is mainly defined by the $T$ parameter, while the role and the respective range of variation of $S$ is milder. One should stress that 
 it is crucial to take into account the correlation between
$S$ and $T$ and  combine limits from these two  parameters. This combination gives a much stronger limit on the parameter 
space, in particular on the $M_{h^+}-M_{h_2}$ mass split, while a much larger splitting would
naively be allowed by looking at the $S$ and $T$ values separately. This can be
seen from  Fig.\ref{fig:s-t-u}(a) and Fig.\ref{fig:s-t-u}(b) respectively.

We also excluded the region defined by the intersection of the conditions below:
\begin{equation}
M_{h_1}<80\mbox{ GeV} \ , \ M_{h_2}<100\mbox{ GeV} \ ,  \ M_{h_2}-M_{h_1}>8 \mbox{ GeV}\,.
\label{eq:dmh12}
\end{equation}
This region is excluded by the LEP data since it would lead to a visible
di-jet or di-lepton signal as demonstrated in~\cite{Lundstrom:2008ai}
where a reinterpretation in the i2HDM of a LEP--II limit of the second neutralino production in the Minimal Supersymmetric Standard Model (MSSM) was presented.
A more detailed analysis of this specific region of the parameter space ---
low $M_{h_1}$ and $M_{h_2}$ with large enough mass gap --- was studied recently~\cite{Belanger:2015kga}.
One should also mention that  $e^+e^-\to h^+h^-$ production at LEP2
sets
\begin{equation}
M_{h^+}>70\mbox{ GeV}
\label{eq:mhcp-lep2}
\end{equation}
as found in~\cite{Pierce:2007ut} as a result of the re-interpretation of LEP--II limits on charginos.

\subsection{Constraints from LHC Higgs data}

The LHC Higgs data further restricts the i2HDM parameters space in the form of constraints on the couplings of the SM-like Higgs boson. A collection of combined fits from the Run I data, for both ATLAS and CMS, can be found in~\cite{Khachatryan:2016vau}.
In the i2HDM, the leading effect is encoded in two observables: the decays of the Higgs into two Dark Matter scalars, $H \to h_1 h_1$, which is kinematically open when $M_{h_1} < M_H/2$; and the contribution of the charged Higgs loops to the $H \to \gamma \gamma$ decay. 
In principle, we would need to do a two-parameter fit of the available Higgs data.  None of the fits presented in~\cite{Khachatryan:2016vau} can therefore be directly applied in our case.

A simpler possibility is, instead, to consider the best possible bound from the available fits on the two parameters.
We follow this simpler procedure, confident that it will lead to a somewhat more conservative estimation of the bounds.
For the invisible Higgs branching ratio, we consider the bound coming from the dedicated ATLAS search~\cite{Aad:2015txa}
\begin{equation}
Br(H\to invisible)  < 28\% 
\label{eq:lhc-higgs-invis}
\end{equation}
at the 95$\%$ CL, which is comparable with a 36\% limit
from the combined CMS analysis~\cite{CMS:2015naa}.\footnote{One could also limit $Br(H\to invisible)$ 
using $Br(H\to BSM)<34\% $ at 95$\%$CL exclusion from Run1  ATLAS-CMS Higgs data analysis~\cite{Khachatryan:2016vau}.
However, here we use the $Br(H\to invisible) < 28\% $ limit from a dedicated ATLAS search
as it is less model dependent.}

For the second observable, the di-photon decay rate, we consider the result from the combined fit on the signal strength in the di-photon channel~\cite{Khachatryan:2016vau}:
\begin{equation}
\frac{Br^{BSM}(H\to \gamma\gamma)}{Br^{SM}(H\to \gamma\gamma)} =\mu^{\gamma\gamma} = 1.14^{+0.38}_{-0.36}\,,
\label{eq:lhc-higgs-aa}
\end{equation}
where we doubled the $1\sigma$ errors given in \cite{Khachatryan:2016vau} to obtain the $\mu^{\gamma\gamma}$ range at the 95\% CL.
A sufficiently light charged Higgs with sufficiently large $\lambda_3$ coupling to the SM Higgs boson,
which would bring the $H\to \gamma\gamma$ decay beyond the quoted limit, is excluded.

It should be noted that we would expect a proper 2-parameter 
fit to lead to stronger constraints that the ones we use, however the qualitative impact of the constraints should be unchanged.
For example, the partial decay width of the Higgs into  DM which is defined by
\begin{equation}
 \Gamma (H \to h_1 h_1) = \frac{1}{8\pi}\frac{\lambda_{345}^2 M_W^2}{g_W^2 M_H}\sqrt{1-4\frac{M_{h_1}^2}{M_H^2}},
\end{equation}
where $g_W$ is the weak coupling constant,  provides the following  bound on $\lambda_{345}$:
\begin{equation}
|\lambda_{345}| < 
\left(
\frac{8\pi g_W^2 \Gamma_{SM} M_H}{M_W^2 \left(\frac{1}{Br^{max}_{invis}}-1\right)\sqrt{1-4\frac{M_{h_1}^2}{M_H^2}}}	
\right)^{1/2},\label{lam345-limit-from-inv}
\end{equation}
where $Br^{max}_{invis}=0.28$ is the current bound on the maximal value of branching ratio
of the Higgs boson decay into invisible mode.
The above limit on $\lambda_{345}$ is $M_{h_1}$ dependent:
for  $M_{h_1}/M_{H} \ll 1$  it is about 0.019, while for  $M_{h_1}$ closer to the threshold,
e.g. 60 GeV, the limit on  $\lambda_{345}$ increases almost by a factor of two and reaches a value of 0.036.
In addition we have included the limit from $H\to h_2 h_2$
when $h_2$ is close in mass to $h_1$, which can be trivally done, taking into account that
$H h_2 h_2$ coupling is equal to $\tilde \lambda_{345}$ in Eq.~(\ref{tildelam345}).
We discuss these limits in more details below, together with the Dark Matter (DM) constraints.

\subsection{Dark Matter relic density and direct/indirect detection}

The results from PLANCK~\cite{Ade:2013zuv,Planck:2015xua} (see also WMAP~\cite{Hinshaw:2012aka}) have further decreased the error on the already quite precise measurement of the dark matter relic density, $\Omega_{\rm DM} h^2$:
\begin{equation}
\Omega_{\rm DM}^{\rm Planck} h^2=0.1184\pm0.0012.
\label{eq:planck-limit}
\end{equation} 
In the i2HDM model, the lightest inert scalar $h_1$ is stable and contributes to this relic density.
In our study we take the upper limit on  $\Omega_{\rm DM} h^2$ as the hard one,
excluding the parameter space points which lead to DM overabundance.
However we do not exclude the i2HDM parameter space regions where $h_1$ is under-abundant,
allowing for other sources of DM coming from an additional new physics sector.

We have evaluated $\Omega_{\rm DM} h^2$ with the
{\texttt{micrOMEGAs 2.4.1}} package~\cite{Belanger:2013oya,Belanger:2006is, Belanger:2010gh}
since it directly reads the model files in CalcHEP format.
{In our analysis we have assumed 10\% theoretical uncertainty  on the DM relic density prediction since it is based on the tree-level calculation. This uncertainty is the dominant one in comparison to about 1\% uncertainty on DM fit from the latest PLANCK results given above and relax the DM relic density limit 
to the following one at 95\% CL:
\begin{equation}
\Omega_{\rm DM}^{\rm limit} h^2=0.1184\pm 2\times 0.1184/10
\simeq 0.118\pm 2\times 0.012
\label{eq:planck-limit-relaxed}
\end{equation}
which we will refer here still as  ``PLANCK limit" .
}

\begin{figure}[htb]
\centering
{\includegraphics[width=0.5\textwidth]{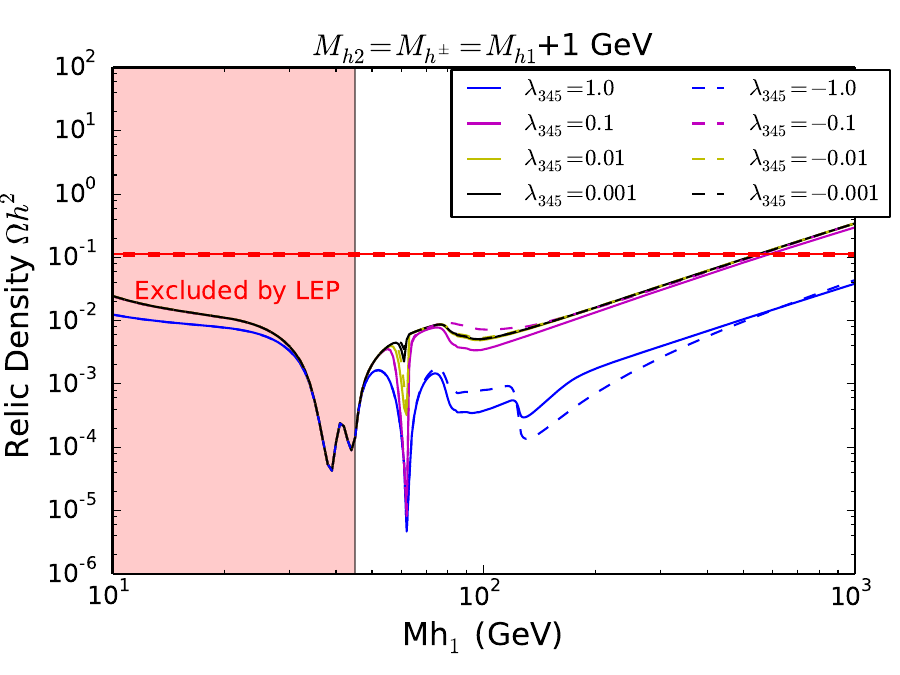}}%
{\includegraphics[width=0.5\textwidth]{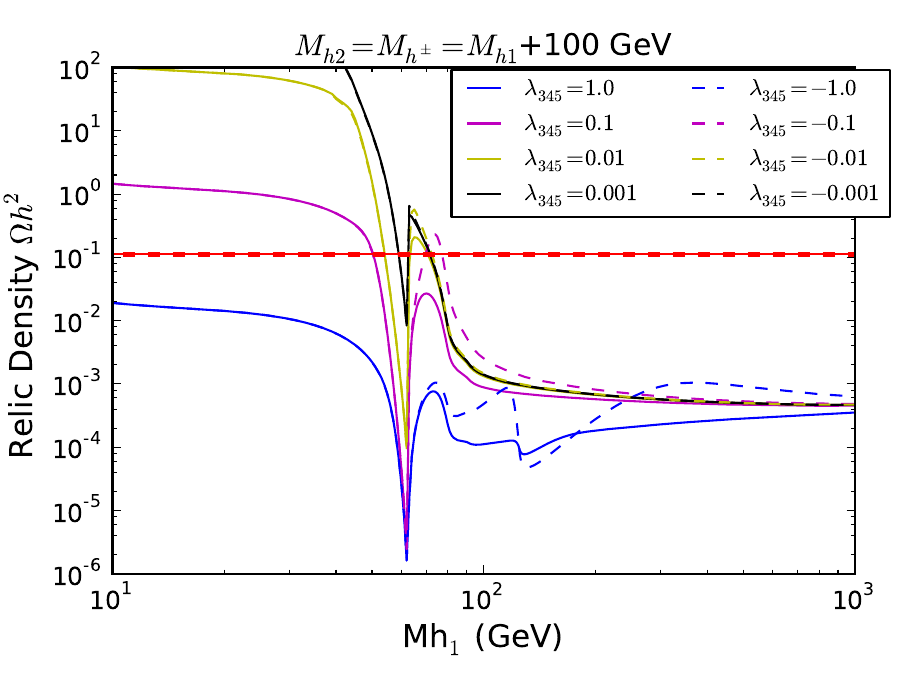}}%
\vskip -0.5cm\hspace*{-3cm}(a)\hspace*{0.48\textwidth}(b)
\caption{The relic density, $\Omega_{\rm DM} h^2$,   as a function of $M_{h_1}$
for various $\lambda_{345}$ parameters. The red-shaded region in the left frame is excluded by the LEP data, 
since in this region $W$ and $Z$ bosons would decay to the light inert scalars. 
The horizontal red line corresponds to the relic density upper limit given by Eq.(\ref{eq:planck-limit-relaxed}).}
\label{fig:1d-mh1-Omega}
\end{figure}

Fig.~\ref{fig:1d-mh1-Omega}(a) shows the relic density in the case of quasi-degenerate $h_1,h_2$ and $h^+$ masses, 
$M_{h_2}=M_{h^+}=M_{h_1}+\Delta M= M_{h_1}+$1~GeV.
This case is qualitatively different from the case with a non-negligible mass splitting
as illustrated in Fig.~\ref{fig:1d-mh1-Omega}(b), where we chose  $M_{h_2}=M_{h^+}=M_{h_1}+\Delta M=M_{h_1}+$100~GeV.
One should also note that scenarios with positive or negative $\lambda_{345}$ values of the same magnitude
are qualitatively similar, except for the effect of interference (see dashed versus solid curves in Fig.~\ref{fig:1d-mh1-Omega}). 
One can observe the following effects and features of the model in Fig.~\ref{fig:1d-mh1-Omega}:
\begin{itemize}
\item The red-shaded region in Fig.~\ref{fig:1d-mh1-Omega}(a) 
is excluded by the LEP data, since in this region $W$ and $Z$ bosons would decay to the light inert scalars. 
Respectively, the effect of the resonant co-annihilation, $h_1 h_2 \to Z$ and $h_1 h^+ \to W^+$, can be seen 
in this region in the first two dips for $M_{h_1} \sim 40$ and $45$~GeV.
These processes are governed by the gauge coupling constant and are independent of $\lambda_{345}$. 
\item
In the case of larger $M_{h_2}-M_{h_1}$  mass split (Fig.~\ref{fig:1d-mh1-Omega}(b)), 
this effect disappears since $M_{h_1}+M_{h_2} > M_Z$ and $M_{h_1}+M_{h^+} > M_W$.
\item
The sharpest dip in the $\Omega_{\rm DM} h^2$ dependence of $M_{h_1}$ is at 65 GeV and 
corresponds to the DM annihilation through the Higgs boson $h_1 h_1 \to H$. It is present in both cases. 
\item
At higher masses, we observe a wider and more shallow  dip at around 80-90 GeV from  $h_1
h_1\to W^+W^-$ and $h_1 h_1\to ZZ$ channels which are merged together. 
\item
Finally, the last dip around 125 GeV corresponds to the
reduction of the DM relic density  due to the opening of the $ h_1 h_1\to HH$ annihilation channel. This dip takes place only for 
large values of $\lambda_{345}$, which provide a high enough rate for the $h_1h_1\to HH$ process via the $s$-channel Higgs boson. 
\item
The pattern of these last three dips is the same for the larger mass split scenario presented in Fig.~\ref{fig:1d-mh1-Omega}(b). 
In both scenarios, the interference effect is sensitive to the sign of $\lambda_{345}$ and appears in this region 
as a result of the positive or negative interference of the $s$-channel Higgs boson exchange
diagram and the rest of annihilation diagrams.
\item
One can also observe qualitative differences in the asymptotic behaviour of the DM relic density for small
and large $M_{h_1}$ values for different $\Delta M$. In the $\Delta M=1$~GeV case with $M_{h_1} < 65$~GeV, the effective
co-annihilation of the inert scalars keeps the DM density always below the PLANCK limit. 
For $\Delta M=100 $~GeV, DM co-annihilation is suppressed and the relic density  is equal or below the experimental
limit only for large values of $\lambda_{345}$ ($\lambda_{345} \gtrsim 0.3$) which are excluded by LHC
limits on the invisible Higgs decay, see Eq.~(\ref{lam345-limit-from-inv}).
\item
For $M_{h_1}$ well above 65 GeV, co-annihilation effects become less important
in comparison with $h_1 h_1$ annihilation into vector bosons,
which opens in this region.
For this annihilation process the quartic couplings of DM with longitudinal vector bosons $h_1h_1 V_L V_L$ play an important role. 
For $h_1 h_1 Z_L Z_L$, it is equal to $\tilde\lambda_{345}$ defined in (\ref{tildelam345}), while for
$h_1 h_1 W_L W_L$ it is given by $\lambda_3 = \lambda_{345} + 2(M_{h^+}^2-M_{h_1}^2)/v^2$.
For small mass splittings $\Delta M_c = M_{h^+}-M_{h_1}$ and $\Delta M_2 = M_{h_2}-M_{h_1}$, 
the correspondingly small values of the $h_1h_1 V_L V_L$ quartic couplings 
generate a low $h_1 h_1$ annihilation cross section $\braket{\sigma v}$, which decreases with growing $M_{h_1}$.
Eventually this leads to comparatively high value of $\Omega_{\rm DM} h^2$
(which increases with $M_{h_1}$ both
due to the decrease of $\braket{\sigma v}$ as well as the increase of the DM mass)
which  reaches  the PLANCK limit for large enough $M_{h_1}$ 
as one can see from  Fig.~\ref{fig:1d-mh1-Omega}(a).
On the contrary, for large $\Delta M_c$ and/or $\Delta M_2$, the mass splittings generate
a high rate for $h_1 h_1$ annihilation into vector bosons, which rises with growing $M_{h_1}$. 
This generates a DM density below the experimental limit even for large values of $M_{h_1}$.
In this scenario the potential increase of  $\Omega_{\rm DM} h^2$ due the large DM mass
is compensated by the respective increase of  $\braket{\sigma v}$
and leads to an approximately flat $\Omega_{\rm DM} h^2$ versus $M_{h_1}$
in the 100--1000 GeV  range.
This makes the asymptotic behaviour of the DM density versus $M_{h_1}$ qualitatively different for $\Delta M = 100$ GeV 
as compared to $\Delta M = 1$ GeV, see Fig.~\ref{fig:1d-mh1-Omega}(b).
These two scenarios with the large and small $\Delta M_c$, $\Delta M_2$ mass splittings 
qualitatively cover the whole parameter space of the i2HDM.
\end{itemize} 

\begin{figure}[htb]
\centering
{\includegraphics[width=0.5\textwidth]{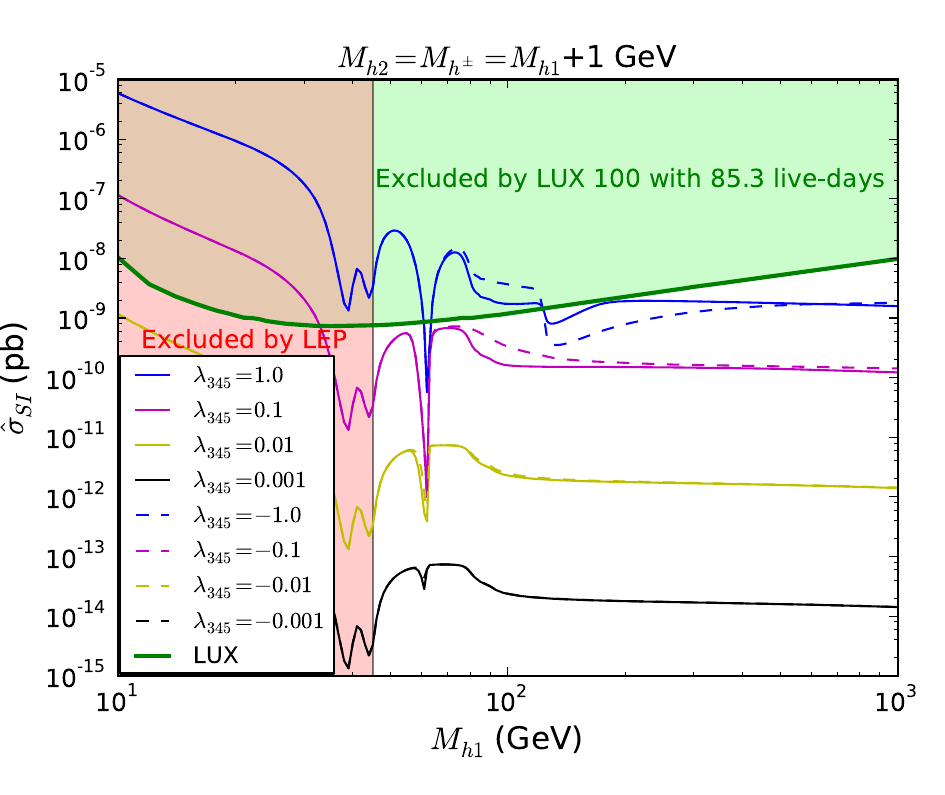}}%
{\includegraphics[width=0.5\textwidth]{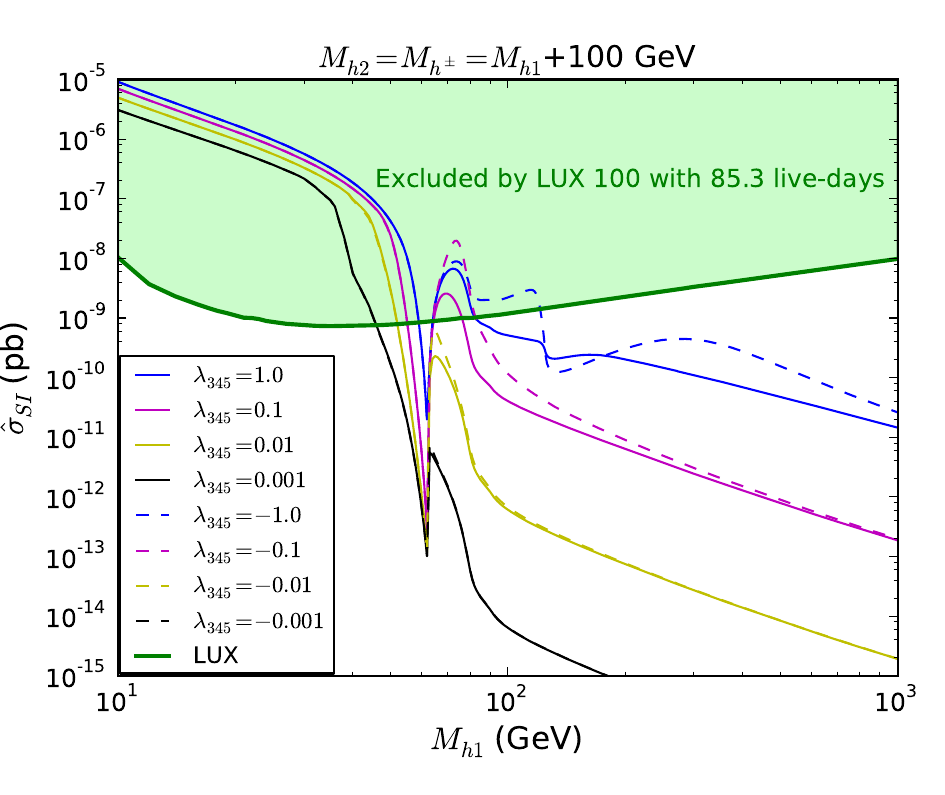}}%
\vskip -0.5cm\hspace*{-3cm}(a)\hspace*{0.48\textwidth}(b)
\caption{Rescaled spin independent direct detection rates $\hat{\sigma}_{SI}$ versus $M_{h_1}$ and the LUX100 constraint.
 The red-shaded region in the left frame is excluded by LEP data.
\label{fig:1d-mh1-DD}}
\end{figure}

We have also checked whether the i2HDM parameter space is consistent with the limits from  DM direct detection (DD) experiments.
We have evaluated the spin-independent cross section of DM scattering off the proton, $\sigma_{SI}$,
 also using the \texttt{micrOMEGAs} package.
In Fig.~\ref{fig:1d-mh1-DD} limits from LUX100 are shown by the shaded green area where the left and right frames
illustrate the small and large $\Delta M$  scenarios as in Fig.~\ref{fig:1d-mh1-Omega}.
To present the results in Fig.~\ref{fig:1d-mh1-DD}, we use the re-scaled DD cross section, $\hat{\sigma}_{SI}= 
R_\Omega\times \sigma_{SI}$, where the scaling factor
$R_\Omega = \Omega_{\rm DM}/\Omega^{\rm Planck}_{\rm DM}$ takes into account the case of $h_1$
representing only a part of the total DM budget, thus allowing for a convenient comparison
of the model prediction with the limits from LUX \cite{Akerib:2013tjd}.

The flat asymptotic of $\hat{\sigma}_{SI}$ in Fig.~\ref{fig:1d-mh1-DD}(a)
for high $M_{h_1}$ means that the decrease of the 
proton-DM scattering cross section ${\sigma}_{SI}$ with increasing $M_{h_1}$ is compensated by the 
growth of the relic density which one can observe in Fig.~\ref{fig:1d-mh1-Omega}(a).
{In Fig.~\ref{fig:1d-mh1-DD}(b), on the other hand, $\hat{\sigma}_{SI}$
drops with large and increasing values of $M_{h_1}$.
This can be understood by observing from Fig.~\ref{fig:1d-mh1-Omega}(b) that in this region
$R_\Omega = \Omega_{\rm DM}/\Omega^{\rm Planck}_{\rm DM}\simeq$ constant, and therefore 
the asymptotic behaviour of $\hat{\sigma}_{SI}$ is the same as for ${\sigma}_{SI}$,
that is, it goes down as $M_{h_1}$ grows due to the reduced $\braket{\sigma v}$.}


A related question is whether the model can be better probed by   indirect detection (ID) experiments, i.e. the detection of   energetic cosmic rays like $e^+$, $\gamma$, $p$ or $\bar{p}$, which may be created by the annihilation of $h_1$ pairs.
We have checked  that the strongest bounds on the i2HDM parameter space
 coming from such experiments are set by gamma ray telescopes: both the Fermi-LAT gamma-ray space telescope~\cite{Ackermann:2011wa} as well as ground based telescopes. Fermi-LAT is sensitive to gamma rays particularly in the low mass range  up to $\mathcal{O}(100\,\mathrm{GeV})$, but the bounds are not competitive with those coming from DD. 
This conclusion is also confirmed by studies in Ref.~\cite{Arhrib:2013ela}.
Another recent work, Ref.~\cite{Eiteneuer:2017hoh}, indirectly confirms that the present Ferm-LAT data do not place additional
strong constraints on the i2HDM parameter space.
The authors of that work looked at the so-called gamma-ray Galactic center excess~\cite{Goodenough:2009gk}
and asked if it can be explained via the DM annihilation in i2HDM.
They indeed found a few possible regions, and estimate that one would need 15 years of Fermi-LAT data 
to conclusively test it. In our work, we stay conservative and do not interpret such signals
as the DM evidence. We can only state that the regions selected in Ref.~\cite{Eiteneuer:2017hoh} as promising are, at best, 
not excluded by Fermi-LAT data. Since the other regions were not favored by the Galactic centre excess in the Fermi-LAT data,
their expected contributions to the indirect detection signals are weaker. 
Finally, we stress that incorporating indirect detection limits into our picture would bring novel uncertainties 
of purely astrophysical origin such as the poorly known DM distribution profile.  
Thus, they cannot be easily translated into a new constraint on the i2HDM parameter space.

\section{Numerical scan of the parameter space\label{sec:num-scan}}

\subsection{Results of the general scan}

To have a complete picture of the properties of i2HDM in the whole parameter space, we have performed a
five-dimensional random scan of the model parameter space with about $10^8$ points, evaluating all relevant
observables and limits mentioned above. The range for 
the model parameters of the scan was chosen according to the Eq.~(\ref{eq:scan-limits}).

When performing the scan, we took into account the constraints mentioned above
in the following succession. First, we applied only theoretical constraints from vacuum stability, 
perturbativity, and unitarity;
second, we applied the collider constraints (LEP, EWPT, LHC Higgs data);
last, we placed the upper bound on the DM relic density at $\Omega_{\rm DM} h^2 \le 0.1184+2\times 0.0012$ 
given by the PLANCK result plus 2 standard deviations, 
and took into account the negative results of the DM DD searches at LUX.

\begin{figure}[tbh]
\begin{center}
\begin{tabular}{p{0.3\textwidth} p{0.4\textwidth} p{0.3\textwidth}}
\hspace*{1cm}theory constraints &+LEP+EWPT,LHC(Higgs) &\hspace*{-1cm}+relic density, LUX
\end{tabular}
\includegraphics[trim={0.5cm   0     2.4cm     0.3cm},clip,height=0.2\textheight]{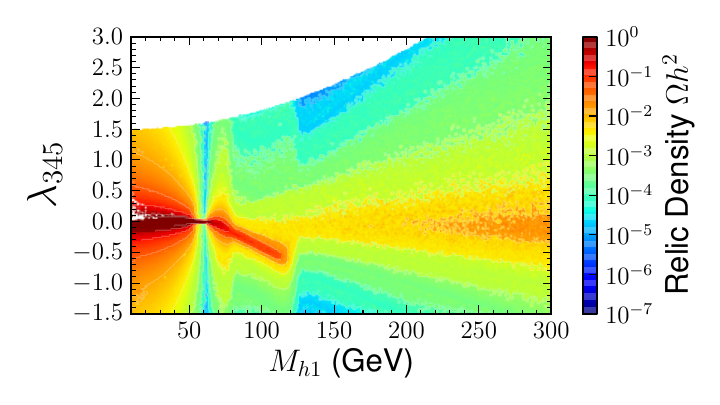}%
\includegraphics[trim={1.3cm 0   2.4cm   0.3cm},  clip,height=0.2\textheight]{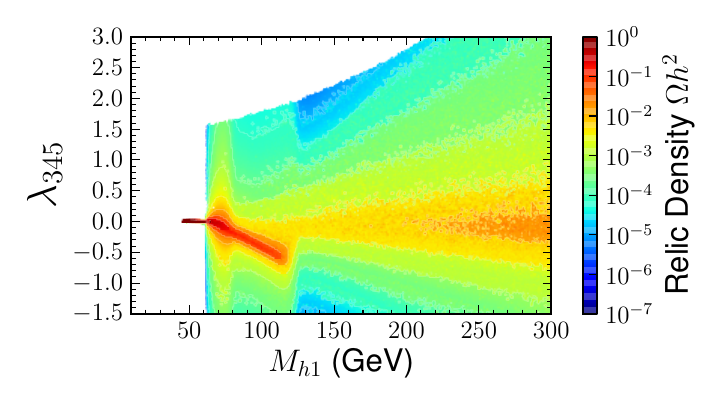}%
\includegraphics[trim={1.3cm 0   0      0.3cm },  clip,height=0.2\textheight]{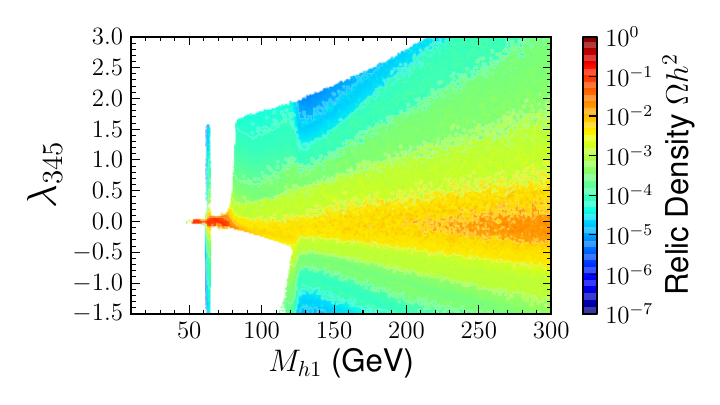}
\caption{Colour maps of DM relic abundance projected on the plane $(M_{h_1},\lambda_{345})$. The three plots correspond to the surviving points after progressively imposing the three layers of constraints described in the text. \label{fig:scan-simplified}} 
\end{center}
\end{figure}

The salient features of the results of this scan,
with all three groups of constraints applied successively, 
are presented in Fig.~\ref{fig:scan-simplified} on the $(M_{h_1},\lambda_{345})$ projection.
The results are presented in the form of color maps, where the color encodes the value of the thermal relic density. The points with higher relic density are always  on top of those with low density.
To make the exposition as clear as possible, we decided to show here only this specific 2D projection. 
The reader can find more information and insights in Appendix~\ref{app:numerical}, which contains more projections and a more detailed description of the effects of each set of constraints.
Our scan highlights the following features:

\begin{itemize}

\item The lower bound of $\lambda_{345}$ corresponds to the theoretical lower limit in Eq.~(\ref{eq:scan-limits}).
The upper bound on $\lambda_{345}$ depends on $M_{h_1}$ and comes from 
the vacuum stability condition given in Eq.~(\ref{eq:l345-vacuum-stab}). 
Taking into account the collider constraints, and in particular the invisible Higgs decay
and the $H\gamma\gamma$ coupling strength, restricts $\lambda_{345}$
to $|\lambda_{345}|\leq 0.02$ at $M_{h_1}<M_H/2$.

\begin{figure}[tbh]
\begin{center}
\includegraphics[trim={0.3cm 0  2.6cm 0.5cm},clip,height=0.26\textheight]{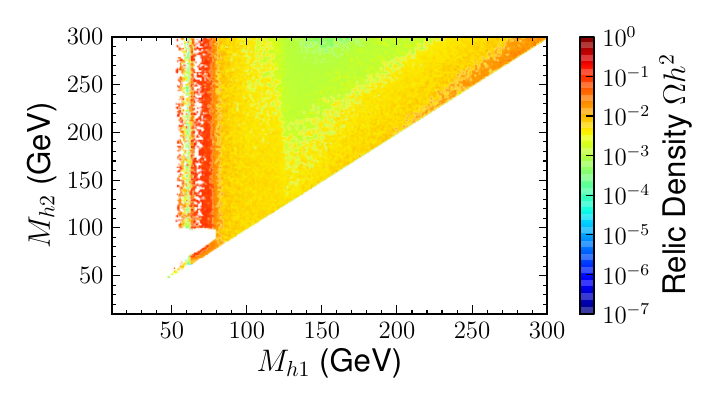}%
\includegraphics[trim={0.3cm 0   0    0.5cm },clip,height=0.26\textheight]{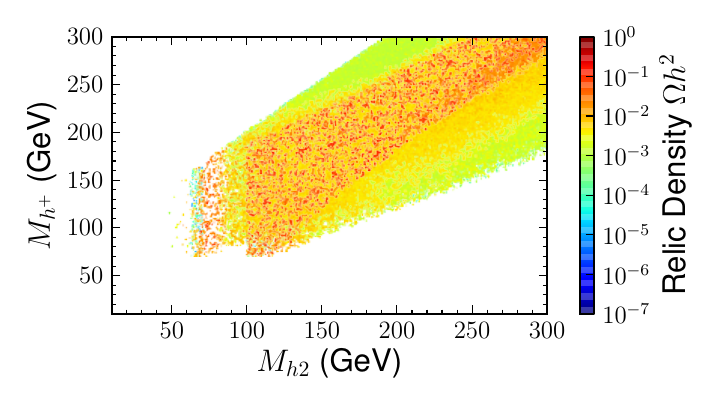}
\caption{Colour maps of DM relic abundance projected on the planes $(M_{h_1}, M_{h_2})$ and $(M_{h_2}, M_{h^+})$, with all the constraints imposed.\label{fig:scan-masses}} 
\end{center}
\end{figure}

\item 
The LEP and LHC data also place constraints on the other inert scalars. 
Charged scalars lighter than 70 GeV as well as $M_{h_2} < M_Z/2$ are generically excluded. 
For $M_{h_2}$ above this value and below approximately 100 GeV, 
the only surviving region is a narrow strip with $M_{h_2} - M_{h_1} < 8$ GeV.
The effect from these constraints can be seen in Fig.~\ref{fig:scan-masses} which shows the points surviving all constraints in a projection in the masses, and more details can be found in Appendix~\ref{app:numerical}.

\item The narrow strip at $M_{h_1}<M_H/2$ surviving after collider data,
which is seen in Fig.~\ref{fig:scan-simplified}(b),
is further cut off once the relic density constraint, in addition,
is taken into account. Indeed, for such a small $M_{h_1}$ and $\lambda_{345}$
(and not too small $M_{h_2}$ to prevent $Z \to h_1 h_2$ decays), 
there remains no mechanism for sufficiently active removal of DM 
in the early universe. The resulting DM relic density turns out too high and is ruled out. We already saw this feature 
in Fig.~\ref{fig:1d-mh1-Omega}b. Values of $M_{h_1} > 45$~GeV are still allowed but they require
a close $M_{h_2}$ for an efficient coannihilation in the early Universe.
This region is well visible as a protrusion in the $(M_{h_1},M_{h_2})$ in Fig.~\ref{fig:scan-masses}.

\item The masses $M_{h_1} > M_H/2$ are not constrained by the relic density,
but the DM DD results from LUX cuts off a part of the parameter space.
This is visible in the last plot in Fig.~\ref{fig:scan-simplified},
for $M_{h_1}\lesssim M_H$ and with large and intermediate $|\lambda_{345}|$.
In this region, the sizable $|\lambda_{345}|$ simultaneously keeps the relic density below the Planck upper bound
and allows the scattering cross section to be enhanced due to the Higgs boson exchange.
The interplay of {\em moderately low} relic density and a sizable cross section leads to a DD signal
which could have been seen by LUX.
For larger DM masses, the direct annihilation into $WW$, $ZZ$, $HH$ pairs opens up,
and the relic density drops further, making LUX insensitive to this region.

\item Finally, we remark that above 200 GeV, EWPT forces $M_{h_2}$ and $M_{h^{+}}$ to stay rather close
to each other, see again Appendix for more details.

\end{itemize}

In summary, after all constraints mentioned here and exposed in more detail in Appendix~\ref{app:numerical},
we found that the parameter space with  
\begin{equation}
M_{h_1},M_{h_2}<45~\mbox{GeV} 
\mbox{\ or\ } M_{h^+}<70~\mbox{GeV} 
\label{eq:exclusion1}
\end{equation}
is completely excluded. Our results  
agree with the conclusions of previous studies on the i2HDM (see, e.g., \cite{Arhrib:2013ela,Ilnicka:2015jba}).
In particular, authors of \cite{Ilnicka:2015jba} have also stated 
the $M_{h_1},M_{h_2}<45~\mbox{GeV}$ limit.
However we would like to stress that the general exclusion 
for $M_{h_1},M_{h_2}$ {\it and for} $M_{h^+}$
given by Eq.~(\ref{eq:exclusion1}) is established here for the first time, to the best of our knowledge. In~\cite{Ilnicka:2015jba}, for example,  the authors demonstrate 
(see Fig.~6 and Eq.~(18) in \cite{Ilnicka:2015jba}) that 
$M_{h^+}$ above $M_H$ is excluded from a  specific scan. Here we find that
$M_{h^+}$ as light as 70 GeV  is allowed by all present constraints, while $M_{h_1}$ and
$M_{h_2}$ are generically allowed to be as light as 45 GeV. One should note that specific regions of the
parameter space can be excluded using di-lepton and missing transverse momentum
signatures: for example, in a recent study~\cite{Belanger:2015kga} the authors showed that values of the masses below
$M_{h_1}\lesssim 50$~GeV and  $M_{h_2}\lesssim 140$~GeV can be excluded using this signature,
provided that the mass gap between $M_{h_2}$ and $M_{h_1}$ is large enough. However, we find that this parameter
space region is already excluded by the upper cut on the relic density, as one can see
from Fig.~\ref{fig:scan-masses}: for  $M_{h_2}>100$~GeV, the entire region $M_{h_1}\lesssim
50$~GeV  is excluded by the relic density cut combined with previous constraints including LEPII limits.

\subsection{Fitting the relic density}

In our analysis, we generically allow the DM relic density to be equal or
below the PLANCK constraints, Eq.~(\ref{eq:planck-limit-relaxed}).
This is the concept of our approach: we assume that in the case of under-abundance
there should be either additional sources of DM or mechanisms other than thermal freeze-out that 
compensate for the DM deficit, such as DM freeze-in scenarios~\cite{Hall:2009bx}.
Keeping this in mind, we exclude in our analysis only those regions of the parameter space 
where the relic density exceeds the PLANCK constraint.

\begin{figure}[htb]
\begin{center}
\includegraphics[width=0.9\textwidth]{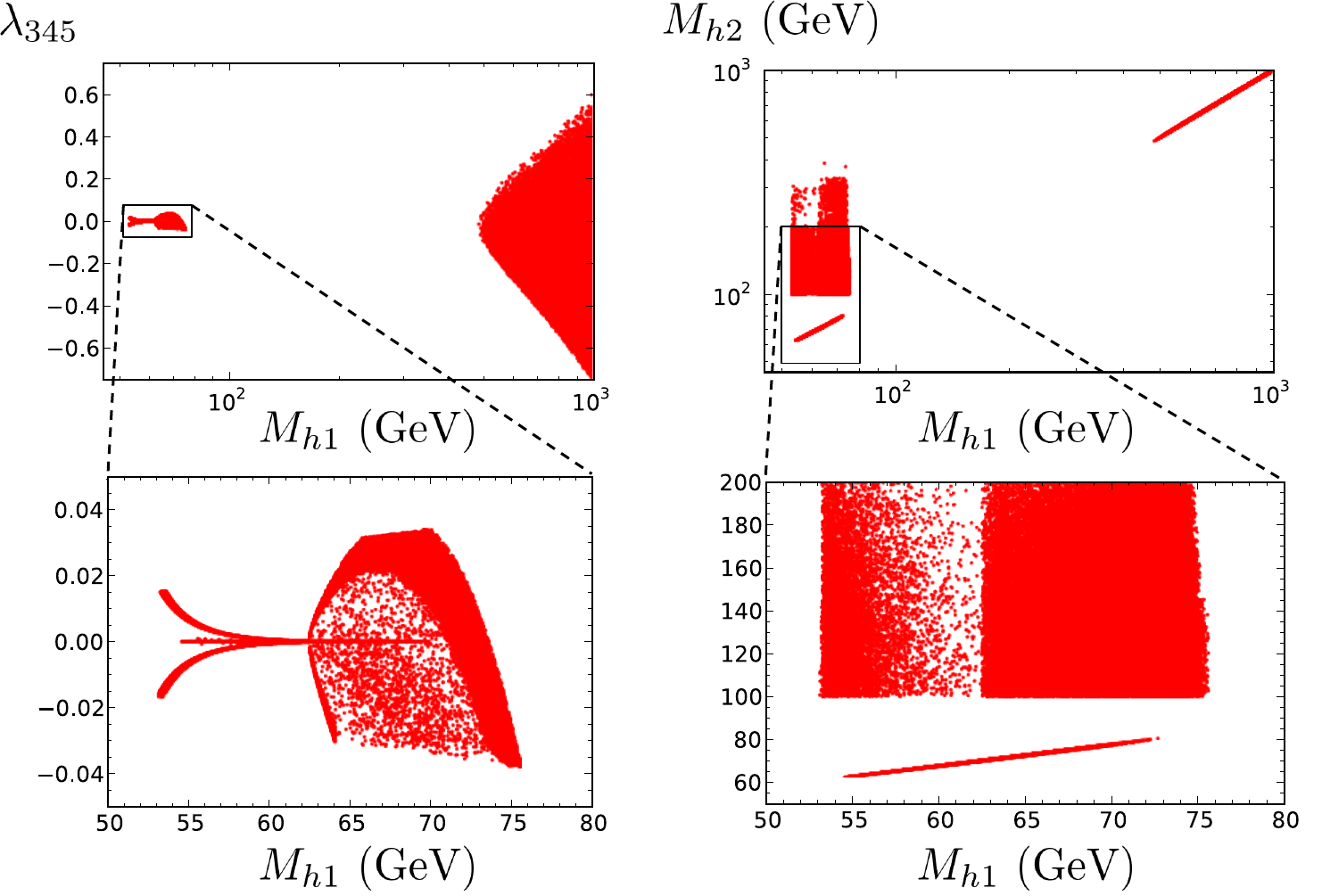}
\caption{Projection on two planes of the scan points passing all constraints, and fitting the PLANCK relic abundance within 2 sigmas. We show both a wide scan with masses between 10 and 1000 GeV, and a zoom on the low mass region.  \label{fig:scan-simplified-fitting}} 
\end{center}
\end{figure}

However, it is also instructive to explore the parameter space where both
the upper and lower PLANCK limits are satisfied.
This parameter space region is presented in Fig.~\ref{fig:scan-simplified-fitting},
for a wider scan 10 GeV $< M_{h_1}, M_{h_2}, M_{h^{+}} < 1000$~GeV and, separately,
for the ``zoomed'' region 10 GeV $< M_{h_1}, M_{h_2}, M_{h^{+}} < 200$~GeV.
Here, we show the two most revealing 2D projections: $(M_{h_1},\lambda_{345})$ and $(M_{h_1},M_{h_2})$. 
Additional plots can be found in Appendix~\ref{app:numerical}.

Many interesting features of the i2HDM parameter space arise once the ``correct'' amount of DM relic density is required. 
One observes two very distinct $M_{h_1}$ regions:  
a low mass region for $53~\mbox{GeV} \lesssim M_{h_1} \lesssim 76~\mbox{GeV}$, shown also in the zoomed panels,
and a high mass region for $M_{h_1}\gtrsim 490$ GeV. Below, we discuss them separately.

\subsubsection{The low mass region}

In the low mass region we clearly distinguish three regimes with specific physical properties:
\begin{itemize}
\item[a)] A thin horizontal line with very small values of $\lambda_{345}$ can be seen, corresponding to  $h_1h_2$ co-annihilation: this region is novel and it has been missed in previous studies. 
It can also be seen in the $(M_{h_1},M_{h_2})$ plots as the thin diagonal strip at low $M_{h_2}$
starting from 54 GeV and extending beyond $M_H/2$ up to about 73 GeV.
%
%

{The width of this strip is defined by 
the maximum allowed value of $\Delta M = M_{h_2}-M_{h_1}=8$~GeV, above which
the parameter space is excluded by LEP di-lepton searches until $M_{h_2} > 100$ GeV (see Eq.~(\ref{eq:dmh12})). 
In this allowed region DM relic density is never above PLANCK limit given by Eq.~(\ref{eq:planck-limit-relaxed}).
The maximum value of $\Omega h^2_{DM}$ reaches
the value of about 0.11 for  $M_{h_2}-M_{h_1} \simeq 8$~GeV
and $\lambda_{345}\simeq 0$, when the only 
${h}_1-{h}_2$ co-annihilation takes place.
For  $\Delta M<8$~GeV
and $M_{h_1}<54$~GeV, the $\Omega h^2$ is below the $0.118-2\times0.012$ limit which use in our study
because ${h}_1-{h}_2$ co-annihilation via $Z$-boson increases 
with  he decrease of ${h}_1,{h}_2$ masses.
On the other hand, the upper edge
at $73$ GeV is defined by the rapid increase of the  $h_1 h_1 \to W W^*$ contribution,
which does not require co-annihilation above this mass. The typical $M_{h_2}-M_{h_1}$
mass split in the co-annihilation region is 7-8 GeV, as required to make the relic density 
consistent with the PLANCK limit.
}

\item[b)] For $M_{h_1}< M_H/2$, two symmetric wings can also be seen, extending for positive and negative $\lambda_{345}$, and 
clearly visible in the zoomed $(M_{h_1},\lambda_{345})$ panel. They correspond to DM annihilation via the Higgs boson exchange. 

\item[c)] In the region $M_H/2<  M_{h_1} \lesssim 76~\mbox{GeV}$,  
large absolute values of  $\lambda_{345}$ are allowed by the LHC Higgs data, however LUX data requires $|\lambda_{345}|$ to be below about 0.04.
In this region, we remark the asymmetric pattern in the ($\lambda_{345},M_{h_1}$) plane
for positive and negative values of $\lambda_{345}$, which is related, respectively, to the 
positive and negative interference of  $h_1 h_1 \to VV$ ($V=Z,W$) annihilation diagrams via Higgs boson exchange
and diagrams with quartic $h_1 h_1 V V$  interactions. 
\end{itemize}

\subsubsection{The high mass region and the LHC sensitivity}

The relic density can also be ``just right'' at large masses $M_{h_1}\gtrsim 490\mbox{ GeV}$, as shown in Fig.~\ref{fig:scan-simplified-fitting}. 
The most salient feature of this high-mass region is the high degree of degeneracy among the three inert Higgs boson masses.
This is clearly seen in the upper right corner of the $(M_{h_1},M_{h_2})$ plane, as well as in
Fig.~\ref{fig:dm-i2hdm-relic} in the Appendix. Numerically we find that the
maximal mass difference among $h_1$, $h_2$ and $h^+$, that we call $\Delta M^{max}$, does not exceed a few GeV. 

Remarkably, the mass split is required to be {\it large enough}, so that the relic density can reach the {\it lower value} of the PLANCK limit: the increase of  the mass
split is, in fact, correlated with the increase of the quartic coupling $h_1 h_1 V_L V_L$ of the DM to longitudinal $Z$ and $W$ bosons,
which enhances the $h_1 h_1$ annihilation cross section, thus bringing
the DM relic density down to  the experimental limits. Due to the connection between the mass split and the $h_1 h_1 V_L V_L$ couplings, 
see Eq.~(\ref{tildelam345}),
this effect is actually stronger than the effect of the
$h_1$, $h_2$ and $h^+$ co-annihilation, which becomes sub-dominant in this high-mass region.
One should also mention that  $\Delta M^{max}$ of the order of few GeV  is {\it generically} not 
small enough to lead to long-lived $h_2$ or $h^+$ at detector level.
However, in the small mass tip, in the interval
$550\mbox{ GeV} \gtrsim M_{h_1}\gtrsim 490\mbox{ GeV}$,  $\Delta M^{max}$ can take values 
about 0.2~GeV. This specific range of the mass split simultaneously provides an $\Omega_{\rm DM} h^2$ consistent with 
PLANCK constraint and a life-time for $h^+$ large enough to travel  about 10 cm or more in the  detector, thus
providing disappearing charged track signatures which have been recently explored  by  CMS~\cite{CMS:2014gxa}
collaboration.

{This small $\Delta M$ region deserves a special discussion which we give below.
For the mass splitting $\Delta M\simeq 0.2  {\rm GeV} = 200 {\rm MeV}$ the $h^+$ will dominantly decay into $\pi^+ h_1$
as soon as $\Delta M = M_{h^+}-M_{h_1} > m_{\pi^+} \simeq 140~{\rm MeV}$. This happens because when the $\Delta M$ is of the order of pion mass, the naive perturbative calculation of $h^+ \to h_1 W^{+*} \to  h_1 u\bar{d}$ underestimates the 
width by about one order of the magnitude and therefore overestimates
 the life-time of $h^+$ by the same amount. 
For proper evaluation of the life time which is crucial for the collider phenomenology we have used the non-perturbative 
$W-\pi$ mixing, 
\begin{equation}
  \mathcal{L}_{W\pi}  = \frac{g f_{\pi}}{2\sqrt{2}}W_{\mu}^{+}\partial^{\mu}\pi^{-} + 
\mathrm{h.c.}
\label{eq:wpi}
  \end{equation}
leading to the effective Lagrangian for $h^+ \to  h_1 \pi^+$ interactions,
which in momentum space reads as:
  \begin{equation}
\mathcal{L}_{h^+h_1\pi^-}  =  
\frac{i g^2 f_{\pi}}{4\sqrt{2}M_W^2} 
(p_{h^+}-p_{h_1})\cdot p_{\pi},
\label{EFT-pion}
\end{equation}
where  $f_{\pi}=130$ MeV is the  pion decay constant. This effective Lagrangian is represented by the  diagram  shown in Figure~\ref{fig:decay} with the virtual $W$ boson line contracted to point-like interaction.
\begin{figure}[htb]
  \centering
  \includegraphics[scale=0.5]{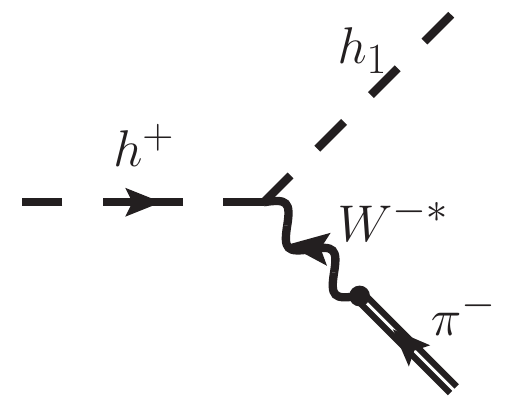}
  \caption{Feynman diagram representing effective 
  $h^+ \to  h_1 \pi^+$ via $W-\pi$ mixing.
  }
  \label{fig:decay}
\end{figure}

From the  Lagrangian above one can find 
the following formula
for the $h^+ \to  h_1 \pi^+$ width in the $\Delta M/M\ll 1$
limit:
\begin{equation}
\Gamma=\frac{g^4 f_{\pi}^2}{64 \pi M_W^4}
\Delta M^2\sqrt{\Delta M^2-m_{\pi^+}^2},
\end{equation}
where $g$ is the weak coupling constant.
\begin{figure}[htb]
  \includegraphics[width=0.5\textwidth]{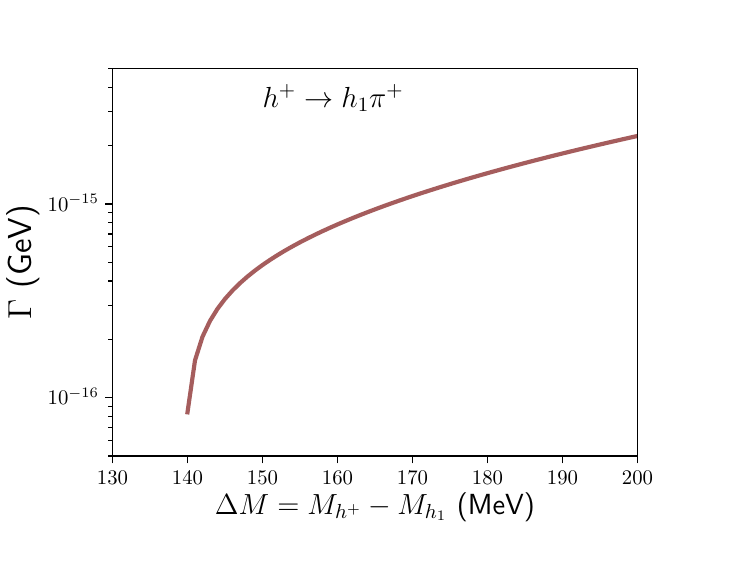}%
  \includegraphics[width=0.5\textwidth]{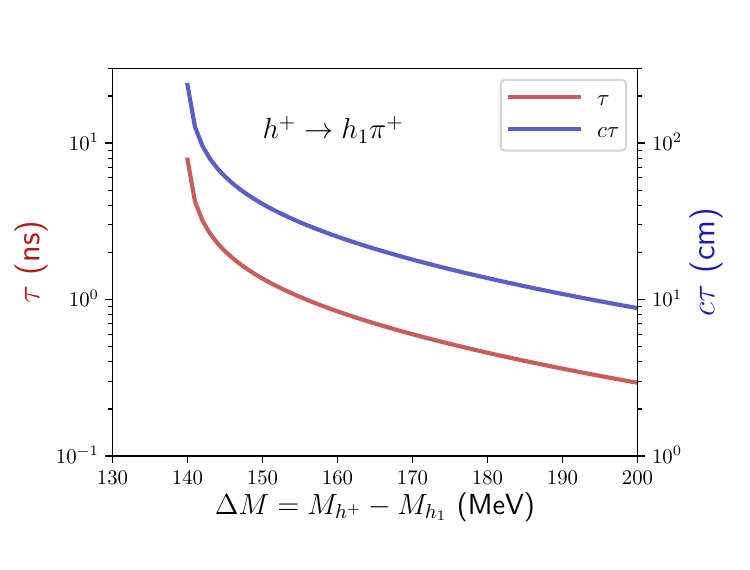}%
  \caption{ The decay width of $h^+$ (left) as well as its lifetime together with the decay length (right).
  }
  \label{fig:gamma}
\end{figure}
In Fig.\ref{fig:gamma} we present the decay width  of $h^+$ (left) as well as lifetime and decay length (right)
which are functions of $\Delta M$ only. One can see that, for $\Delta M$ in the range 140-200 MeV,
$h^+$ will provide disappearing charged track signature with the length of 100-10 cm respectively.
For $\Delta M$ below the pion mass the width is defined by the $h^+\to W^{+*}h_1 \to e^+\nu_e h_1$
process and drops to the level of $10^{-18}$~GeV or below,
meaning that $h^+$ becomes collider stable
and goes trough the whole detector. For such a small mass split, the width of  $h^+$ is proportional to  $\Delta M^5/M_W^4$.

\begin{figure}[htb]
\begin{center}
  \includegraphics[width=0.7\textwidth]{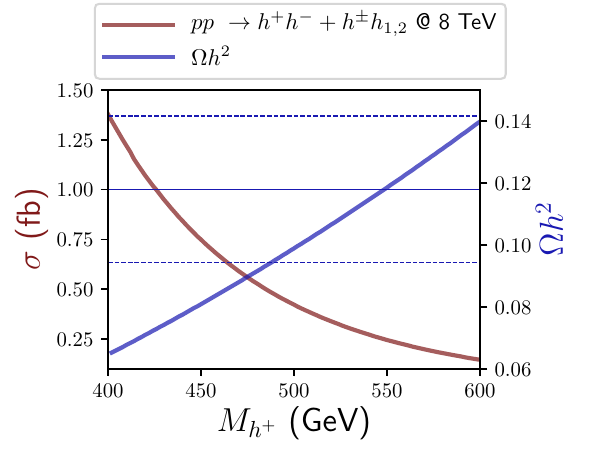}%
  \caption{The $\Omega h^2$ value for $\Delta M/M\ll 1$ and $\lambda_{345}$ parameter space and the $pp\to h^+h^- + h^\pm h_{1,2}$ 
production rate  evaluated for the LHC@8TeV}
  \label{fig:sigmah+}
\end{center}
\end{figure}
The next step is to check which $M_{h^+}$ mass range 
in the $\Delta M/M\ll 1$ and $\lambda_{345}$ parameter space 
is consistent with the relic density constraints
given by Eq.(\ref{eq:planck-limit-relaxed}).
Let us stress that in the  $\Delta M/M\ll 1$ and $\lambda_{345}\simeq 0$ regime 
the relic density is maximised for $M_{h^+}\simeq 500$~GeV or above
(this happens because the boost of DM annihilation into the gauge bosons
 dominates the reduction of co-annihilation effects
with $\Delta M/M$ increase).
In Fig.~\ref{fig:sigmah+} we present 
the $\Omega h^2$ value for $\Delta M/M\ll 1$ and $\lambda_{345}\simeq 0$ parameter space
which grows with $M_{h^+}$ and becomes consistent with upper and lower limits for relic density constraints in the range of $490< M_{h^+} <600$ GeV  
masses.

At the same time the figure presents the $pp\to h^+h^- + h^\pm h_{1,2}$ 
production rate which we have evaluated for the LHC@8TeV
with NNPDF23LO (\verb|as_0119_qed|) 
PDF set~\cite{Ball:2012cx}
and QCD scale chosen to be equal to the averaged transverse mass of the
final state particles.

After applying efficiency for the  disappearing charged track signatures provided by  CMS~\cite{CMS:2014gxa}
as a function of charged track transverse momentum as well as efficiency for distance travelled by the charge particle,
we have estimated  that CMS@8TeV with 19.5fb$^{-1}$ data excludes  $h^+$  in the 490-550 GeV mass range for   $\Delta M=140-200$~MeV.
For example, for $M_{h^+}=500$~GeV the sum of the cross section of $pp\to{h^+}{h^-}$ and  $pp\to{h^\pm}{h_{1,2}}$
is about 0.4 fb, and the product of this cross section, the luminosity, and the above efficiencies gives about 2.5 events
which are above 2 event exclusion level.

One should also note that with increasing DM mass, the required split between $h_1$, $h_2$ and $h^+$
increases.
At about 20 TeV for $M_{h_1}$, the DM relic density constraint together with 
requirement of unitarity and perturbativity which are saturated by  $\Delta M^{max}\simeq 10$~GeV,
close the i2HDM parameter space.
}



\section{Probing Dark Matter signals from i2HDM at the LHC}

The i2HDM exhibits various signatures that are potentially accessible at the LHC.
They can be generically described as ``mono-object production'', that is,
production of several final states in association with large missing transverse momentum.
In this section, we undertake a detailed exploration of such processes which goes 
beyond the previously published state-of-the-art.
We will first list the relevant processes, then produce a cumulative plot which
helps us compare their rates. 
With this knowledge, we will formulate convenient benchmark points which
represent various qualitatively distinct regimes of i2HDM, and finally go into a more detailed
calculation of monojet production.

\subsection{Dark Matter signatures: diagrams and features} \label{sec:DMsignatures}

\subsubsection{Mono-jet production}

\begin{figure}[htb!]
\begin{center}
\includegraphics[width=0.9\textwidth]{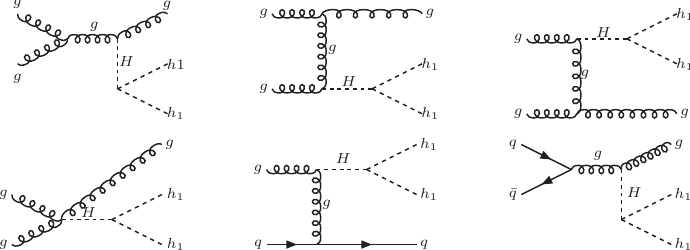} 
\caption{Feynman diagrams for the $pp\to h_1 h_1j$ process
contributing to a mono-jet signature.}
\label{fig:fd-monojet1}
\end{center}
\end{figure}

The mono-jet signature originates from the $pp\to h_1 h_1j$ process,
the Feynman diagrams for which are presented in Fig.~\ref{fig:fd-monojet1}.
For this process, the relevant non-trivial  parameter space is  one dimensional:
it is just the DM mass, $M_{h_1}$,  since the second parameter, $\lambda_{345}$,
simply scales the production cross section which is proportional to  $(\lambda_{345})^2$
for $M_{h_1}>M_H/2$.
One should note  that the mediator mass for this signature is the Higgs mass, $M_H = 125$ GeV, thus
the Effective Field Theory (EFT) approach is not applicable for this process. Also, the recent limits by ATLAS
\cite{Diehl:2014dda} and CMS \cite{Chatrchyan:2012me, Khachatryan:2014rra} collaborations  are not directly
applicable for this process since they have been obtained for a different spin of the mediator and different
spin of DM. 

\begin{figure}[htb!]
\begin{center}
\includegraphics[width=0.9\textwidth]{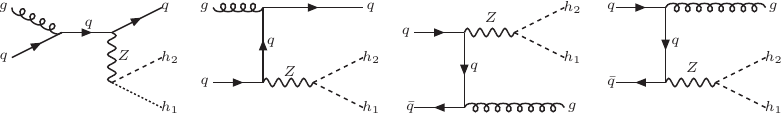} 
\caption{Feynman diagrams for $q\bar{q}\to h_1 h_2g$ ($gq\to h_1 h_2q$) process 
contributing to mono-jet signature.}
\label{fig:fd-monojet2}
\end{center}
\end{figure}

There is one more process, namely $q\bar{q}\to
h_1 h_2g$ ($gq\to h_1 h_2q$) (see diagrams in 
Fig.~\ref{fig:fd-monojet2}),  that can contribute to a mono-jet signature in the special case of a small mass split between $h_1$ and $h_2$.
In this scenario $h_2$ decays to $h_1$ plus soft 
jets and/or leptons. 
The essential parameter space for this process is the
two-dimensional  ($M_{h_1},M_{h_2}$) plane
which fixes its cross section. This channel is particularly relevant in the $\Lambda_{345}\sim 0$ region at low mass that we discovered in this study.

\subsubsection{Mono-$Z$ production}

\begin{figure}[h!]
\begin{center}
\includegraphics[width=0.9\textwidth]{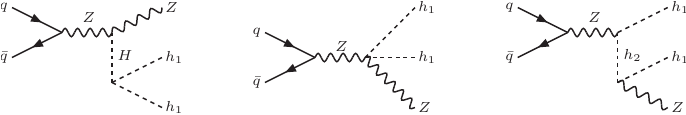} 
\caption{Feynman diagrams for $q\bar{q}\to h_1 h_1Z$  process 
contributing to mono-Z signature.}
\label{fig:fd-mono-Z}
\end{center}
\end{figure}

Besides mono-jets, the i2HDM gives rise to a mono-$Z$ signature, the 
diagrams for which are presented in Fig.~\ref{fig:fd-mono-Z}.
The first diagram scales with $\lambda_{345}$ while the other two are fixed by electroweak interactions.\footnote{For the second diagram, the $ZZh_1h_1$ vertex for transverse $Z$-bosons is fixed by the weak coupling,
while for longitudinal $Z$-boson it scales with with $\tilde\lambda_{345}$ in Eq.~(\ref{tildelam345}). 
When this coupling is small, the strength of the  $ZZh_1h_1$ vertex therefore
is fixed by the gauge interactions.}
In general, non-trivial interference takes place between the three different topologies represented by each of three 
diagrams, so  this process cannot be approximated by a simplified model.
However, we found that when $|\lambda_{345}|\gtrsim 0.02$ with $M_{h_1}< M_H/2$ (below the Higgs boson threshold)
or $|\lambda_{345}|\gtrsim 1$ with $M_{h_1}> M_H/2$ (above the Higgs boson threshold),
the first diagram is dominant and defines the event kinematics. 
So for these values of $\lambda_{345}$ and $M_{h_1}$, a simplified
model with the Higgs boson as the mediator is sufficient to set the LHC limits.

One should also note that for values of $|\lambda_{345}|$
below $0.02$ the contribution from diagrams scaling with $|\lambda_{345}|$
drops below $1\%$. In this case the
$Z$ boson will be the only mediator to probe the i2HDM model at the LHC,
with the mono-$Z$ process being the leading signature
for this purpose (and not only as a probe complementary to the mono-jet signature). 
This signature will be especially pronounced if $M_{h_2}-M_{h_1} > M_Z$,
so that the cross section of the mono-$Z$ signature
is essentially defined by the cross section of the $2 \to 2$ process,
$pp\to h_1 h_2 \to h_1 h_1 Z$.
For small mass split, like it happens in the $\lambda_{345} \sim 0$ region, the mono-$Z$ will complement the signal in mono-jet described above.
The parameter space for this process is
the two-dimensional $(M_{h_1},M_{h_2})$ plane.

\subsubsection{Mono-Higgs production}

\begin{figure}[h!]
\begin{center}
\includegraphics[width=0.9\textwidth]{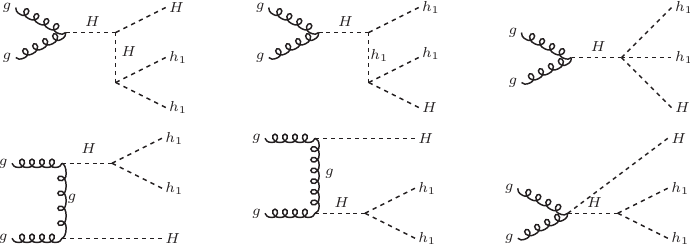} 
\caption{Feynman diagrams for $gg\to h_1 h_1H$  process 
contributing to mono-Higgs signature.}
\label{fig:fd-mono-H1}
\end{center}
\end{figure}

\begin{figure}[h!]
\begin{center}
\includegraphics[width=0.9\textwidth]{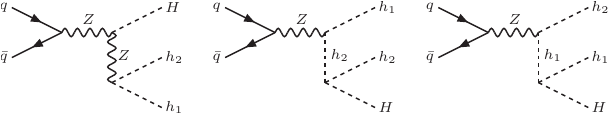} 
\caption{Feynman diagrams for $q\bar{q}\to h_1 h_2H$  process 
  contributing to mono-Higgs signature.}
\label{fig:fd-mono-H2}
\end{center}
\end{figure}

The i2HDM could also provide a mono-Higgs signature
via $gg\to  h_1 h_1 H$ and $q\bar{q} \to  h_1 h_2H$,
whose diagrams are presented in Fig.~\ref{fig:fd-mono-H1}
and Fig.~\ref{fig:fd-mono-H2} respectively.
The only mediator for $gg\to  h_1 h_1 H$ is the Higgs boson,
and the respective cross section scales as $(\lambda_{345})^2$
for small values of $\lambda_{345}$ 
and $(\lambda_{345})^4$ for large values of $\lambda_{345}$
because of the second diagram.
On the other hand, the $q\bar{q}\to h_1 h_2H$  process 
takes place via either a $Z$-boson or an $h_2$ as a mediator:
the first diagram does not scale with $\lambda_{345}$,
while the last two do. Therefore for large $\lambda_{345}$,
the $(\lambda_{345})^2$  scaling takes place for $q\bar{q}\to h_1 h_2H$ process.
In fact,  the contribution from the second and 
the third  diagrams of  $q\bar{q}\to h_1 h_2H$
to the total cross section
drops below 1\% only for $\lambda_{345}<0.002$,
below which the process kinematics and the cross section 
are determined by the first diagram with two $Z$-boson propagators.

\subsubsection{Vector boson fusion}

\begin{figure}[h!]
\begin{center}
\includegraphics[width=0.9\textwidth]{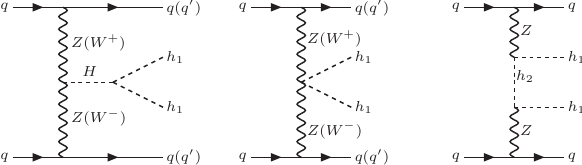} 
\caption{Diagrams for $qq\to q^{(')}q^{(')} h_1 h_1$ DM production in vector boson
  fusion process.}
\label{fig:fd-vbf}
\end{center}
\end{figure}
Finally, one should mention the production of DM via vector boson fusion, $pp\to h_1 h_1 jj$, the diagrams for which are
presented in Fig.~\ref{fig:fd-vbf}.
Similarly to the mono-$Z$ process, there are three diagrams with different
topologies and mediators which contribute to this process; thus,
it cannot be described by  just one simplified model.
The first two diagrams scale with $\lambda_{345}$.
To be accurate, the $Z_L Z_L h_1 h_1$ coupling in the second diagram is proportional to 
$\tilde\lambda_{345}$, see Eq.~(\ref{tildelam345}), which is approximately equal to $\lambda_{345}$ for small 
$M_{h_2}-M_{h_1}$.
They give the dominant contribution to the  $pp\to h_1 h_1 jj$ process for $\lambda_{345}\simeq 1$,
but their contribution is negligible with very small $\lambda_{345}$.
On the other hand, for large $h_1-h_2$ and $h_1-h^+$ splittings,
they get stronger even with small $\lambda_{345}$ and enhance the VBF process.
This opens a new  perspective for the exploration of the i2HDM model 
which we plan to perform in the near future.

\subsection{Mono-object production: rates and comparison}

\subsubsection{Implementation and cuts}

When calculating the cross sections of mono-object production at the LHC,
we used the following setup for the process evaluation:
\begin{itemize}
\item the QCD renormalisation and factorisation scales $Q$ were chosen to be equal to the
transverse momentum of the pair of DM particles,  i.e. missing transverse momentum, \MET{}
for all processes;
\item the PDF and the strong coupling constant are as provided by the
NNPDF23LO (\verb|as_0119_qed|) PDF set~\cite{Ball:2012cx};
\item for all processes a cut on the minimal value of missing transverse momentum of 100 GeV 
was applied;
\item the VBF cross section has been evaluated with the following additional cuts:
  \begin{align}
    P_T^j>30\mbox{ GeV}, \hspace{2mm} \Delta\eta_{jj}>4, \hspace{2mm} E_j> 400\mbox{ GeV}.
  \end{align}
  \end{itemize}
Below we present plots and numbers for cross sections (in the text and table)
with three significant digits corresponding to the accuracy of the MC phase space integration.
But we would like to note that when $Q$ is varied in the range $\MET/2$ to $2\times \MET$,
the QCD scale uncertainty is around 20-30\% for the tree-level cross sections presented,
dominating over PDF uncertainties which are below 10\%.
The presentation and detailed discussion of
these uncertainties is out of the scope of this paper.

\subsubsection{Production rates}

\begin{figure}[htb]
\hskip 3cm\includegraphics[height=0.4\textheight]{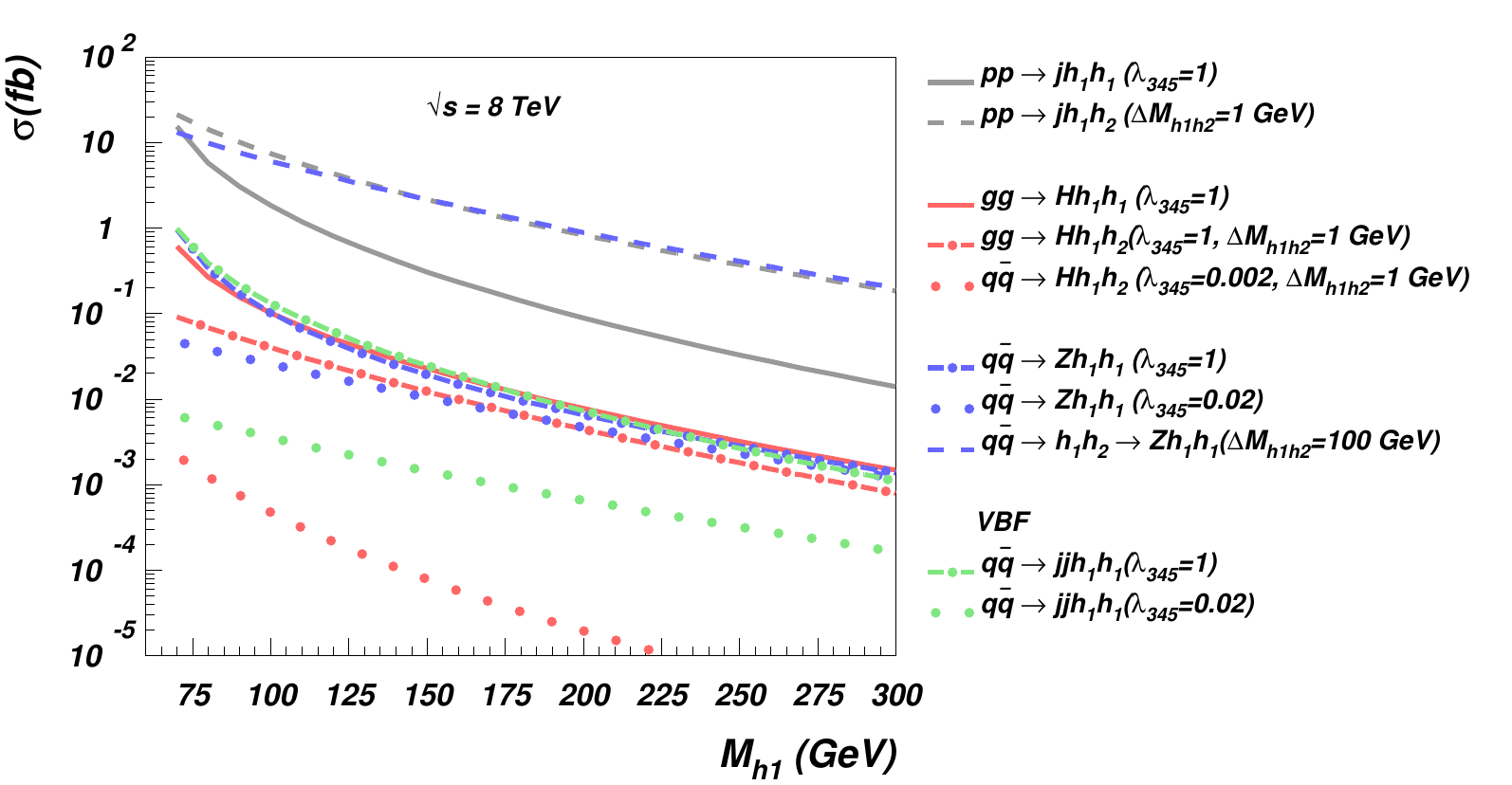} 
\vskip -0.2cm
\hskip 3cm\includegraphics[trim={0 0 10.5cm 0},clip,height=0.4\textheight]{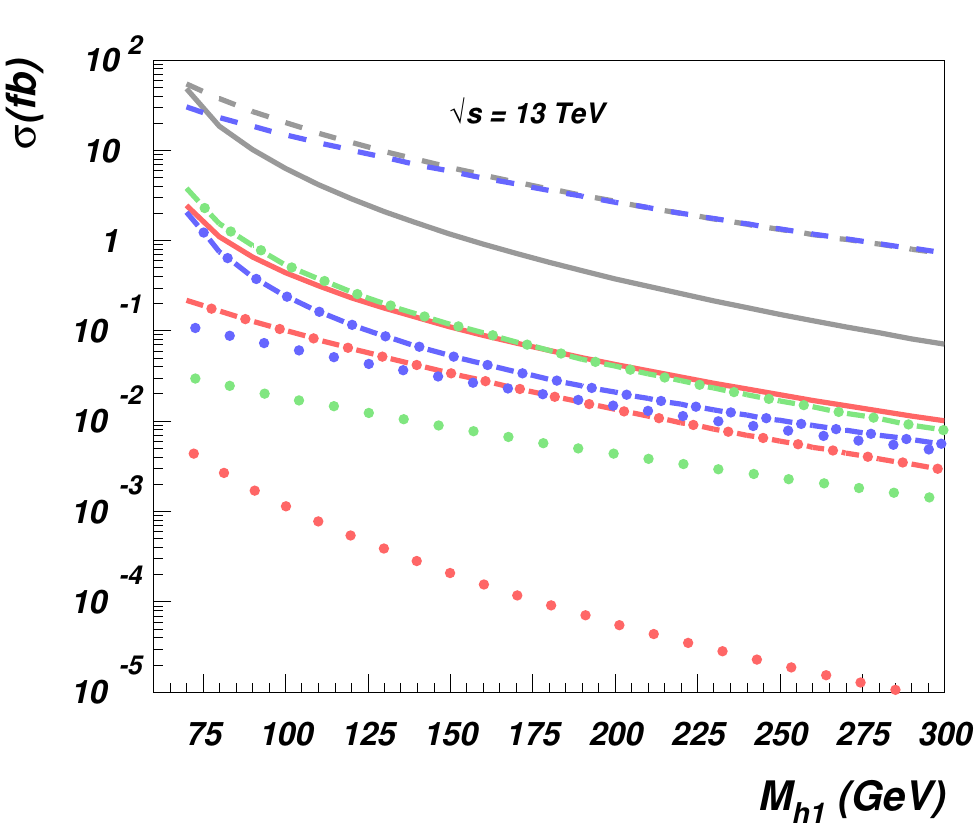} 
\vskip -0.2cm
\caption{ Cross sections versus
Dark Matter mass, $M_{h_1}$, for  processes
contributing to mono-jet, mono-Z, mono-Higgs and VBF signatures
for the LHC@8 TeV and  LHC@13 TeV.}
\label{fig:cs}
\end{figure}

In Fig.~\ref{fig:cs} we present a summary of the cross sections versus
Dark Matter mass $M_{h_1}$ for all the processes mentioned above, 
which contribute to the mono-jet, mono-$Z$, mono-Higgs and VBF signatures
for the LHC at 8 and 13 TeV.
In the plot we chose some particular values of the mass split $M_{h_2} - M_{h_1}$
and of the coupling $\lambda_{345}$, however the cross sections in other cases can be easily inferred
by referring to the scaling properties mentioned in Section~\ref{sec:DMsignatures}.
It is evident form the plot that the dominant cross section always tend to be mono-jet processes (in grey), and the mono-$Z$ signature coming from the on-shell decay of $h_2$.


{It is also worth focusing on the new $\lambda_{345}\sim 0$ region at small masses: the relevant cross sections are related to the mono-jet process $p p \to j h_1 h_2$ at small mass split (dashed grey in Fig.~\ref{fig:cs}) and mono-Higgs production (dotted red).
In the low $M_{h_1}$ region, below $75$~GeV, the former channel provides cross sections of several tens of fb at $13$~TeV, which may be probed in mono-jet searches. This is a very important point, as the $\lambda_{345}\sim 0$ region is very difficult to probe in DD experiments due to the suppressed coupling to the Higgs.

The plot also shows that the mono-Higgs channels does not yield detectable rates.

The process $p p \to j h_1 h_2$ is also relevant for larger mass splits: while the cross section drops, for splits of few tens of GeV, the $h_2$ decays can provide visible leptons and thus produce SUSY-like signatures. Finally, when the mass split grows above the $Z$ mass, on-shell decays of $h_2$ allow for mono-$Z$ signatures, as shown by the blue-dashed line.
}

\subsection{Benchmark points}

\begin{table}[htb]
\centering
\begin{tabular}{|c||c|c|c|c|c|c|c|}
\hline
 {\bf BM}                       &  {\bf 1}  		& {\bf 2}  	   	   & {\bf 3}  		& {\bf 4}  		    & {\bf 5}     	    &  {\bf 6}  \\
  \hline\hline 
  $M_{h_{1}}$ (GeV)     	& 55      		& 55 			& 50 		      & 70		       & 100	      	      &100 \\
  \hline
  $M_{h_{2}}$ (GeV)     	& 63      		& 63 			&150  		      & 170	       	       & 105		      &105 \\
  \hline
  $M_{h_{+}}$ (GeV)   		& 150     		& 150 			&200  	  	      & 200		       & 200	     	      &200 \\
  \hline
  $\lambda_{345}$       	& $1.0\times10^{-4}$ 	 & $0.027$  		 & $0.015$ 		& $0.02$ 		& $1.0$  		& $0.002$	   \\
  \hline
  $\lambda_{2}$         	&  1.0    		& 1.0		        & 1.0	    	      & 1.0		       & 1.0		      & 1.0 \\ 
  \hline	
  $\Omega_{\rm DM} h^2$          	& $9.2 \times 10^{-2}$  & $1.5 \times 10^{-2}$   & $9.9 \times 10^{-2}$ & $9.7 \times 10^{-2}$	& $1.3 \times 10^{-4}$  &  $1.7 \times 10^{-3}$ \\
  \hline 
  $\sigma_{SI}^p$ (pb)    	& $1.7 \times 10^{-14}$ &  $1.3 \times 10^{-9}$  & $4.8 \times 10^{-10}$& $4.3 \times 10^{-10}$ & $5.3 \times 10^{-7}$  &  $2.1 \times 10^{-12}$ \\
  $R_{SI}^{LUX}$     	        & $1.6\times 10^{-5}$   &  $0.19$		 & $0.51$  		& $0.37$  		& $0.48$  		& $2.5  \times 10^{-5} $ \\
 \hline 
  $Br(H\to h_1 h_1)$     	& $5.2\times 10^{-6}$ 	&  $0.27$  		& $0.13$  		& $0.0$  		& $0.0$		 	&  $0.0$ \\
 \hline 
  $\sigma_{LHC8}$ (fb) &&&&&&\\
  ${h_1 h_1 j}$  		& $5.44\times 10^{-3}$ 	&  $288$  		& $134$  		& $6.05\times 10^{-3}$ 	&  $1.80$ 		&  $7.23\times 10^{-6}$ \\
  ${h_1 h_2 j}$  		& $36.7$      		&  $36.7$ 		& $6.48$  		& $3.90$ 		&  $6.93$ 		&  $6.93$ \\
  ${h_1h_1 Z}$  		& $6.14\times 10^{-2}$  &  $21.4$  		& $30.7$  		& $12.2$		&  $0.101$ 		&  $2.52\times 10^{-2}$ \\
  ${h_1h_1 H}$  		& $1.70\times 10^{-4}$  &  $8.98$  		& $4.21$  		& $2.19\times 10^{-4}$ 	&  $0.100$ 		&  $3.33\times 10^{-7}$ \\
  ${h_1h_2 H}$  		& $5.35\times 10^{-3}$  &  $6.31\times 10^{-3}$ & $9.80\times 10^{-3}$  & $7.54\times 10^{-3}$ 	&  $3.86\times 10^{-2}$ &  $5.51\times 10^{-4}$ \\
  ${h_1 h_1 jj}$		& $2.39\times 10^{-2}$	&  $17.2$  		& $8.11$  		& $4.44\times 10^{-2}$ 	&  $0.212$		&  $1.62\times 10^{-2}$ \\
 \hline 
  $\sigma_{LHC13}$ (fb) &&&&&&\\
  ${h_1 h_1 j}$  		& $1.67\times 10^{-2}$ 	&  $878$  		& $411$  		& $1.93\times 10^{-2}$ 	&  $6.25$ 		& $2.50\times 10^{-5} $ \\
  ${h_1 h_2 j}$  		& $92.4$      		&  $92.4$  		& $17.8$  		& $11.1$ 		&  $19.1$ 		&  $19.1$ \\
  ${h_1h_1 Z}$  		& $0.153$       	&  $46.2$  		& $66.9$  		& $28.3$ 		&  $0.241$ 		&  $6.47\times 10^{-2}$ \\
  ${h_1h_1 H}$  		& $6.69\times 10^{-4}$  &  $35.3$  		& $16.5$  		& $9.08\times 10^{-4}$ 	&  $0.441$ 		&  $1.51\times 10^{-6}$ \\
  ${h_1h_2 H}$  		& $1.18\times 10^{-2}$  &  $1.40\times 10^{-2}$ & $2.47\times 10^{-2}$  & $1.99\times 10^{-2}$ 	&  $9.82\times 10^{-2}$ &  $1.34\times 10^{-3}$ \\
  ${h_1 h_1 jj}$ 		& $0.101$	     	&  $62.7$  		& $29.6$  		& $0.189$ 		&  $0.904$ 		&  $7.49\times 10^{-2}$ \\
  \hline\hline
\end{tabular}
\caption{Benchmarks (BM) from the i2HDM parameter space
together with corresponding observables: DM relic density ($\Omega_{\rm DM} h^2$),
 spin-independent  DM scattering rate on the proton ($\sigma_{SI}^p$)
accompanied with its ratio to the experimental limit from LUX following re-scaling with the relic density: 
$R_{SI}^{LUX} =(\sigma_{SI}^p/\sigma_{SI}^{LUX})\cdot (\Omega_{DM}/\Omega_{DM}^{Planck})$,
and the LHC cross sections for mono-jet, mono-$Z$, and mono-$H$ signatures with a $\MET > 100$ GeV cut applied.
\label{tab:i2HDMbenchMarks}}
\end{table}


The experience we have gained so far, both in relic density and mono-object cross section calculations,
allows us to discern several qualitatively distinct regimes
of i2HDM and find their representative benchmark points.
In Table~\ref{tab:i2HDMbenchMarks} we present six benchmarks (BM) from the i2HDM  parameter space
together with corresponding observables: DM relic density ($\Omega_{\rm DM} h^2$),
 spin-independent  DM scattering rate on the proton ($\sigma_{SI}^p$)
accompanied with its ratio to the experimental limit from LUX following re-scaling with the relic density: 
$R_{SI}^{LUX} =(\sigma_{SI}^p/\sigma_{SI}^{LUX})\cdot (\Omega_{DM}/\Omega_{DM}^{Planck})$.
We also present the LHC cross sections for the mono-jet, mono-$Z$ and mono-$H$ signatures
discussed above with a $\MET > 100$ GeV cut applied.
All of these benchmarks are allowed by the present experimental data.
{
In this table we do not give specific benchmarks 
for long-lived $h^+$ scenario discussed above since in Fig.~\ref{fig:sigmah+}
we have effectively provided the whole parameter space for this scenario
($\Delta M = 140-200$~Mev, $\lambda_{345}\simeq 0$, $M_{h^+}=490-600$~GeV).
}

The first two benchmarks have small and medium values of $\lambda_{345}$ and correspond to the 
scenario when $M_{h_1}$ is below $M_H/2$, and the mass split $\Delta M = M_{h_2}-M_{h_1}$ is small.
BM1 has a very small value of  $\lambda_{345}=10^{-4}$ and is therefore characterised 
by having a small $Br(H\to h_1 h_1)$ value and a very low DM direct detection rate, $\sigma^p_{SI}$,
whilst the relic density is consistent with the Planck limit {due to co-annihilation}.
The $h_1h_1j$ mono-jet signature rate at the LHC scales with $(\lambda_{345})^2$ and is therefore very low,
while the $\lambda_{345}$-independent $h_1h_2j$ signature cross section 
is about 36.7 fb (LHC@8 TeV) and 92.4 fb (LHC@13 TeV).

BM2 differs from BM1 only by the value of $\lambda_{345}=0.027$, which is chosen as the maximum value 
allowed by the Higgs invisible branching ratio. For this $\lambda_{345}$,
the $h_1 h_1 j$ mono-jet production rates are 288 fb (LHC@8 TeV) and 878 fb (LHC@13 TeV).

BM3 and BM4 correspond to the scenarios where $\Delta M > M_Z$
with $M_{h_1}$ below and above $M_H/2$ respectively,
{with the other parameters chosen such that the} relic density is consistent with Planck data.
In comparison to BM3, BM4 has a very low $h_1h_1j$ production cross section because
the SM Higgs boson is produced off mass shell.
At the same time the $h_1 h_1 Z$ cross section 
is of the same order for both benchmarks: 6.48 fb and 3.90 fb for LHC@8 TeV,
and 17.8 fb and 11.1 fb for LHC@13 TeV, respectively.

Finally, BM5 and BM6 represent the cases with a small (5 GeV) mass split and $M_{h_1}=100$~GeV.
The only difference in the input parameters is the value of $\lambda_{345}$: 
large $\lambda_{345}=1$ for BM5 and small $\lambda_{345}=0.002$ for BM6.
For both benchmarks, the DM relic density is well below the PLANCK limit, and therefore an
additional source of Dark Matter is required. Even for BM6 which has a small
value of $\lambda_{345}$, the DM relic density is
of the order of $10^{-3}$ because the DM effectively annihilate via 
$h_1 h_1 \to VV$  and $h_1 h_1 \to HH$ channels.
They are open for this value of DM mass
and are defined essentially by the weak coupling,
the contribution from $h_1 h_1 \to V_L V_L$ being small because of the small $h_1-h_2$ mass split
and the contribution from co-annihilation being subdominant for this value of mass split.
For both of these benchmarks, the $h_1 h_2 j$ channel
which has cross-sections of 6.93 fb (LHC@8 TeV) and 19.1 fb (LHC@13 TeV)
looks the most promising.

From Table~\ref{tab:i2HDMbenchMarks} one can see that different mono-object signatures are 
very complementary for these suggested benchmarks,
especially the $h_1 h_1 j$ and $h_1 h_2 j$ processes which are
the main focus of the collider study presented below.

\subsection{Limits from LHC@8TeV and projections for LHC@13TeV}

In the previous subsections, we calculated the mono-object production cross sections at the LHC
as a function of DM mass $M_{h_1}$ for a selection of parameters.
In this subsection, we invert the problem: we examine the limits on the parameter space which follow 
from the current 8 TeV and projected 13 TeV LHC data. 
We concentrate on limits from mono-jet processes, as these are the mono-object signatures
with the highest cross sections, as shown in Fig.~\ref{fig:cs}.
For mono-jet signals we consider two different processes: $pp\rightarrow h_1h_1j$ and $pp\rightarrow h_1h_2j$.
The cross section of the former depends on the two parameters only, the dark matter mass $M_{h_1}$ and $\lambda_{345}$. 
For the latter, all the vertices depend only on the gauge constants. The only two parameters that shape its cross section 
are the inert scalar masses $M_{h_1}$ and $M_{h_2}$, or equivalently $M_{h_1}$ and $\Delta M = M_{h_2} - M_{h_1}$. 

\subsubsection{Implementation and the LHC data used}

In order to calculate the limits from the LHC at 8 TeV, we used the {\tt CheckMATE}
\cite{Drees:2013wra,deFavereau:2013fsa,Cacciari:2011ma,Cacciari:2005hq,Cacciari:2008gp,Read:2002hq,Lester:1999tx,Barr:2003rg,Cheng:2008hk} framework, which allows an easy
application of the implemented search analyses. This tool takes a given sample of Monte Carlo events in the HEP or HepMC format after parton showering and hadronisation,
for which we used {\tt Pythia-6} \cite{Sjostrand:2006za}, and performs a detector simulation on these events using Delphes-3 \cite{deFavereau:2013fsa}. Subsequently {\tt CheckMATE} can
apply any of its pre-programmed and validated analyses to the generated signal events and uses the resulting efficiencies along with published information, such as the 95\%
confidence level limit on signal count, to produce results from which we can find the cross-section limit placed on our model by each analysis.

The signature of both processes that we consider, $pp\rightarrow h_1h_1j$  and $pp\rightarrow h_1h_2j$,
is a high-$p_T$ jet and a large missing transverse momentum, $\MET{}$. 
In the case of $pp\rightarrow h_1h_2j$, the $h_2$ will decay via a $h_1$ and a $Z^{(*)}$-boson. When $\Delta M$ is very small, the decay
products of the $Z$ will generally be too soft to be reconstructed in the detector. Therefore in this case $pp\rightarrow h_1h_2j$ will give a mono-jet + $\MET{}$
signature. Using {\tt CheckMATE} and HepMC files created with the i2HDM model implemented in {\tt CalcHEP}, we calculated the limits given by all of the mono-jet +
$\MET{}$ analyses currently implemented in {\tt CheckMATE} \cite{ATLAS:2012zim,Aad:2014nra,Aad:2015zva,Khachatryan:2014rra} (3 ATLAS and 1 CMS analysis).

For both processes considered, we found that the lowest cross section limits for each benchmark point considered were provided by one of the ATLAS mono-jet +
$\MET{}$ analysis \cite{Aad:2015zva}. These are the limits presented in this section. This analysis requires a leading jet with a $p_T > 120$ GeV and $|\eta|
<2.0$, and the leading jet $p_T/\MET > 0.5$. Furthermore, to reduce multijet background where the large $\MET{}$ originating mainly from the jet energy
mismeasurement, we place a requirement on the azimuthal separation $\Delta \phi (\text{jet},p_T^{\text{miss}}) > 1.0$ between the direction of the missing transverse momentum
and that of each jet. A number of different signal regions are considered with increasing $\MET{}$ thresholds from 150 GeV to 700 GeV. Full details are available in
the ATLAS paper \cite{Aad:2015zva}.

In order to project these limits for increased luminosity and to 13 TeV, we use Monte Carlo events to estimate the efficiencies for the signal and background at 13 TeV, which
is a function of $M_{h_1}$ and depends on the best analysis signal region for each mass. We make the assumption that the analysis cuts for 13 TeV data will be the same as for
8 TeV data, which does not take into account improvements in the signal to background ratio which would likely occur with new analysis cuts at 13 TeV. Therefore our projected
limits will be slightly conservative.

\subsubsection{Limits from $pp\rightarrow h_1h_1 j$}

\begin{figure}[h!]
\centering
  \subfigure[$M_{h_1}$ vs cross section at 8 TeV .]{\includegraphics[width=0.5\textwidth]{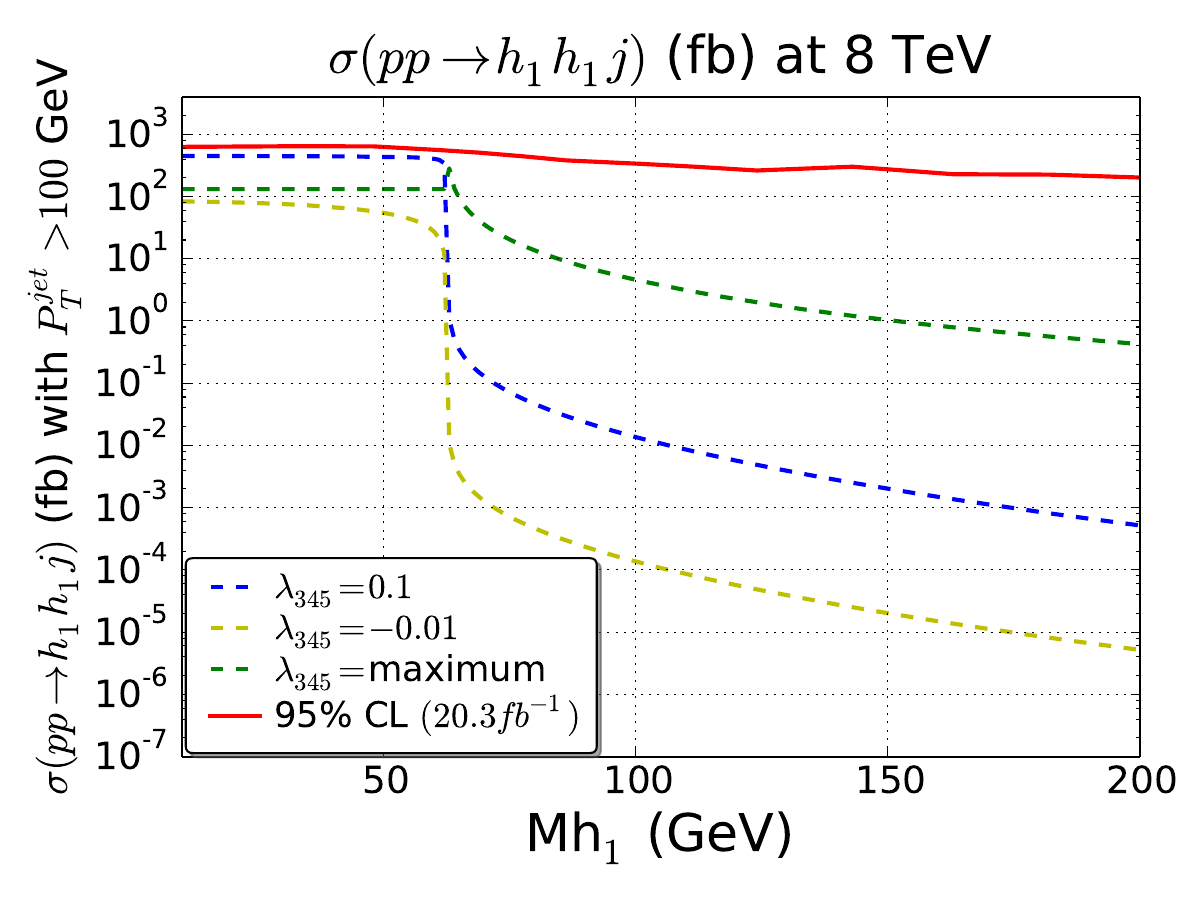}}%
  \subfigure[$M_{h_1}$ vs cross section at 13 TeV.]{\includegraphics[width=0.5\textwidth]{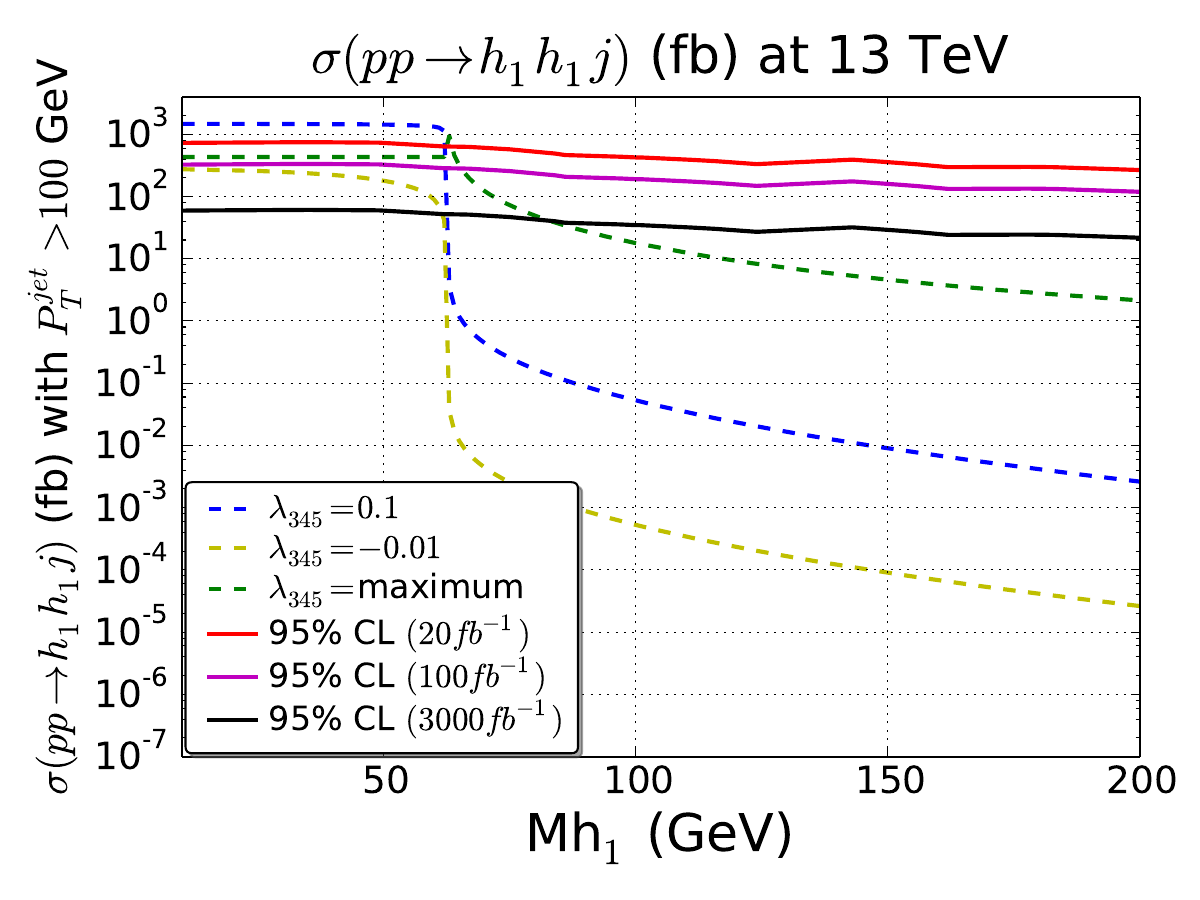}}
\caption{Cross sections and 95\% CLs for $pp \to h_1 h_1 j$ versus $M_{h_1}$ at 8 TeV and 13 TeV. In both cases, the cross sections are shown for 3 different values of $\lambda_{345}$: (i) $\lambda_{345}=0.1$ ({\bf \blue blue} dashed), (ii) $\lambda_{345}= -0.01$ ({\bf \yellow yellow} dashed), (iii) the maximum $\lambda_{345}$  value ({\bf \green green} dashed) allowed by constraints (described in text). (a) Results for 8 TeV, with limits (solid {\bf \red red}) calculated using the ATLAS analysis \cite{Aad:2015zva}. (b) Results for 13 TeV, with projected limits for the ATLAS analysis \cite{Aad:2015zva} with luminosities of 20 $fb^{-1}$, 100 $fb^{-1}$ and 3000 $fb^{-1}$ ({\bf \red red}, {\bf \magenta magenta}, {\bf black} solid lines) at 13 TeV.} \label{cc_limit_h1h1}
\end{figure}

The results for the process $pp\rightarrow h_1h_1 j$ are shown for 8 TeV in Fig.~\ref{cc_limit_h1h1} (a) with projections to 13 TeV and higher luminosities in Fig.~\ref{cc_limit_h1h1} (b). The limits are denoted by the solid lines, whilst the cross sections for the i2HDM for different values of $\lambda_{345}$ are shown by the dashed lines. For $M_{h_1} < M_H/2$, the maximum allowed value of $\lambda_{345}$ is given by the bound on the invisible Higgs branching ratio in Eq.~(\ref{eq:lhc-higgs-invis}) (this constraint has not been applied on the dashed blue $\lambda_{345} = 0.1$ curve), whilst when $M_{h_1} > M_H/2$ the maximum allowed value is calculated using the constraints of Eq.~(\ref{eq:l345-vacuum-stab}). The cross section with this maximum value of $\lambda_{345}$ is denoted by the dashed green line. We see in Fig.~\ref{cc_limit_h1h1} (a), that the 8 TeV LHC mono-jet + $\MET{}$ searches do not constrain the i2HDM via the $pp\rightarrow h_1h_1 j$ process. However at 13 TeV, shown in Fig.~\ref{cc_limit_h1h1} (b), with around 100 $fb^{-1}$ of data (purple solid), we would be able to set limits on $\lambda_{345}$ for $M_{h_1}$ up to 66 GeV, and for 3000 $fb^{-1} $ (black solid) LHC data would set limits on $\lambda_{345}$ for $M_{h_1}$ up to 83 GeV. 
It should be remarked that the spike in cross section on the green dashed line at $M_{h_1} \sim M_H/2$ is due to the release of the ($H \to invisible$) bound on $\lambda_{345}$ once the decay of the Higgs into DM is kinematically closed.


We should note that a similar projection of CMS mono-jet limits \cite{Khachatryan:2014rra} at 14 TeV has been studied previously \cite{Arhrib:2013ela}, where the projected limits were slightly stronger than in Fig.~\ref{cc_limit_h1h1} (b). Their projection was able to limit $M_{h_1}$ for values of $\lambda_{345}$ as small as $\lambda_{345} = 0.01$, while we require slightly larger values of $\lambda_{345}$ in order to limit $M_{h_1}$. 
We would like to note that in our paper the limits are based on the fast detector simulations rather than parton level 
ones used in \cite{Arhrib:2013ela} done for  14 TeV. Taking this into account we consider our results as more realistic
projection of the future LHC data potential.

\begin{figure}[h!]
\centering
 \includegraphics[width=0.65\textwidth]{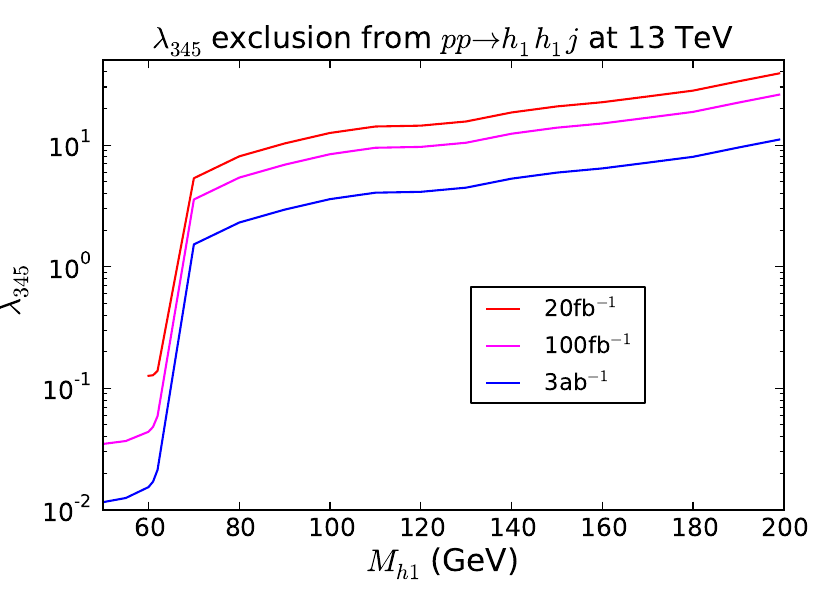}
\caption{Projected limit on $\lambda_{345}$ from  $pp \to h_1 h_1 j$ at 13 TeV
derived from the analysis presented in Fig.~\ref{cc_limit_h1h1} \label{lam345_coll_limit}}
\end{figure}

In Fig.~\ref{lam345_coll_limit} we provide the  limit on $\lambda_{345}$ versus $M_{h_1}$ for different projected luminosities at the LHC@13TeV. This limit is derived from the analysis presented in Fig.~\ref{cc_limit_h1h1} and could be more practical 
for comparison with limits on $\lambda_{345}$ from different experiments.

\subsubsection{Limits from $pp\rightarrow h_1h_2 j$}

\begin{figure}[h!]
\centering
  \subfigure[$M_{h_1}$ vs cross section at 8 TeV. ]{\includegraphics[width=0.5\textwidth]{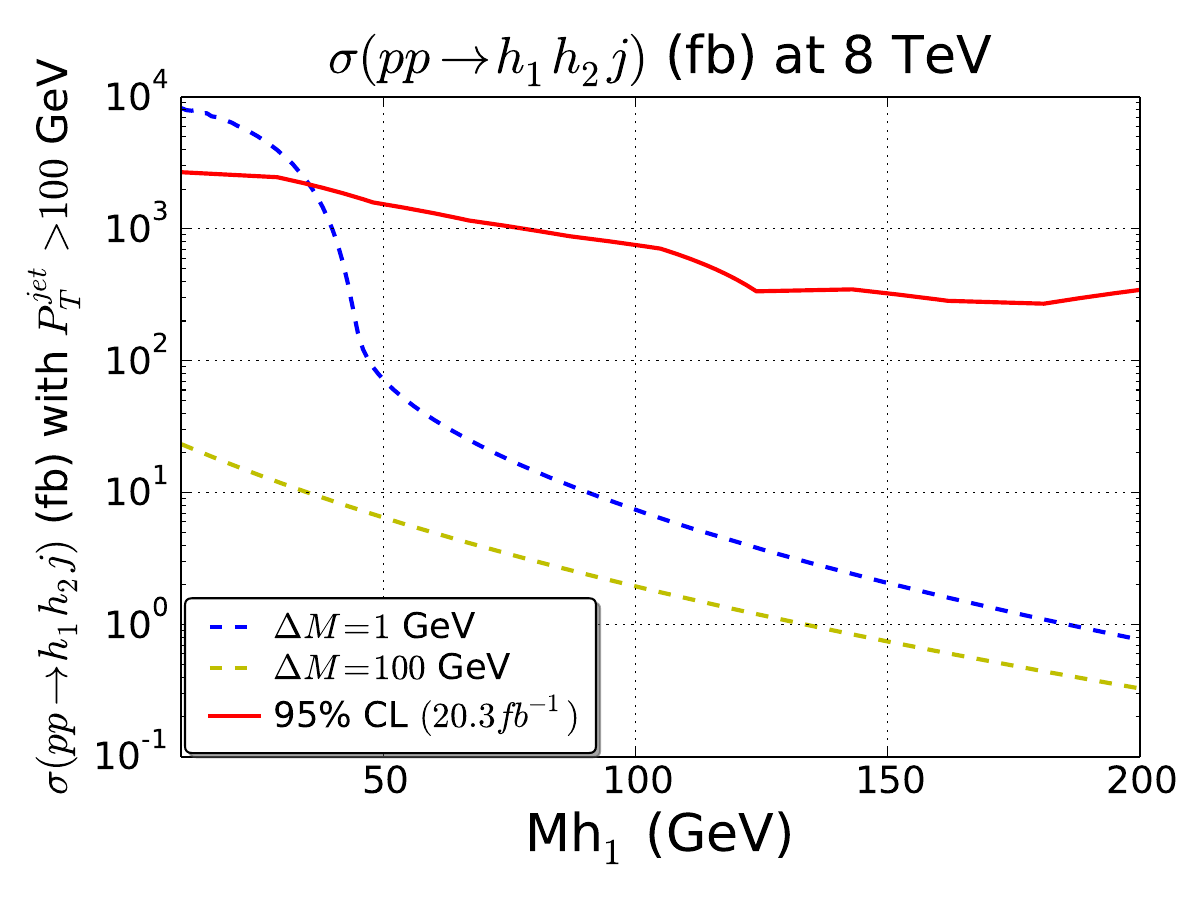}}%
  \subfigure[$M_{h_1}$ vs cross section at 13 TeV.]{\includegraphics[width=0.5\textwidth]{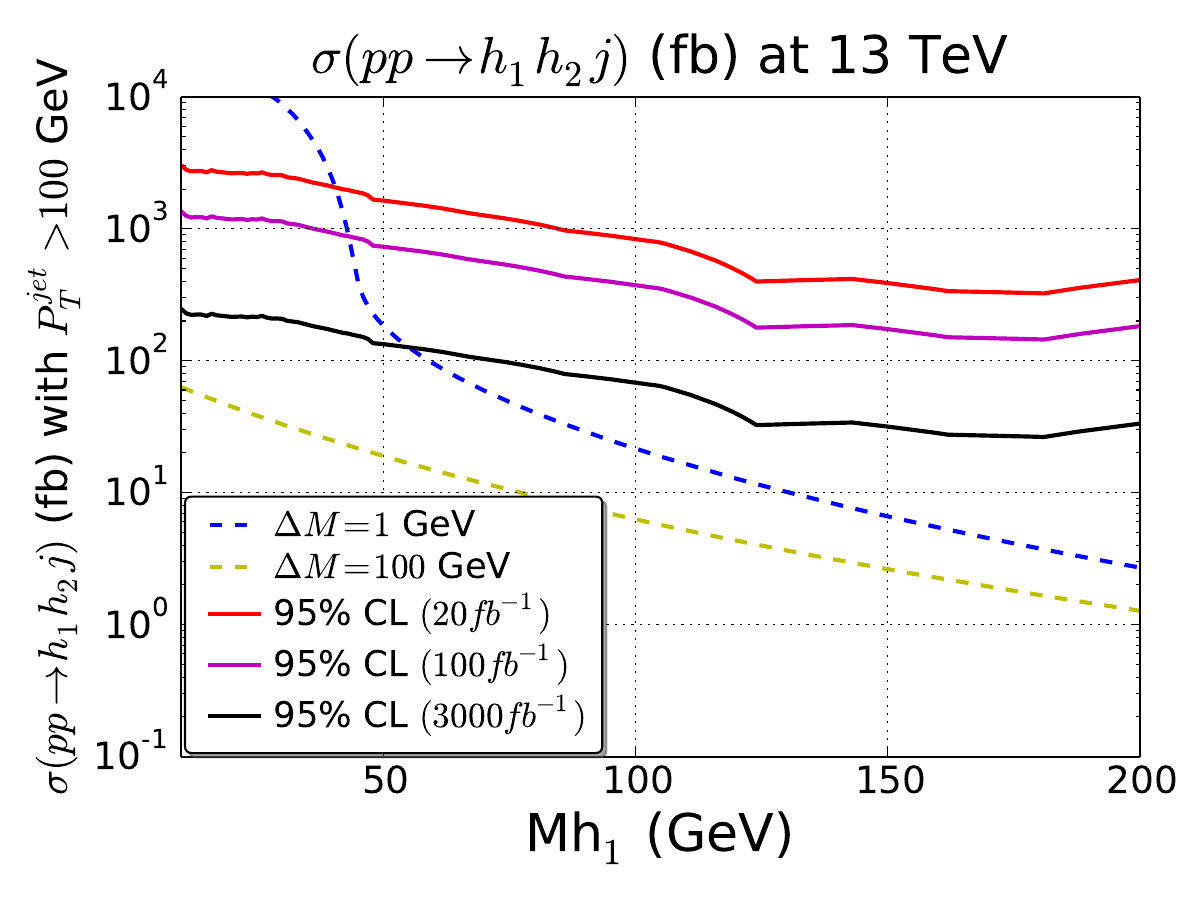}}
\caption{Cross sections and 95\% CLs for $pp \to h_1 h_2 j$ versus $M_{h_1}$ at 8 TeV and 13 TeV. In both cases, the cross sections are shown for 2 different values of $\Delta M = M_{h_2} - M_{h_1}$: (i) $\Delta M = 1$ GeV ({\bf \blue blue} dashed), (ii) $\Delta M = 100$ ({\bf \yellow yellow} dashed). (a) Results for 8 TeV, with limits (solid {\bf \red red}) calculated using the ATLAS analysis \cite{Aad:2015zva}. (b) Results for 13 TeV, with projected limits for the ATLAS analysis \cite{Aad:2015zva} with luminosities of 20 $fb^{-1}$, 100 $fb^{-1}$ and 3000 $fb^{-1}$ ({\bf \red red}, {\bf \magenta magenta}, {\bf black} solid lines) at 13 TeV.} \label{cc_limit_h1h2}
\end{figure}

For $pp\rightarrow h_1h_2 j$, the results are shown in Fig.~\ref{cc_limit_h1h2}(a) for 8 TeV and in Fig.~\ref{cc_limit_h1h2}(b) for 13 TeV. We consider two scenarios with a small ($\Delta M =1$ GeV in blue) and large ($\Delta M =100$ GeV in yellow) mass split. The projected cross section limits are again denoted by the solid lines. When $\Delta M = 1$ GeV, the current LHC Run I results are able to rule out $M_{h_1} < 35$ GeV. In this case, it should be emphasised that as the couplings of the relevant diagrams (see Fig.~\ref{fig:fd-monojet2}) are fixed by the gauge couplings, this limit on $M_{h_1}$ is independent of all parameters other than $\Delta M$. At 13 TeV, and at higher luminosities, this lower limit on $M_{h_1}$ in this degenerate mass scenario is improved slightly to 41 GeV, 43 GeV and 55 GeV for $20$ fb$^{-1}$ (solid red), $100$ fb$^{-1}$ (solid magenta) and $3000$ fb$^{-1}$ (solid black) of integrated luminosity respectively, as is shown in Fig.~\ref{cc_limit_h1h2} (b). For $\Delta M = 100$ GeV, the production cross section is much smaller and the model is not constrained via mono-jet and $\MET{}$ signatures from the $pp\rightarrow h_1h_2 j$ process. However, in this region other collider signatures such as dilepton + $\MET{}$ from the decay $h_2 \to h_1Z$ are available and will provide stronger limits as studied for example in \cite{Belanger:2015kga}.

%
\section{Constraining i2HDM: future projections}

Taking into consideration these collider limits and also adding the projections of the Direct Detection XENON1T experiment,
we are able to impose the complete set of constraints on the i2HDM parameter space. 
It is worth stressing that, as before, we  present the limits using  the re-scaled DD cross section 
$\hat{\sigma}_{SI}= R_\Omega\times \sigma_{SI}$, where $R_\Omega = \Omega_{\rm DM}/\Omega^{\rm Planck}_{\rm DM}$, which allows us to take into account additional sources that could contribute to the DM relic density. 

\subsection{Highlighting the ``always allowed'' regions}

\begin{figure}[htb]
\vskip -0.5cm
\subfigure[Coloured points - always allowed points (AA)]%
{\includegraphics[trim={0 0.5cm 0 0},clip,width=0.49\textwidth]{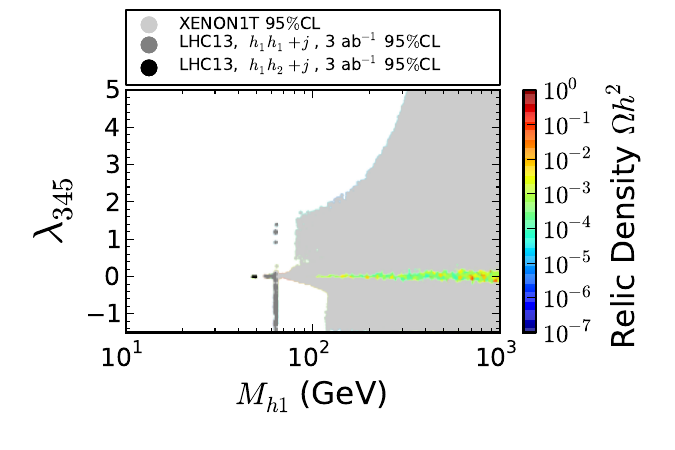}}%
\subfigure[Zoomed  AA region]%
{\includegraphics[trim={0 0.5cm 0 0},clip,width=0.49\textwidth]{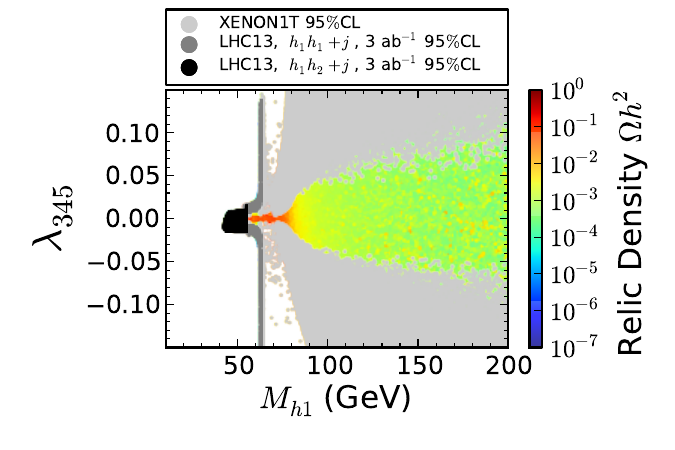}}\\
\subfigure[Zoomed AA region only for XENON1T constraints]%
{\includegraphics[trim={0 0.5cm 0 0},clip,width=0.49\textwidth]{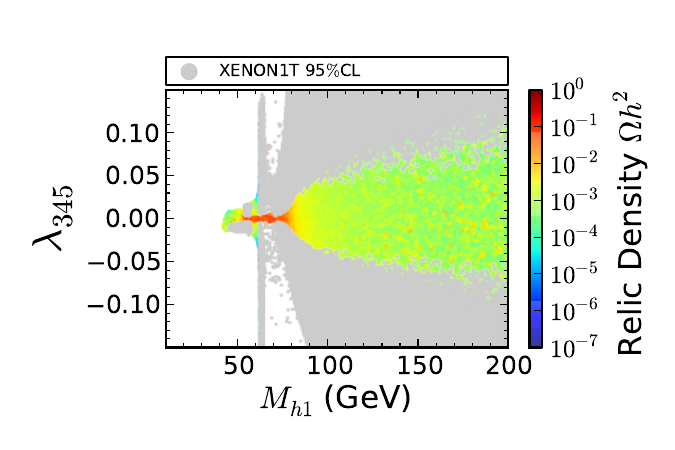}}%
\subfigure[Zoomed AA region only for the LHC13 constraints from $h1h1+j$ signature]%
{\includegraphics[trim={0 0.5cm 0 0},clip,width=0.49\textwidth]{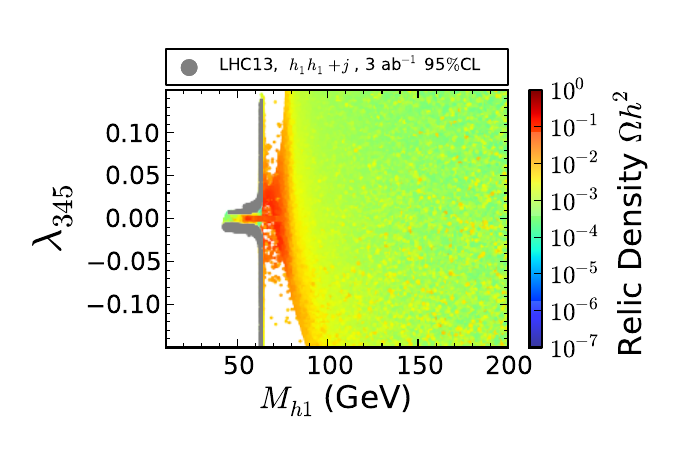}}
\vskip -0.3cm
\caption{Projection of the 5D random scan of the i2HDM into ($M_{h_1},\lambda_{345}$)
plane and the expected  reach of the LHC@13TeV with 3$ab^{-1}$ of integrated luminosity 
using  $h_1 h_2 j$  and  $h_1 h_1 j$ channels as well as for XENON1T experiment
indicated by black,
dark grey and light grey colours respectively.
Excluded points are plotted on the top of the allowed points   demonstrating the coloured region of the parameter space which will be always allowed (AA):
(a) and (b)
present AA parameter space for the combined constraints for 
large and zoomed ($M_{h_1},\lambda_{345}$) regions respectively;
(c) and (d)
present AA regions for separate  XENON1T and $h_1h_1+jet$ LHC13 constraints
respectively.
\label{collider-XENON1T-constraint}} 
\end{figure}
The results of the constraints are presented in Fig.~\ref{collider-XENON1T-constraint} as the color map of DM relic density
in the ($M_{h_1},\lambda_{345}$)  plane together with the projected sensitivity of the LHC@13TeV with 3 ab$^{-1}$ of integrated luminosity 
using  $h_1 h_2 j$  and  $h_1 h_1 j$ channels, as well as a projection for the XENON1T experiment
95\%CL exclusion regions. These constraints are indicated by black, dark grey and light grey colours, respectively.
In this figure we plot {\it excluded points on the top of the allowed points}   demonstrating the coloured region of the parameter space which will be always allowed (AA).
Fig.~\ref{collider-XENON1T-constraint}(a) and Fig.~\ref{collider-XENON1T-constraint}(b)
present AA parameter space for the combined constraints (black on the top of dark grey and dark grey on the top of light grey) for 
large and zoomed ($M_{h_1},\lambda_{345}$) regions respectively,
while  Fig.~\ref{collider-XENON1T-constraint}(c) and  Fig.~\ref{collider-XENON1T-constraint}(d)
present AA regions for separate  XENON1T and $h_1h_1+jet$ LHC13 constraints
respectively. From Fig.\ref{collider-XENON1T-constraint}a-c 
one can see how constraints from the LHC and XENON1T are complementary to each other.
One can see that XENON1T will exclude large $M_{h_1}$ masses
for large enough values of  $\lambda_{345}$ while the LHC will probe the region of smaller values of  $\lambda_{345}$ for $M_{h_1}$ below the $M_H/2$ threshold using the
$h_1h_1j$ channel, and will cover all values of $\lambda_{345}$
using the $h_1h_2j$ channel for  $M_{h_1}$ below 55 GeV.

\subsection{Highlighting the ``always excluded'' regions}

\begin{figure}[htb]
\vskip -0.5cm
\subfigure[Grey-black points - always excluded points (AE)]%
{\includegraphics[trim={0 0.5cm 0 0},clip,width=0.49\textwidth]{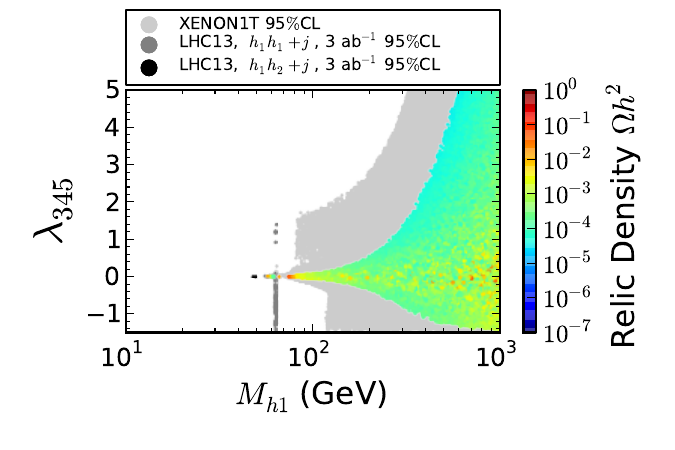}}%
\subfigure[Zoomed region with AE points]%
{\includegraphics[trim={0 0.5cm 0 0},clip,width=0.49\textwidth]{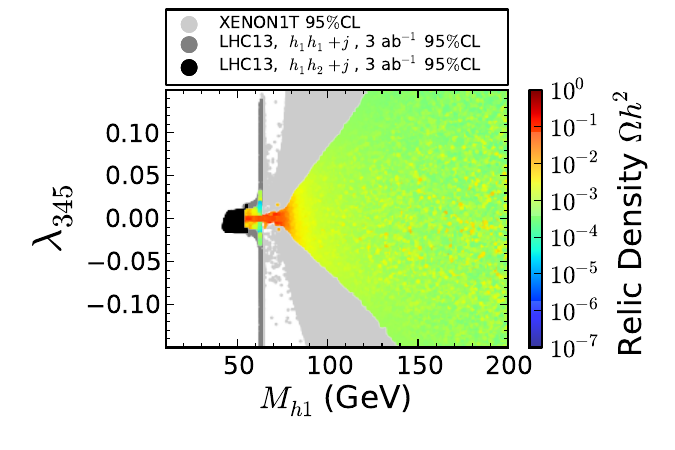}}
\vskip -0.3cm
\caption{Projection of the 5D random scan of the i2HDM into ($M_{h_1},\lambda_{345}$)
plane and the expected  reach of the LHC@13TeV with 3$ab^{-1}$ of integrated luminosity 
using  $h_1 h_2 j$  and  $h_1 h_1 j$ channels as well as for XENON1T experiment.
Allowed points are on the top of the excluded ones, therefore
the black and grey colours
present  the region which will be always excluded (AE)
or probed by the above experiments.
\label{collider-XENON1T-constraint-AE}} 
\end{figure}

Besides the AA region it is informative to find and analyse the region
with {\it allowed points on the top of excluded points}, therefore the
black and grey colours
present the region which will be always 
probed---and in the case of negative results, always excluded (AE)---by the above experiments. 
Such region is presented in Fig.~\ref{collider-XENON1T-constraint-AE}(a,b)
in exact analogy to  Fig.~\ref{collider-XENON1T-constraint}(a,b).

When comparing Fig.~\ref{collider-XENON1T-constraint}(a,b) and 
Fig.\ref{collider-XENON1T-constraint-AE}(a,b)---i.e. the plots with AA versus AE points,---one observes
a big difference between the order of the overlay of the excluded and allowed points.
This is related to the fact that the $\Omega_{\rm DM} h^2$ can substantially vary:
even for  fixed $M_{h_1}$ and  $\lambda_{345}$ values,
a large $M_{h^+}$ or $M_{h_2}$  can provide respectively large quartic couplings $h_1 h_1 W_L W_L$ and  $h_1 h_1 Z_L Z_L$, see Eq.~(\ref{tildelam345}), 
which lead in their turn to an effective $h_1 h_1 \to VV$  annihilation. This brings the relic density down and 
avoids the XENON1T constraints (once we use DD rates re-scaled to relic density).
In the ($\lambda_{345},M_{h_1}$) plane, for example, these points overlap with the points 
where the quartic couplings mentioned above are small and the  $\Omega_{\rm DM} h^2$ (and respectively exclusion) is driven only by $\lambda_{345}$. 
So the most complete picture comes from the combination of AA and AE plots:
the most conservative allowed region comes from AA plots of Fig.~\ref{collider-XENON1T-constraint}, while the most conservative exclusion region
is presented by AE plots of Fig.\ref{collider-XENON1T-constraint-AE}.

\begin{figure}[htb]
\subfigure[]{\includegraphics[trim={0 0.5cm 0 0},clip,width=0.5\textwidth]{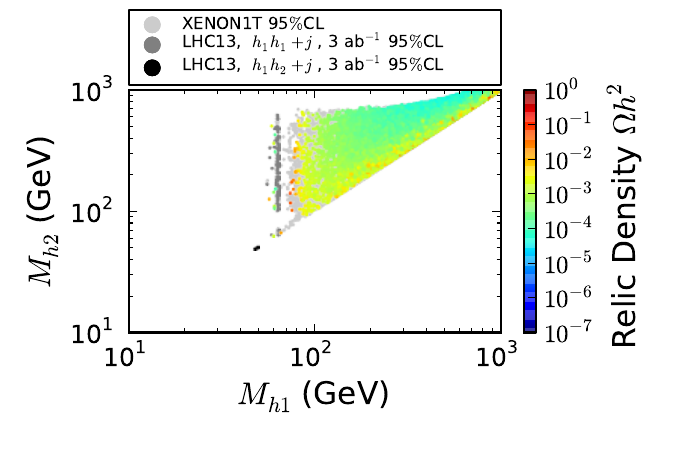}}%
\subfigure[]{\includegraphics[trim={0 0.5cm 0 0},clip,width=0.5\textwidth]{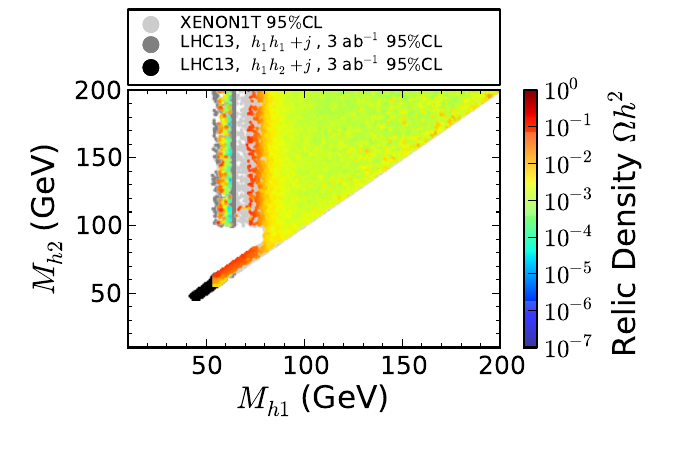}}\\
\subfigure[]{\includegraphics[trim={0 0.5cm 0 0},clip,width=0.5\textwidth]{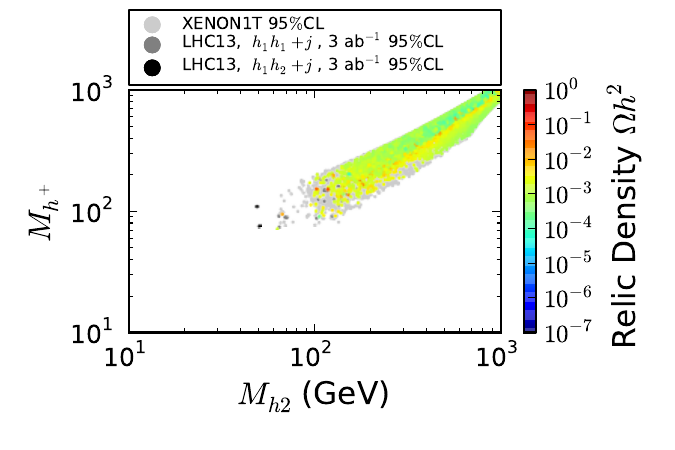}}%
\subfigure[]{\includegraphics[trim={0 0.5cm 0 0},clip,width=0.5\textwidth]{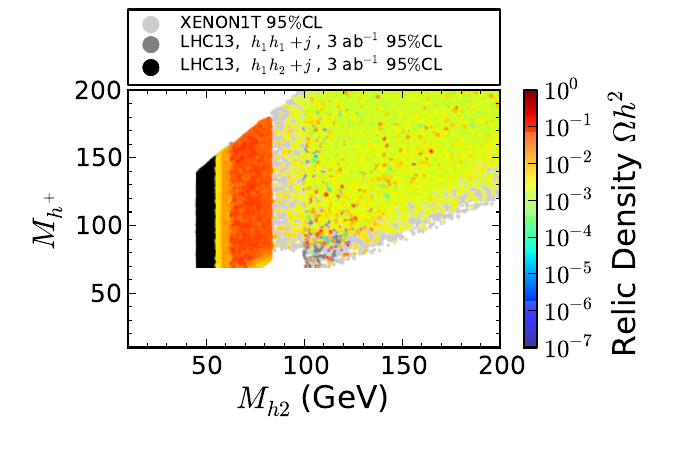}}\\
\caption{\label{collider-XENON1T-constraint-AE-other}
Projection of the 5D random scan of the i2HDM into ($M_{h_1},M_{h_2}$) (a,b)
and ($M_{h_2},M_{h^+}$) (c,d) planes
and the expected  reach of the LHC@13TeV with 3$ab^{-1}$ of integrated luminosity 
using  $h_1 h_2 j$  and  $h_1 h_1 j$ channels as well as for XENON1T experiment.
Allowed points are on the top of the excluded ones,
presenting AE points. The left panels (a,c) present a bigger region of the parameter space, while the
right ones (b,d) present a zoomed region with AE points.} 
\end{figure}
From Figs.\ref{collider-XENON1T-constraint} and \ref{collider-XENON1T-constraint-AE}
one can see  that imposition of the XENON1T constraint reduces substantially the parameter space, greatly expanding the previous limits imposed by LUX. This effect is not so evident in the other planes, presented in Fig.\ref{collider-XENON1T-constraint-AE-other}
in analogy to Fig.\ref{collider-XENON1T-constraint-AE},
because the spin-independent cross section $\hat{\sigma}_{SI}$ for DD is driven by the $t$-channel Higgs boson exchange 
and therefore is proportional to $\lambda_{345}^2$.
\begin{figure}[htb]
\subfigure[]{\includegraphics[trim={0 0.5cm 0 0},clip,width=0.5\textwidth]{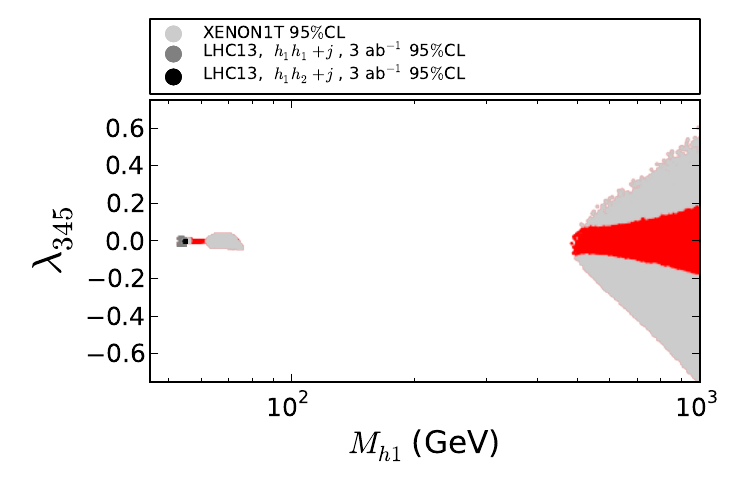}}%
\subfigure[]{\includegraphics[trim={0 0.5cm 0 0},clip,width=0.5\textwidth]{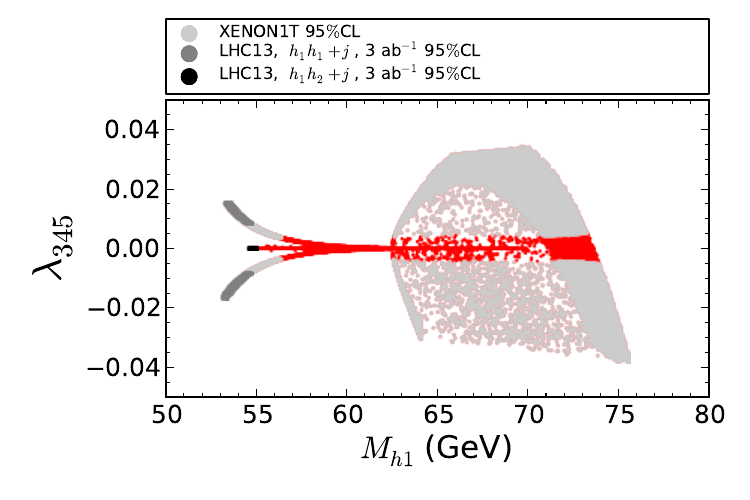}}\\
\subfigure[]{\includegraphics[trim={0 0.5cm 0 0},clip,width=0.5\textwidth]{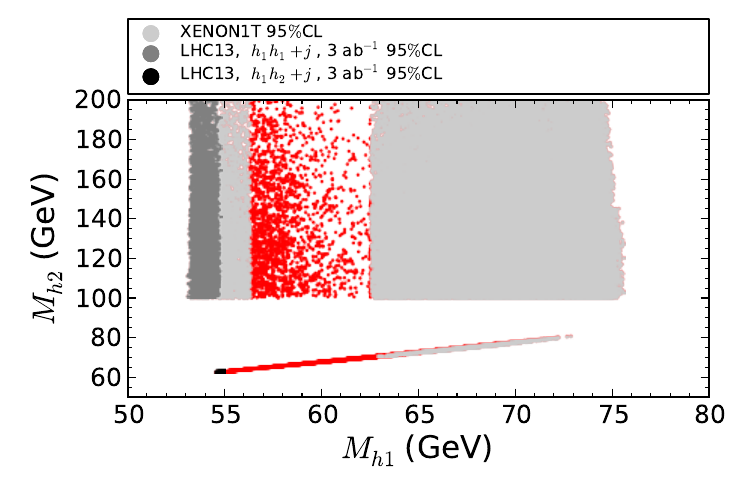}}%
\subfigure[]{\includegraphics[trim={0 0.5cm 0 0},clip,width=0.5\textwidth]{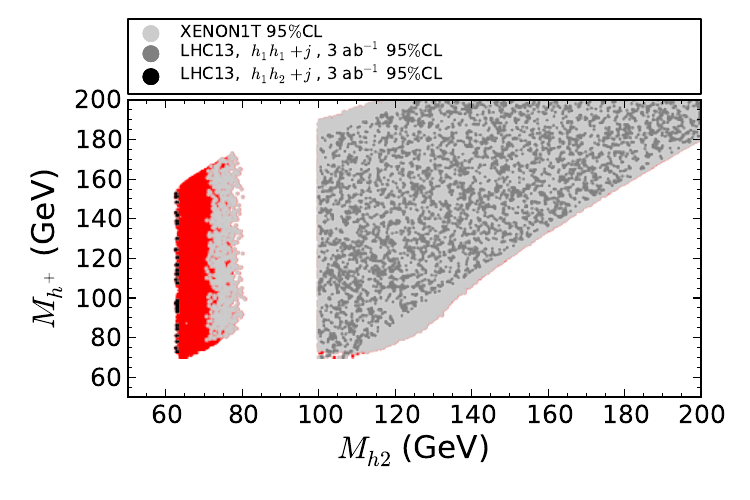}}
\caption{2D projections of the 5D random scan of the i2HDM satisfying all constraints (Cut-1 to Cut-4) considered above for Fig.(\ref{fig:dm-i2hdm},\ref{fig:dm-i2hdm-small}) plus the lower limit on the constraint on relic density given by Eq.(\ref{eq:planck-limit-relaxed}), taking in consideration the collider limits of mono-jet signatures at 13 TeV with 3$ab^{-1}$ of integrated luminosity and the projections of the DD XENON1T experiment. In the first row we present the parameter space of the plane ($M_{h_1},\lambda_{345}$) in the range $\subset $[10~GeV--1000~GeV]). In the second row we present the planes ($M_{h_1},\lambda_{345}$) and ($M_{h_1},M_{h_2}$) in the range $ \subset $[50~GeV--80~GeV]).} \label{fig:dm-i2hdm-LHC-DD}
\end{figure}

One should also stress again the  importance of the   $pp \to h_1h_2+j$ process,
using which  one can exclude $M_{h_1}<55$ GeV when  $M_{h_2}-M_{h_1}$ is small
for all values of $\lambda_{345}$.
 This is shown clearly  with the black dots in the Fig.~\ref{collider-XENON1T-constraint-AE-other}(b,d) where the (co)annihilation and respective mass degeneracy between $M_{h_1}$ and $M_{h_2}$ take place. Finally the $pp \to h_1 h_1 + j$ process imposes an extra constraint for the zone with low relic density corresponding to the $h_1h_1 \to H$ resonant annihilation just above $M_{h_1}=M_H/2$. In this case the invisible Higgs decay $H \to h_1h_1$ is closed and there is no restriction on $|\lambda_{345}|$, as we can see in Fig.~\ref{collider-XENON1T-constraint-AE-other}(b,d) represented by the dark grey points.

\subsection{Fitting the PLANCK data: future projections}

We have also found the  projected limits from colliders of mono-jet signatures  and the XENON1T DD experiment for the i2HDM points which satisfy both the upper and lower PLANCK limits, Eq.(\ref{eq:planck-limit-relaxed}). In this case, the scattering cross section $\sigma_{SI}$ is not re-scaled, because we are in the zone with the right amount of DM relic density. The results of these constraints are presented in Fig.~\ref{fig:dm-i2hdm-LHC-DD} as a scatter plot where the red zones represent the right amount of DM relic density. In the first row we show the parameter space of the plane ($M_{h_1},\lambda_{345}$) in the full
mass range from 10 to 1000~GeV. In the second row we present the planes ($M_{h_1},\lambda_{345}$) and ($M_{h_1},M_{h_2}$) but in a narrow mass range 
between 50 and 80~GeV.

As we can see, the incorporation of the DD constraint sets important restrictions on the parameter space. Still, in Fig.~\ref{fig:dm-i2hdm-LHC-DD}(a) there is a zone in the upper mass range that is not ruled out. Also in the low mass range there is a region between 55 GeV $< M_{h_1} <$ 74 GeV
which survives the restrictive constraint for small values of $\lambda_{345}$. We zoom into the surviving low mass region in Figs.~(\ref{fig:dm-i2hdm-LHC-DD}(b,c). Because of the improved limits of the DD experiment, the parameter $\lambda_{345}$ is very sensitive to scattering cross section, 
which sets a limit of $|\lambda_{345}|<0.01$ for this mass range.

The $pp \to h_1h_2+j$ process sets the exclusion limit for $M_{h_1}<55$ GeV (black dots) at the beginning of the $h_1 h_2$ coannihilation region represented by the thin horizontal strip for very small values of $\lambda_{345}$ in Fig.~\ref{fig:dm-i2hdm-LHC-DD}(b), which is also seen in the lower part of Fig.~\ref{fig:dm-i2hdm-LHC-DD}(c). The $pp \to h_1 h_1 + j$ process imposes an extra constraint on the lower mass zone where the DM annihilates through Higgs boson exchange and is visible in Fig.~\ref{fig:dm-i2hdm-LHC-DD}(b) in the shape of two symmetric wings for negative and positive values of $\lambda_{345}$. This excludes the $M_{h_1}<55$ GeV region. XENON1T will improve this constraint and exclude the $M_{h_1}<56.5$ GeV region.

\section{Concluding remarks}


The i2HDM is a clear example of a minimal consistent DM model which is
very well motivated by theoretical considerations.
At the same time this model could provide
 mono-jet, mono-$Z$, mono-Higgs and VBF+$\MET{}$
signatures at the LHC
complemented by
signals in direct and indirect DM search experiments.

The model is implemented into the CalcHEP and micrOMEGAs packages
and is publicly available at the HEPMDB database together with the
LanHEP model source. It is ready for further
exploration in the context of the LHC, relic density and DM direct detection.

In this paper we have performed 
detailed analysis of the   constraints in the full 5D  parameter space of the i2HDM from perturbativity, unitarity,
electroweak precision data, Higgs data from the LHC,
DM relic density, direct/indirect DM detection and the LHC mono-jet analysis as well as 
implications of experimental LHC studies on disappearing
charged tracks relevant to high DM mass region.
The LHC mono-jet analysis for the i2HDM model has been performed at the fast detector simulation level
and provides new results together with limits from disappearing
charged tracks at the LHC.
{Our results on non-LHC constraints are summarised in Figs.~\ref{fig:scan-simplified}-\ref{fig:scan-simplified-fitting} in Section\ref{sec:num-scan}
as well as in more detailed Figs.\ref{fig:dm-i2hdm}-\ref{fig:dm-i2hdm-relic}
in Appendix}
which show the effect of consequent application of constraints
from:  a) vacuum stability, perturbativity and unitarity;
b) electroweak precision data, LEP constraints and the LHC Higgs data;
c) relic density constraints constraints as well as constraints from LUX on DM from direct detection.
In this paper we have explored for the first time 
the parameter space where 
DM from the i2HDM is underabundant implying an additional source of DM,
using above constraints complemented by the  collider searches.
We have also explored  the parameter space in which the DM candidate of i2HDM represents 
100\% of the total DM budget of the Universe.
We found that the parameter space with  
$M_{h_1},M_{h_2}<45~\mbox{GeV} 
\mbox{\ or\ } M_{h^+}<70~\mbox{GeV}$
is completely excluded, confirming the first limit found previously
complemented the second one found in this study.

Though in general the parameter space of the i2HDM is 5-dimensional,
the parameter space relevant to the LHC mono-jet signature is only 1 or 2 dimensional,
so the model can be easily explored at the LHC.
There are two qualitatively different and complementary channels in mono-jet searches:
$pp\to h_1 h_1j$ and $pp\to h_1 h_2 j$,
with the second one being relevant to mono-jet signature
when the mass gap between $h_2$ and $h_1$
is of the order of a few GeV.
In the case of $h_1 h_2$ degeneracy, the rate for $pp\to h_1 h_1j$
will be effectively doubled since $g_{Hh_2 h_2} = g_{Hh_1h_1}$, see Eq.~(\ref{tildelam345}),
and this can be easily taken into accounts for the estimation of constraints
in the respective region of the parameter space.
For a fixed $M_{h_1}$, the strength of the first  process 
depends only on $\lambda_{345}$ because the Higgs boson is the only mediator,
while the  strength of the second  process 
is fixed by the weak coupling since the $Z$-boson is the  only mediator
for this process.
The last process is important to cover the $h_1 h_2$ co-annihilation region
available for 54 GeV $< M_{h_1} <73$ GeV,
where the relic density agrees with the PLANCK data.
The results on this process and on this region are new to our best knowledge.
Therefore these two processes complement each other in covering the parameter space:
for large values of $\lambda_{345}$, $pp\to h_1 h_1j$ would be the dominant LHC
signature, while for small or vanishing values of $\lambda_{345}$, the
$pp\to h_1 h_2 j$ process will cover additional parameter space
as demonstrated in Fig.~\ref{collider-XENON1T-constraint}--\ref{fig:dm-i2hdm-LHC-DD}.

Talking about quantitative results,
the LHC has rather limited potential to probe $M_{h_1}$
with the mono-jet signature. Even for the projected luminosity of 3 ab$^{-1}$,
we have found that the LHC could set a limit on $M_{h_1}$  up to 83 GeV
from the $pp\to h_1 h_1j$ process with the maximal value allowed for  $\lambda_{345}$
and  only up to 55 GeV from  $pp\to h_1 h_2j$
for any value of  $\lambda_{345}$, covering just the tip of the $h_1 h_2$ co-annihilation
region. Such a weak sensitivity of the LHC is related to
the similarity between the shapes of the \MET{} distribution of the dominant $Zj\to \nu\nu j$ background
and that of the signal which has the same $Z$-boson mediator, while the DM mass is not very different from $M_Z/2$,
which as shown in \cite{Belyaev:2016pxe} is the reason for such a similarity in \MET{} shape.
At the same time, the potential of the LHC using a search for disappearing charged tracks
is quite impressive  in probing  $M_{h_1}$ masses up to about 500 GeV
already at 8 TeV with 19.5 fb$^{-1}$ luminosity
as we have found in our study.

We have also explored the projected potential of XENON1T to probe the i2HDM
parameter space and have found that it is quite impressive,
confirming results of previous studies.
In our study we have presented  ``absolutely allowed"  and ``absolutely excluded" points
in different projections of the i2HDM 5D space demonstrating different features of the models
and the potential of current and future experiments.
In general, DM DD experiments and collider searches complement each other:
the $pp\to h_1 h_1j$ process covers in the region with large $\lambda_{345}$ coupling
where DM DD rates are low because of the low relic density re-scaling,
while the $pp\to h_1 h_2j$ process is sensitive to the parameter space with low $\lambda_{345}$
where DM DD rates are low because of the low rate of DM scattering off the nuclei.

\section*{Acknowledgements}

AB thanks Tania Robens and Alexander Pukhov for useful discussions.
AB and MT acknowledge partial support from the STFC grant number  ST/L000296/1
and the NExT Institute, FAPESP grant 2011/11973-4 for funding their visit to ICTP-SAIFR as well as SOTON-FAPESP collaboration grant.
AB thanks the Royal Society Leverhulme Trust Senior Research Fellowship LT140094.
MT acknowledges support from an STFC STEP award. 
GC acknowledges partial support from the Labex-LIO (Lyon Institute of Origins) under grant ANR-10-LABX-66 and FRAMA (FR3127, F\'ed\'eration de Recherche ``Andr\'e Marie Amp\`ere").
IPI acknowledges funding from \textit{Fun\-da\-\c{c}\~{a}o para a Ci\^{e}ncia e a Tecnologia} (FCT)
through the Investigator contract IF/00989/2014/CP1214/CT0004
under the IF2014 Program and through FCT contracts UID/FIS/00777/2013 and CERN/FIS-NUC/0010/2015,
which are partially funded through POCTI, COMPETE, QREN, and the European Union. 

\appendix

\section{Numerical scan: detailed discussion} \label{app:numerical}

To have a complete picture of the properties of i2HDM in the whole parameter space, we have performed a
five-dimensional random scan of the model parameter space with about $10^8$ points, evaluating all relevant
observables and limits mentioned above. The range for 
the model parameters of the scan was chosen according to the Eq.~(\ref{eq:scan-limits}).


To better delineate the impact of each constraint, we have imposed different cuts on the parameter space sequentially, 
following the classification below:
\begin{itemize}
\item[Cut-1:] theoretical constraints on the potential from vacuum stability 
[Eq.(\ref{eq:scalar-pot1}-\ref{eq:scalar-pot2}) and (\ref{eq:l345-vacuum-stab})], perturbativity  and
unitarity [Eq.(\ref{eq:unit}-\ref{eq:l345max})];
\item[Cut-2:] constraints from LEP [Eq.~(\ref{eq:constr-widths}) and (\ref{eq:dmh12})], EWPT
[Eq.~(\ref{eq:ewpt})] and the LHC Higgs data
[Eq.~(\ref{eq:lhc-higgs-invis}-\ref{eq:lhc-higgs-aa})];
\item[Cut-3:] constraint on the relic
density  [$\Omega_{\rm DM} h^2 \le 0.1184+2\times 0.012$], where we consider only the upper bound within 2 standard deviations;
\item[Cut-4:] constraints from DM DD searches from LUX.
\end{itemize}

\begin{figure}[htb]
\vskip -1.0cm
\hspace*{-0.3cm}\includegraphics[width=0.41\textwidth]{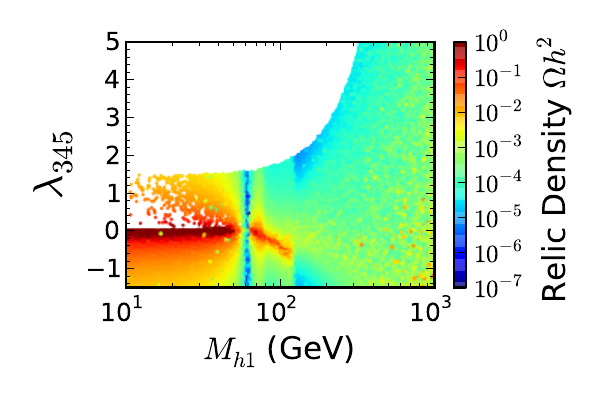}%
\hspace*{-1.78cm}\includegraphics[width=0.41\textwidth]{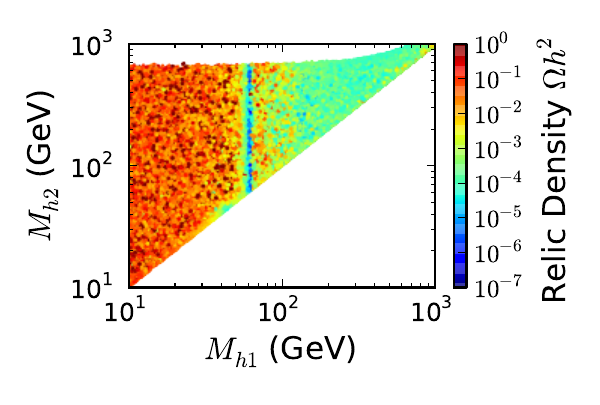}%
\hspace*{-1.77cm}\includegraphics[width=0.41\textwidth]{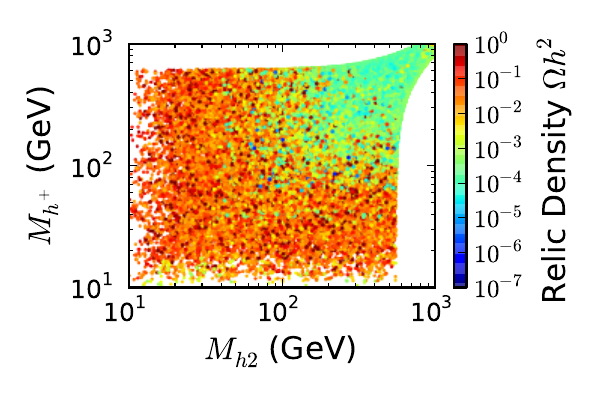}%
\vskip -1.1cm
\hspace*{1.2cm}(a)\hspace*{0.35\textwidth}\hspace*{-1.3cm}(b)\hspace*{0.35\textwidth}\hspace*{-1.2cm}(c)
\vskip 0.0cm
{\hspace*{-0.3cm}\includegraphics[width=0.41\textwidth]{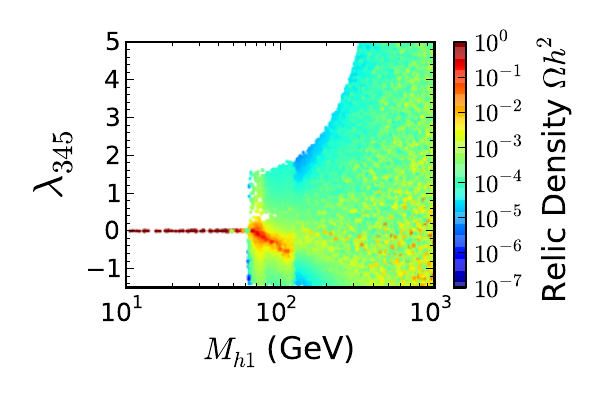}}%
{\hspace*{-1.78cm}\includegraphics[width=0.41\textwidth]{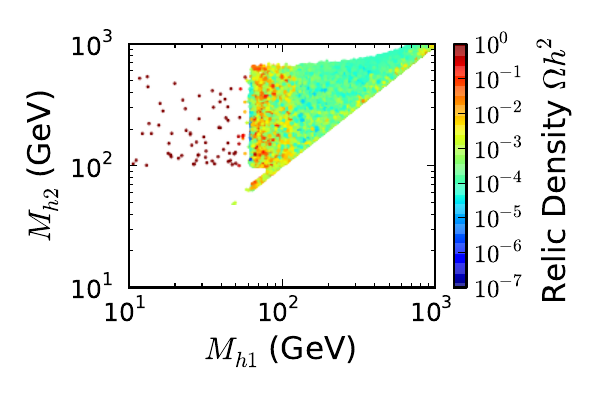}}%
{\hspace*{-1.77cm}\includegraphics[width=0.41\textwidth]{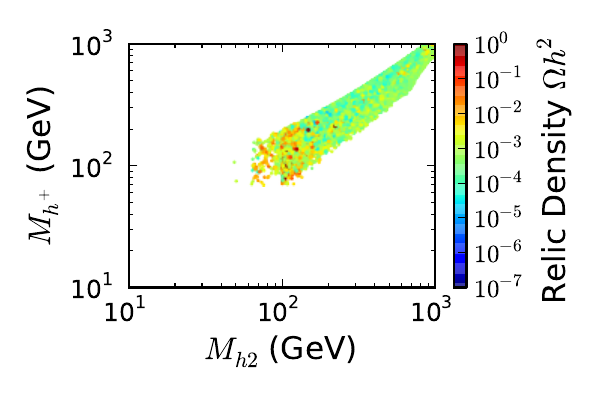}}%
\vskip -1.1cm
\hspace*{1.2cm}(d)\hspace*{0.35\textwidth}\hspace*{-1.3cm}(e)\hspace*{0.35\textwidth}\hspace*{-1.2cm}(f)
\vskip 0.0cm
{\hspace*{-0.3cm}\includegraphics[width=0.41\textwidth]{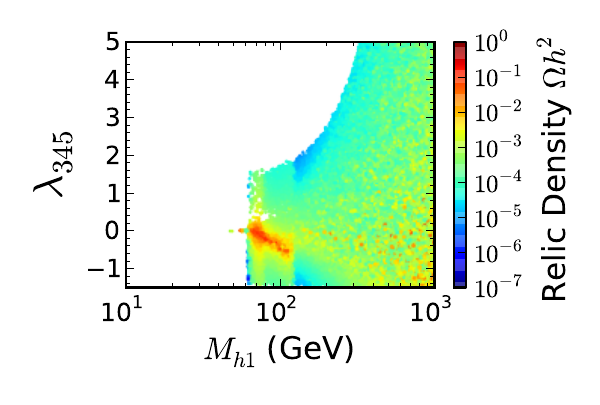}}%
{\hspace*{-1.78cm}\includegraphics[width=0.41\textwidth]{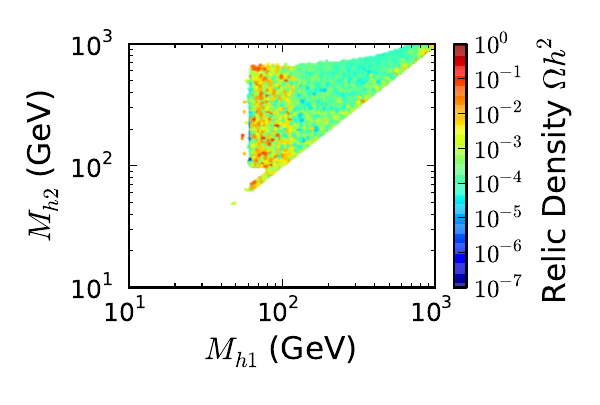}}%
{\hspace*{-1.77cm}\includegraphics[width=0.41\textwidth]{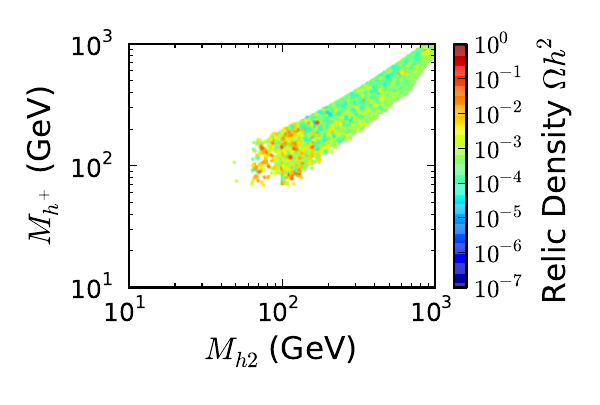}}%
\vskip -1.05cm
\hspace*{1.2cm}(g)\hspace*{0.35\textwidth}\hspace*{-1.3cm}(h)\hspace*{0.35\textwidth}\hspace*{-1.2cm}(i)
\vskip 0.0cm
{\hspace*{-0.3cm}\includegraphics[width=0.41\textwidth]{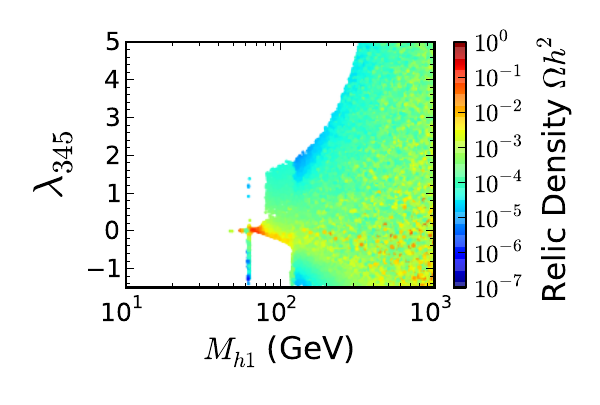}}%
{\hspace*{-1.78cm}\includegraphics[width=0.41\textwidth]{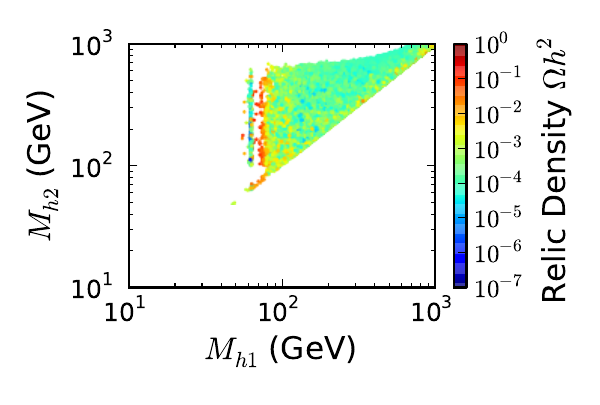}}%
{\hspace*{-1.77cm}\includegraphics[width=0.41\textwidth]{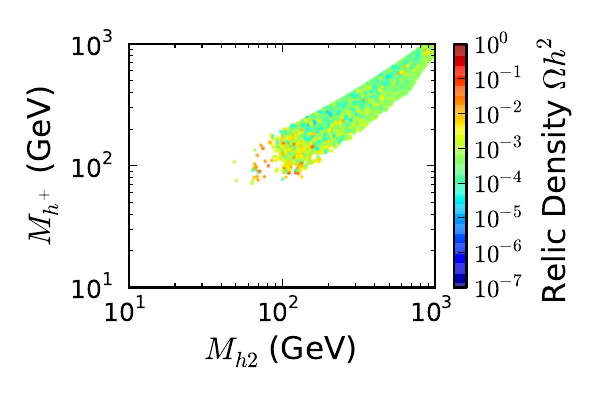}}%
\vskip -1.05cm
\hspace*{1.2cm}(j)\hspace*{0.35\textwidth}\hspace*{-1.3cm}(k)\hspace*{0.35\textwidth}\hspace*{-1.2cm}(l)
\caption{Colour maps of DM relic density for 2D projections of the 5D random scan of the i2HDM:  
each row demonstrates the effect of consequent application
of the experimental and theoretical constraints in the $(M_{h_1},\lambda_{345})$, $(M_{h_1},M_{h_2})$ and 
$(M_{h_1},M_{h^{+}})$ planes. Each row correspond to the Cut-1-4, described in the text: Cut-1 for (a-c) [Eqs.~(\ref{eq:scalar-pot1}-\ref{eq:l345min})]; Cut-2 for (d-f) [Eqs.~(\ref{eq:constr-widths}),(\ref{eq:dmh12}),(\ref{eq:ewpt}),(\ref{eq:lhc-higgs-invis}-\ref{eq:lhc-higgs-aa})]; Cut-3 for (g-i) [$\Omega_{\rm DM}^{\rm Planck} h^2 \le 0.1184+2\times 0.012$]; Cut-4 for (j-l) [LUX].
%
\label{fig:dm-i2hdm}} 
\end{figure}

\clearpage
\begin{figure}[htb]
\vskip -1.0cm
\hspace*{-0.3cm}\includegraphics[width=0.4\textwidth]{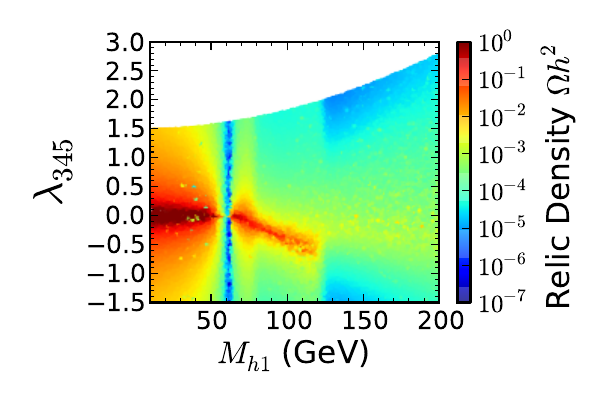}%
\hspace*{-1.55cm}\includegraphics[width=0.4\textwidth]{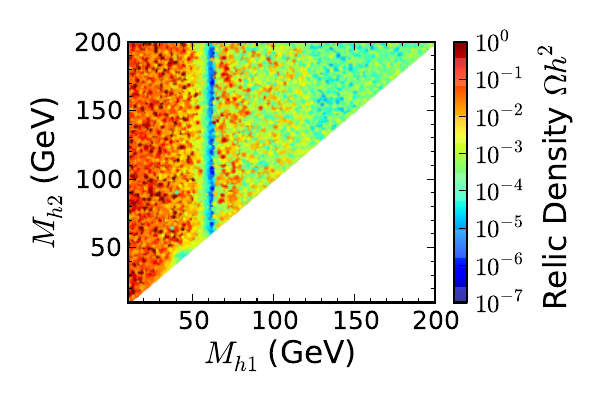}%
\hspace*{-1.55cm}\includegraphics[width=0.4\textwidth]{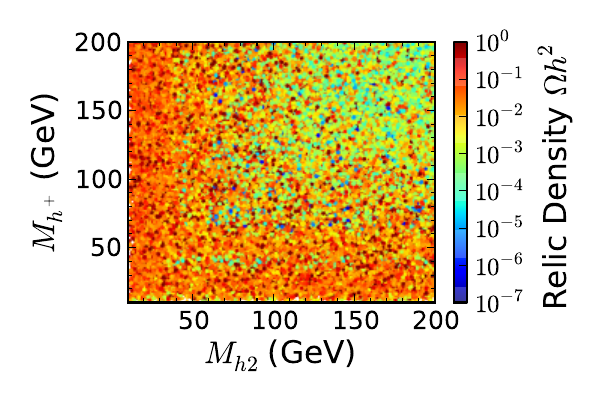}%
\vskip -5.2cm
\hspace*{4.9cm}\includegraphics[width=0.55cm,height=4.cm]{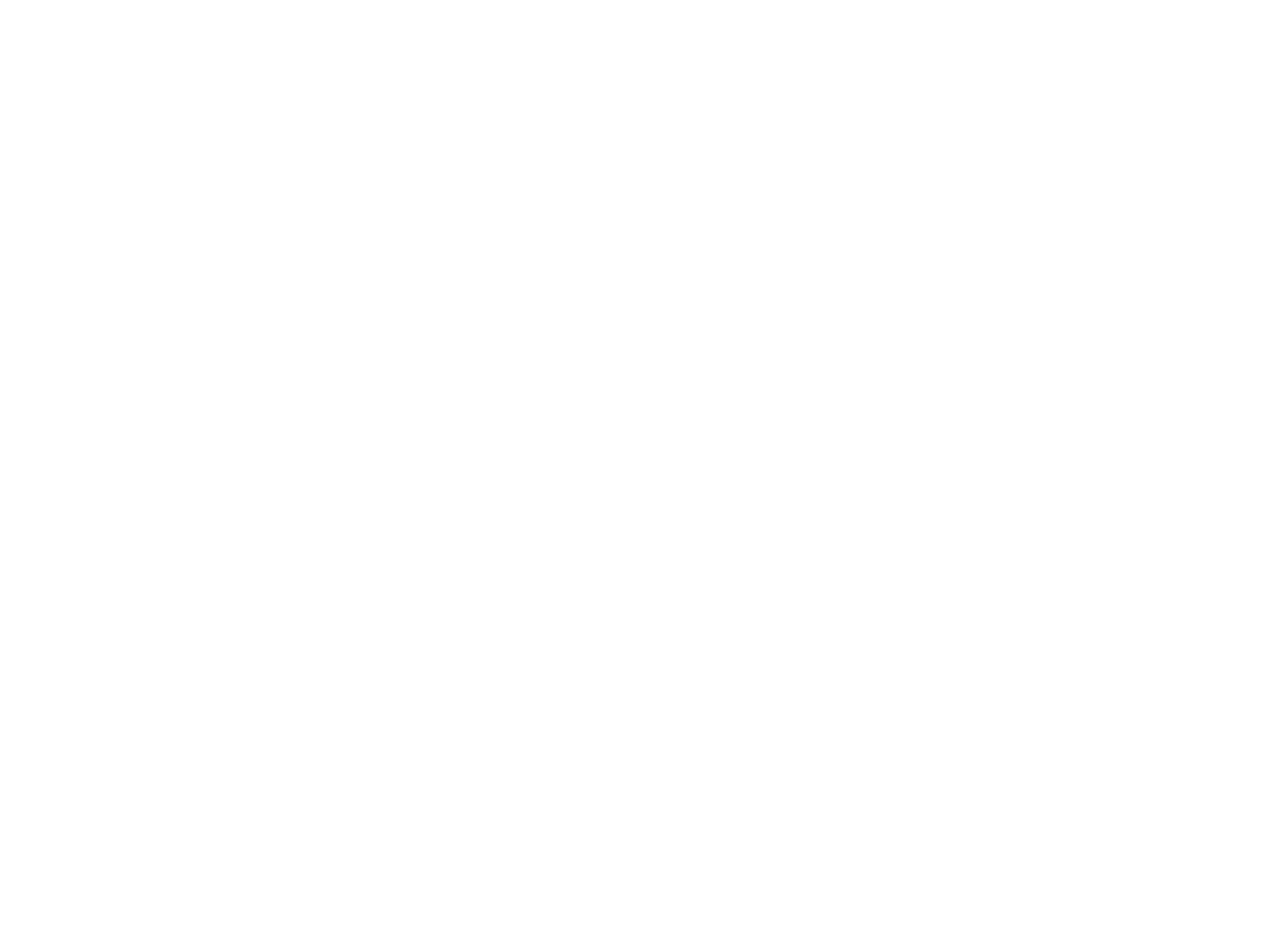}%
\hspace*{4.9cm}\includegraphics[width=0.55cm,height=4.cm]{blank.pdf}%
\vskip 0.2cm
\hspace*{1.4cm}(a)\hspace*{0.35\textwidth}\hspace*{-1.4cm}(b)\hspace*{0.35\textwidth}\hspace*{-1.5cm}(c)
\vskip  0.0cm
{\hspace*{-0.3cm}\includegraphics[width=0.4\textwidth]{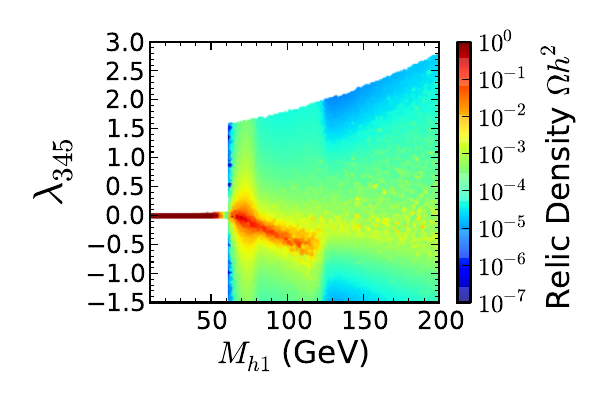}}%
{\hspace*{-1.55cm}\includegraphics[width=0.4\textwidth]{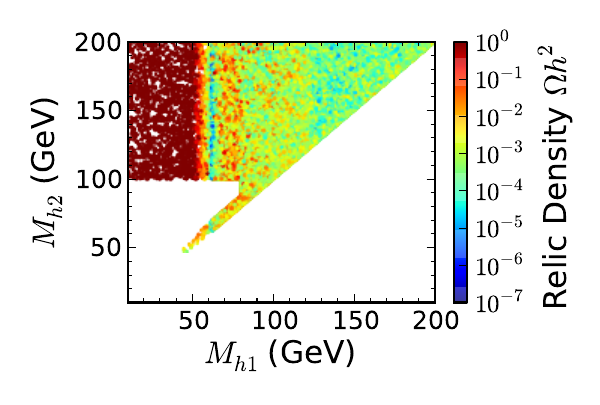}}%
{\hspace*{-1.55cm}\includegraphics[width=0.4\textwidth]{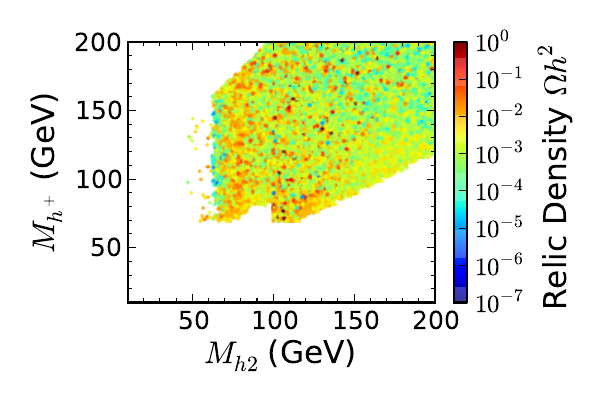}}%
\vskip -5.2cm
\hspace*{4.9cm}\includegraphics[width=0.55cm,height=4.cm]{blank.pdf}%
\hspace*{4.9cm}\includegraphics[width=0.55cm,height=4.cm]{blank.pdf}%
\vskip 0.2cm
\hspace*{1.4cm}(d)\hspace*{0.35\textwidth}\hspace*{-1.4cm}(e)\hspace*{0.35\textwidth}\hspace*{-1.5cm}(f)
\vskip 0.0cm
{\hspace*{-0.3cm}\includegraphics[width=0.4\textwidth]{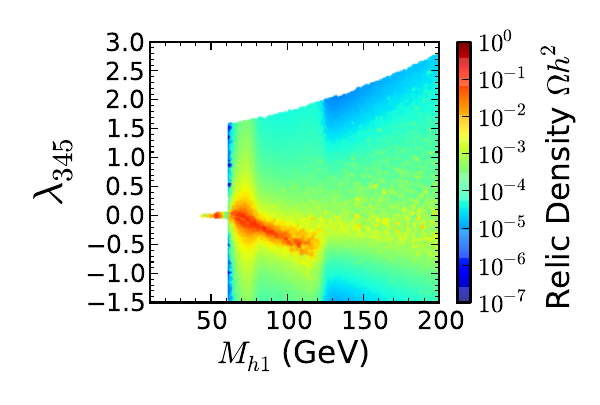}}%
{\hspace*{-1.55cm}\includegraphics[width=0.4\textwidth]{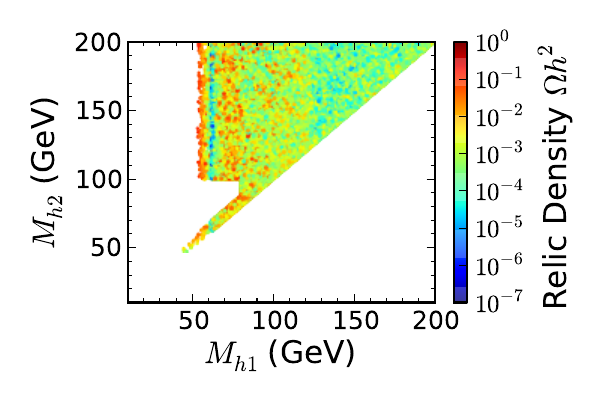}}%
{\hspace*{-1.55cm}\includegraphics[width=0.4\textwidth]{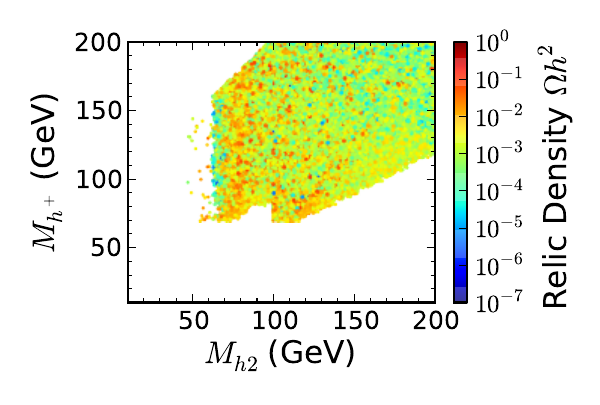}}%
\vskip -5.2cm
\hspace*{4.9cm}\includegraphics[width=0.55cm,height=4.cm]{blank.pdf}%
\hspace*{4.9cm}\includegraphics[width=0.55cm,height=4.cm]{blank.pdf}%
\vskip 0.2cm
\hspace*{1.4cm}(g)\hspace*{0.35\textwidth}\hspace*{-1.5cm}(h)\hspace*{0.35\textwidth}\hspace*{-1.6cm}(i)
\vskip 0.0cm
{\hspace*{-0.3cm}\includegraphics[width=0.4\textwidth]{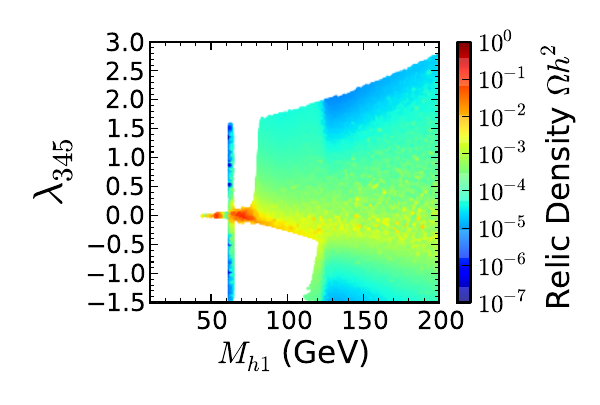}}%
{\hspace*{-1.55cm}\includegraphics[width=0.4\textwidth]{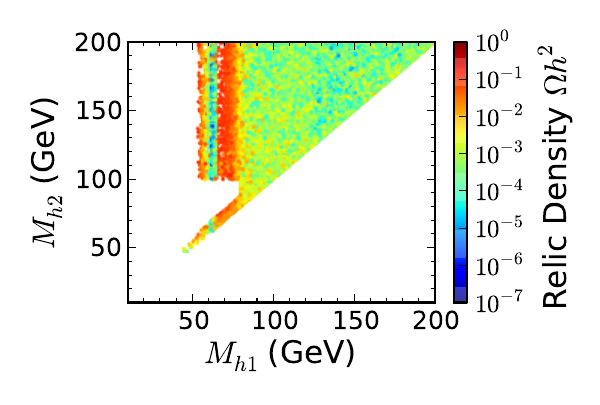}}%
{\hspace*{-1.55cm}\includegraphics[width=0.4\textwidth]{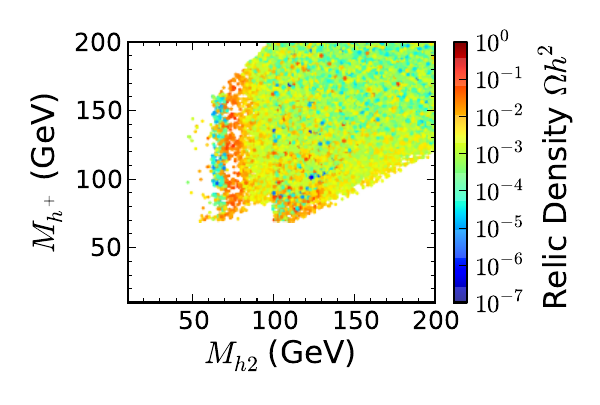}}%
\vskip -5.2cm
\hspace*{4.9cm}\includegraphics[width=0.55cm,height=4.cm]{blank.pdf}%
\hspace*{4.9cm}\includegraphics[width=0.55cm,height=4.cm]{blank.pdf}%
\vskip 0.2cm
\hspace*{1.4cm}(j)\hspace*{0.35\textwidth}\hspace*{-1.5cm}(k)\hspace*{0.35\textwidth}\hspace*{-1.6cm}(l)
\caption{Colour maps of DM relic density for 2D projections of the 5D random scan of the
i2HDM for the parameter space restricted to (10 GeV - 200 GeV) for $M_{h_1},M_{h_2}$ and
$M_{h^{+}}$. As for Fig.~\ref{fig:dm-i2hdm} each row correspond to the Cut-1-4, described in the text: Cut-1 for (a-c) [Eqs.~(\ref{eq:scalar-pot1}-\ref{eq:l345min})]; Cut-2 for (d-f) [Eqs.~(\ref{eq:constr-widths}),(\ref{eq:dmh12}),(\ref{eq:ewpt}),(\ref{eq:lhc-higgs-invis}-\ref{eq:lhc-higgs-aa})]; Cut-3 for (g-i) [$\Omega_{\rm DM}^{\rm Planck} h^2 \le 0.1184+2\times 0.012$]; Cut-4 for (j-l) [LUX].\label{fig:dm-i2hdm-small}}
\end{figure}

The results of the  scan are  presented in Fig.~\ref{fig:dm-i2hdm} in the form of a colour map of
DM relic density, projected on two-dimensional planes: $(M_{h_1},\lambda_{345})$ in the first,
$(M_{h_1},M_{h_2})$ in the second, and $(M_{h_2},M_{h^{+}})$ in the third column, respectively.
The four rows reproduce the effect of the progressive application of the four Cuts defined above.
In Fig.~\ref{fig:dm-i2hdm-small} we also present, in the same format, the results of a finer scan, zoomed to the region of low masses, where the range has been restricted to $10$--$200$ GeV for the three masses $M_{h_1},M_{h_2}$ and $M_{h^{+}}$. The latter is the most relevant corner of parameter space for the LHC phenomenology that we will discuss in the next section. Note that the lower bound of $\lambda_{345}$ presented in these plots corresponds to the lowest limits allowed by unitarity, perturbativity and scalar potential constraints (see Fig.~\ref{fig:par-space1}).


One can see from Figs.~\ref{fig:dm-i2hdm}-\ref{fig:dm-i2hdm-small}(a) that $\lambda_{345}$ is limited from above,
and the dependence which defines the  shape of this limit as a function of $M_{h_1}$ comes from 
the vacuum stability condition given by Eq.(\ref{eq:l345max}). One can also see from
Figs.~\ref{fig:dm-i2hdm}-\ref{fig:dm-i2hdm-small}(a), (b), and analogous figures in the rows below, that the relic density
is too high for small $M_{h_1}$ values and small $\lambda_{345}$.
Therefore, the relic density
constraint combined with the LHC Higgs data constraints (limiting the invisible decays of the Higgs) restricts $M_{h_1}$ to be above 45 GeV,
as it can be clearly seen from 
Figs.~\ref{fig:dm-i2hdm}-\ref{fig:dm-i2hdm-small}(g) and (h).
For example, the range 45 GeV $<M_{h_1}<$ 50 GeV is allowed but it requires $h_1h_2$ co-annihilation
and respective mass degeneracy, as one can see from Figs.~\ref{fig:dm-i2hdm}-\ref{fig:dm-i2hdm-small}(h) and (k). From
Figs.~\ref{fig:dm-i2hdm}-\ref{fig:dm-i2hdm-small}(a), (b) and analogous ones in the rows below, one can see a clear vertical blue
pattern of low relic density corresponding to the $h_1 h_1 \to H$ resonant annihilation.  For $M_{h_1}>M_H/2$
the pattern of DM relic density follows the pattern of  $WW$, $ZZ$ and $HH$ thresholds 
presented earlier in Fig.~\ref{fig:1d-mh1-Omega}.

One can also observe that the effect of Cut-1 plus Cut-2 is quite dramatic: a)
$Br(H\to h_1 h_1)<0.28$ and $\mu^{\gamma\gamma} = 1.14^{+0.38}_{-0.36}$ constraints
require $\lambda_{345}\leq 0.02$ for $M_{h_1}<M_H/2$
[Figs.~\ref{fig:dm-i2hdm}-\ref{fig:dm-i2hdm-small}(d)]; b) LEP constraints require $M_{h_2}\gtrsim 100$ GeV if 
$M_{h_2}-M_{h_1}>8$~GeV [Figs.~\ref{fig:dm-i2hdm}-\ref{fig:dm-i2hdm-small}(e)]; c) LEP and LHC Higgs data constraints 
require  $M_{h^+}>70$~GeV, while $M_{h_2}$ is generically excluded below $M_Z/2$ [Figs.~\ref{fig:dm-i2hdm}-\ref{fig:dm-i2hdm-small}(f)].
The effect from adding the (upper) cut from relic density (Cut-3)
is shown in Figs.~\ref{fig:dm-i2hdm}-\ref{fig:dm-i2hdm-small}(g-i): one can see that this cut
(combined with the previous ones)
excludes $M_{h_1}<M_Z/2$
for the whole i2HDM parameter space [Figs.~\ref{fig:dm-i2hdm}-\ref{fig:dm-i2hdm-small}(g,h)],
but does not have a visible effect in  $(M_{h_2},M_{h^{+}})$ plane [Fig.\ref{fig:dm-i2hdm}-\ref{fig:dm-i2hdm-small}(i)].
Actually the  region with $M_{h_1}<M_Z/2$ is excluded due to the interplay of several constraints.
In the  $M_{h_1}<M_H/2$ region with $|\lambda_{345}|\lesssim 0.02$ as required by LHC Higgs data,
the only possibility for relic density of $h_1$ to be sufficiently low to satisfy the PLANCK constraints
is the $h_1 h_2$ co-annihilation channel:
potentially this co-annihilation could provide low enough relic density 
for $M_{h_1}$ down to about 20 GeV. However,  for $M_{h_1}+M_{h_2}< M_Z$
the $Z\to h_1 h_2$ decay is open and contributes significantly to the invisible
Z-boson decay, that is strongly limited by LEP. As the $Z$-boson partial width for this decay channel
is defined just by $M_{h_1}$ and $M_{h_2}$, since $Z h_1 h_2$ coupling is fixed by the gauge invariance,
the $M_{h_1}+M_{h_2}< M_Z$ parameter space is completely excluded. 
For  $h_1 h_2$ co-annihilation region, this exclusion is equivalent to   $M_{h_1}, M_{h_2} \gtrsim M_Z/2$.
The $h_1 h_2$ co-annihilation corridor which provides relic density
below or equal to PLANCK limit is clearly visible in Figs.~\ref{fig:dm-i2hdm}-\ref{fig:dm-i2hdm-small}(e,h,k).

\begin{figure}[htb]
\hspace*{-0.2cm}\includegraphics[width=0.50\textwidth]{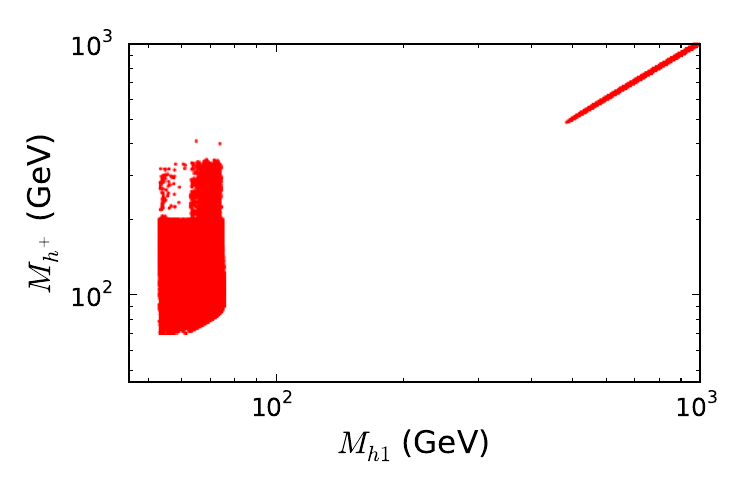}%
\hspace*{-0.2cm}\includegraphics[width=0.50\textwidth]{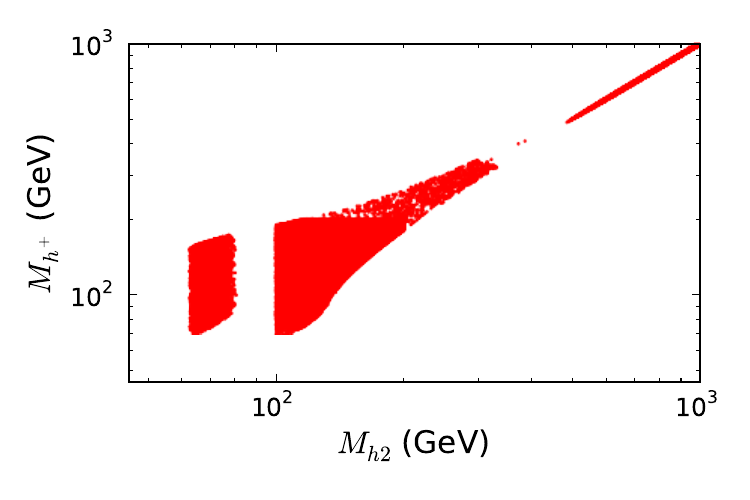}
\vskip -1.0cm
\hspace*{1cm}(a)\hspace*{0.55\textwidth}\hspace*{-1.5cm}(b)
\\
\\
\hspace*{-0.2cm}\includegraphics[width=0.50\textwidth]{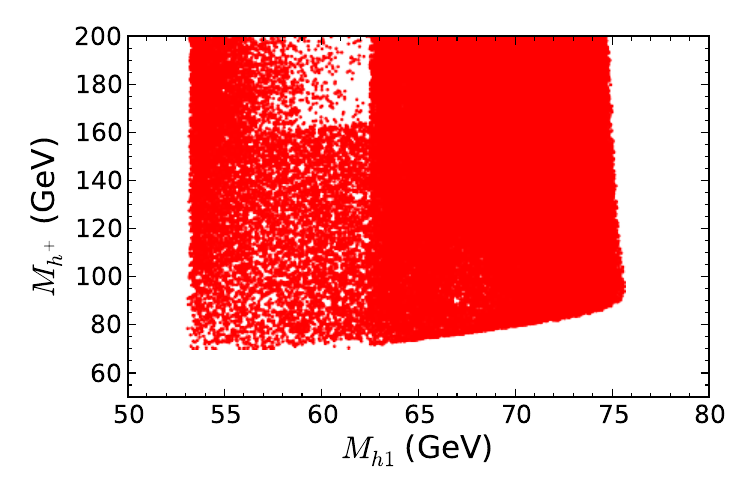}%
\hspace*{-0.2cm}\includegraphics[width=0.50\textwidth]{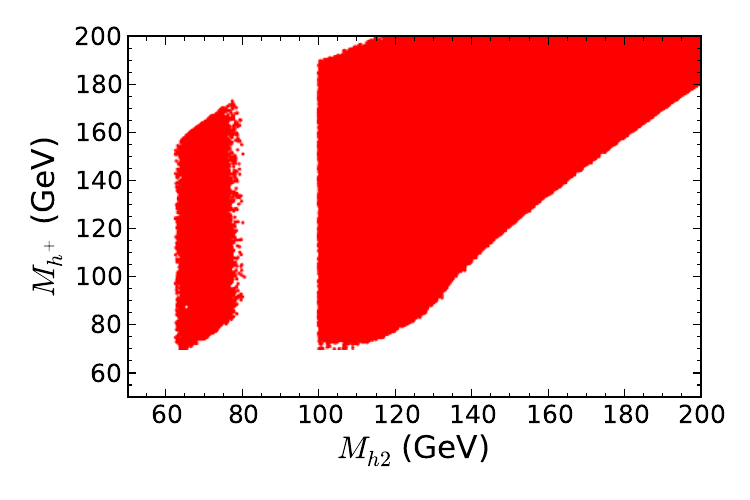}
\vskip -0.8cm
\hspace*{1cm}(c)\hspace*{0.55\textwidth}\hspace*{-1.5cm}(d)
\caption{2D projections of the random scan of the i2HDM for points satisfying all constraints and providing the correct relic abundance within 2 sigma of the PLANCK result.
 The top row corresponds to the ``full'' parameter space 10 GeV $< M_{h_1}, M_{h_2}, M_{h^{+}} < 1000$~GeV,
 while the lower row refers to the ``zoomed'' parameter space
10 GeV $< M_{h_1}, M_{h_2}, M_{h^{+}} < 200$~GeV.}
\label{fig:dm-i2hdm-relic}
\end{figure}

The additional constraint from DM DD from LUX (Cut-4)
removes a substantial portion of the  parameter space for large and intermediate $|\lambda_{345}|$
values for $M_{h_1}\lesssim M_H$  [Figs.~\ref{fig:dm-i2hdm}-\ref{fig:dm-i2hdm-small}(j)].
In this excluded parameter space the scattering cross section of $h_1$ on the proton 
is quite large due to the Higgs boson exchange enhanced by $|\lambda_{345}|$, while the relic density is respectively low,
again due to the large value  $|\lambda_{345}|$, but it is not low enough to suppress
the DM detection rate below the experimental exclusion.
So, the LUX cut removes the low relic density region, 
and one can see this clearly in Figs.~\ref{fig:dm-i2hdm}-\ref{fig:dm-i2hdm-small}(k-l)
by the enhanced yellow-red colour in the $M_{h_1}\lesssim M_H$ region 
in comparison to the respective Figs.~\ref{fig:dm-i2hdm}-\ref{fig:dm-i2hdm-small}(h-i)
where the DM DD cut was not applied.
For $\lambda_{345} \gtrsim 0.2$ the parameter space is excluded
for $M_H/2<M_{h_1}<M_W$ while for  $\lambda_{345} \lesssim -0.2$
it is excluded for $M_H/2<M_{h_1}<M_H$ as illustrated in Figs.~\ref{fig:dm-i2hdm}-\ref{fig:dm-i2hdm-small}(j).
Once the $h_1 h_1 \to W^+ W^-$ channel is open for positive  $\lambda_{345}$,
or $h_1 h_1 \to H H$ channel is open for negative $\lambda_{345}$,
the relic density drops substantially below the PLANCK limit, which makes the rescaling factor 
low enough to avoid limits from LUX searches.
The difference between the positive and negative $\lambda_{345}$ cases
is related to the respective positive and negative interference of
$h_1 h_1 \to H \to X X$ channel with non-Higgs-exchange diagrams.
This asymmetry between positive and negative $\lambda_{345}$ cases was seen initially
in Fig.~\ref{fig:1d-mh1-Omega}, where the $h_1$ relic density was presented as a function 
of $M_{h_1}$ for different $\lambda_{345}$ values.

We would also like to point to some features of the scan for  the region of $M_{h_1},M_{h_2}$ and
$M_{h^{+}}$ above 200 GeV, presented  in Fig.~\ref{fig:dm-i2hdm}. From 
Figs.~\ref{fig:dm-i2hdm}(f),(i),(l), one can see that EWPT constraints require a very  modest
mass split between $M_{h_2}$ and $M_{h^{+}}$ since this mass split is directly related to
values of the $M_{h_2}$ and $M_{h^{+}}$ couplings to  the SM Higgs as well as to the couplings
to longitudinal components of the W and Z-bosons. Therefore constraints from $S$ and $T$
parameters leave only a rather narrow corridor in the  ($M_{h^{+}},M_{h_2}$) plane.

Finally, for the case, when the relic density is required to fit the PLANCK result within 2 sigma,
 2D projections on the $(M_{h_1}, M_{h^+})$ and $(M_{h_2}, M_{h^+})$ are shown in Fig.~\ref{fig:dm-i2hdm-relic}. The plots in the top row show results for the ``full'' scan 
10 GeV $< M_{h_1}, M_{h_2}, M_{h^{+}} < 1000$~GeV, while in the lower row we present the ``zoomed'' scan 10 GeV $< M_{h_1}, M_{h_2}, M_{h^{+}} < 200$~GeV.
These plots complement the information on the surviving regions given by Fig.~\ref{fig:scan-simplified-fitting}
in the main text.

\bibliography{bib}
\bibliographystyle{bib}

\end{document}